\documentclass[12pt]{article}

\usepackage{amsmath,amssymb,graphicx,jheppub,comment}
\usepackage{tikz}
\usetikzlibrary{decorations.pathmorphing}
\title{\Large Towards celestial CFT dual of 4d conformal gravity}

\author{Nirmal Ghorai, Partha Paul and Nemani V. Suryanarayana}
\affiliation[]{The Institute of Mathematical Sciences, \\ IV Cross Road, CIT Campus, Taramani, Chennai, India
600113. \\ \&}

\affiliation[]{Homi Bhabha National Institute, \\ Anushakti Nagar, Mumbai 400094, India. \\}
\emailAdd{nirmalg@imsc.res.in}
\emailAdd{parthapaul@imsc.res.in}
\emailAdd{nemani@imsc.res.in}

\abstract{We compute tree-level celestial operator product expansions (OPE) in a bosonic sub-sector of the Berkovits-Witten conformal supergravity from the scattering amplitudes in the MHV configuration. While the OPE between a leading soft graviton current for a positive helicity graviton and any of the  primary operators exhibits the same singularity structure as in a gravitational theory with two-derivative kinetic terms, the OPE of a subleading soft graviton current with a positive helicity hard graviton primary operator receives corrections, as a consequence of the non-universal nature of the subleading soft graviton theorem in the bulk. Remarkably, the subleading soft graviton terms remain consistent with the Ward identity of the chiral $\mathfrak{sl}(2,\mathbb{R}) $ current algebra, albeit with a different realisation where particle-changing operators play a role. Our analysis suggests that the dual celestial CFT continues to enjoy at least the chiral $\mathfrak{bms}_4$ symmetry, though in a non-trivial way, and possibly a conformal extension of it. }

\begin{document}
\maketitle

\section{Introduction}
Soft theorems in the context of scattering amplitudes in theories with massless particles such as photons, gluons and gravitons \cite{Low:1954kd,Low:1958qz,Weinberg:1965kv,Cachazo:2014kd,Sen:2017xjn,Sen:2017gln,Elvang:2016qvq,Laddha:2017vfh,Laddha:2017kpo,Laddha:2018kbm,Strominger:2013lka,Casali:2014xpa} have lead to significant understanding towards the symmetries of those theories  \cite{Strominger:2017zoo,He:2014laa,Kapec:2014opa,He:2014cra,Lysov:2014csa,Campiglia:2014yka,Kapec:2014zla,Avery:2015gxa,Strominger:2015bla,Campiglia:2015lxa,Avery:2015rga,Banerjee:2020zlg,Banerjee:2021dlm,Dumitrescu:2015fej,Avery:2015iix,Lysov:2015jrs,Campiglia:2016hvg,Campiglia:2016efb,He:2017fsb,Campiglia:2017dpg,Laddha:2017vfh,Liu:2021dyq,Campiglia:2016jdj,Gupta:2021cwo}. In particular, in the context of gravitational theories that are Einstein-type (that is, their actions take the form of Einstein-Hilbert one with correction terms) it is established (at least at tree-level in 4$d$ and in general at higher $d$) that the leading and subleading \cite{Weinberg:1965kv,Cachazo:2014kd,Sen:2017xjn,Sen:2017gln} soft terms are universal. However, there are interesting gravitational theories that are not of the Einstein-type. One such class of these includes conformal gravities in 4$d$. Although they are not believed to be good theories (because of the presence of ghost degrees of freedom), they may exhibit good ultraviolet (UV) behaviour and are counted (see, for instance, \cite{Fradkin:1985am,Hamada:2009hb}) among examples of renormalisable theories of gravity. While their UV properties have been considered before, the infrared (IR) sector of these theories remains largely unexplored. It is, therefore, important to investigate the soft behaviour of tree-level MHV scattering amplitudes of such theories and the symmetries responsible for them, with the aim to extract some of the essential features of their holograms.

In particular, we focus on the Berkovits-Witten (BW) theory \cite{Berkovits:2004jj}, a superconformal gravity whose field content arises from a specific twister-string theory. The tree-level scattering amplitudes of this theory have been studied in \cite{Berkovits:2004jj,Johansson:2017srf,Johansson:2018ues}. By focusing on a particular bosonic sub-sector of the BW-theory, the authors of \cite{Johansson:2017srf,Johansson:2018ues} showed that the tree-level scattering amplitudes can be obtained from the double copy of two gauge theories. The gauge theories are (super-) Yang-Mills theory and a gauge theory with a four-derivative kinetic term of the form $(DF)^2$. We consider this particular sector of the BW theory.\footnote{Though we work with this bosonic sub-sector of the BW-theory, for brevity we will continue to refer to it simply as the `BW-theory'.}
We perform the leading and subleading soft graviton expansion of the tree-level MHV amplitudes of the BW theory and show that they still follow as a consequence of the chiral supertranslations and chiral $\mathfrak{sl}(2,{\mathbb R})$ current algebra symmetries \cite{Banerjee:2020zlg}. However, somewhat interestingly, the realisation of the $\mathfrak{sl}(2, {\mathbb R})$ current algebra is quite different from the usual, and involves new representations that use particle changing operators of the type seen by the authors in \cite{Elvang:2016qvq,Laddha:2017vfh} in quite different contexts. 

Another motivation for studying this theory comes from celestial holography. The conjecture for celestial holography states that any quantum theory of gravity in an asymptotically flat spacetime is dual to a conformal field theory (CFT) on the celestial sphere at null infinity, referred to as the celestial CFT \cite{Strominger:2017zoo,Kapec:2014opa,Kapec:2016jld,He:2017fsb,Ball:2019atb,Pasterski:2016qvg,Pasterski:2017kqt,deBoer:2003vf,Cheung:2016iub}. The correlation functions of primary operators in the celestial CFT, known as celestial amplitudes (sometimes called Mellin amplitudes), recast the $S$-matrix elements in a basis of boost eigenstates. For massless scattering, this change of basis is achieved by Mellin transformation with respect to the energies of the external massless states \cite{Pasterski:2017kqt,Banerjee:2018gce}. 
A useful way to study various aspects of a celestial CFT is through the construction of celestial operator product expansions (OPE) \cite{Banerjee:2020zlg,Pate:2019lpp,Banerjee:2020kaa,Banerjee:2021cly,Himwich:2021dau,Banerjee:2021dlm,Guevara:2021abz,Strominger:2021lvk,Banerjee:2023zip,Banerjee:2020vnt,Ebert:2020nqf,Banerjee:2023rni,Adamo:2022wjo,Ren:2023trv,Hu:2021lrx,Bhardwaj:2022anh,Krishna:2023ukw}. Usually in a generic CFT, the OPE coefficient that multiplies a primary operator cannot be determined using the conformal symmetry. However, what is remarkable about a celestial CFT is that in some cases one can determine these OPE coefficients using symmetry considerations alone. More specifically, let us consider two primary operators of conformal weights $(h_1, \bar h_1)$ and $ (h_2, \bar h_2) $ in the celestial CFT. The contribution to the OPE between these two primary operators from any other primary with conformal weights $(h_p, \bar h_p)$ is given schematically by,
\begin{equation}
    \mathcal{O}_{h_1, \bar h_1}(z_1, \bar z_1)  \mathcal{O}_{h_2, \bar h_2}(z_2, \bar z_2) \sim \sum_{p} C_{12p} \, z_{12}^{h_p-h_1-h_2} \bar z_{12}^{\bar h_p -\bar h_1-\bar h_2} \mathcal{O}_{h_p, \bar h_p}(z_2, \bar z_2) 
\end{equation}
where the sum is over all primary operators in the theory. Using the symmetry algebra, the leading singular structure in the above OPE can be completely fixed in some cases. For example, the leading singular term in the OPE between two graviton primary operators in the MHV-sector of Einstein gravity can be completely determined using the chiral supertranslations and chiral $\mathfrak{sl}(2,{\mathbb R})$ current algebra symmetries \cite{Banerjee:2020zlg}, and is given by,
\begin{equation} \label{ein_grav}
G^{++}_{\Delta_1}(z_1,\bar z_1) G^\sigma_{\Delta_2}(z_2,\bar z_2) \sim - \frac{\bar z_{12}}{z_{12}}  B(\Delta_1-1,\Delta_2-\sigma+1) G^\sigma_{\Delta_1+\Delta_2}(z_2,\bar z_2) 
\end{equation}
where $G^\sigma_\Delta(z, \bar z) $ is a spin-2 (graviton) primary operator with helicity $\sigma$ and dimension $\Delta$ inserted at the point $(z, \bar z)$ on the celestial sphere.\footnote{The celestial OPE for MHV sector of Einstein gravity is actually consistent with a bigger symmetry algebra, namely the chiral $\mathfrak{bms}_4$ \cite{Banerjee:2021dlm,Gupta:2021cwo}, that is generated by a chiral stress tensor $T(z)$ along with the chiral supertranslation charges and the chiral $\mathfrak{sl}(2, {\mathbb R})$ currents of \cite{Banerjee:2020zlg}.} The OPE \eqref{ein_grav} can also be obtained by Mellin transforming the collinear singularities of the gravitational scattering amplitudes in the bulk Einstein-type gravity, {\it i.e.} the theories with $p^{-2}$ propagators and the bulk scaling dimension of the three-point vertex equal to $5$         \cite{Pate:2019lpp,Himwich:2021dau}. However, does it necessarily imply that the converse is always true? That is, does the OPE \eqref{ein_grav} always imply that the corresponding bulk theory must be a two-derivative theory of gravity, even if the symmetry algebra remains the same? 
This is an important question, as answering this would allow one to differentiate between an Einstein-type theory and a higher derivative (and potentially non-unitary) theory in the bulk by looking at the celestial OPE. 

One possible place this diagnostic deviation can arise is in the OPE between soft graviton currents and other primary operators, and whether the conformal soft graviton theorems have been modified or not. Conformal soft theorems, for celestial amplitudes, are obtained by Mellin transforming the momentum space soft theorems where the poles in the soft energy translates to poles in the conformal weight \cite{Donnay:2018neh,Pate:2019mfs,Fan:2019emx,Nandan:2019jas,Adamo:2019ipt,Puhm:2019zbl,Guevara:2019ypd,Pano:2021ewd}. Thus, any change in the conformal soft graviton theorems on the celestial sphere will indicate modifications in the momentum space soft theorems in the bulk. Now, the arguments for the universality of the leading and subleading soft graviton theorems in any unitary effective field theory, including Einstein gravity, use the fact that the graviton propagator goes as $p^{-2}$  \cite{Weinberg:1965kv,Cachazo:2014kd,Sen:2017xjn, Sen:2017gln, Elvang:2016qvq}. However, if the graviton propagator in a theory behaves differently (i.e, $\nsim p^{-2}$) and there are operators with three-point interactions  that can change the particle nature in the lower point amplitude at leading and subleading orders of the soft expansion, then it is not necessary that the leading\footnote{Please see the discussion section for some additional comments.} and subleading soft graviton theorems continue to hold.  Therefore, exploring the soft behaviour of tree-level MHV scattering amplitudes of the BW theory whose propagator goes as $p^{-4}$ provides a crucial example in this regard. 

For this purpose we use the known expressions of the MHV amplitudes in BW theory from \cite{Berkovits:2004jj,Johansson:2017srf,Johansson:2018ues} and compute the celestial OPE between two different primary operators (a graviton or a scalar), and between a soft current and a primary operator in the BW theory. We find that the OPE between a leading soft graviton current (for positive helicity) and a hard graviton/scalar primary operator maintains the same singularity structure as in an Einstein-type theory. However, the OPE involving a subleading soft graviton current and a hard graviton primary operator receives corrections via some additional terms. These corrections modify the conformal subleading soft graviton theorem indicating that the bulk subleading soft graviton theorem is altered by additional terms, that can be recast in terms of the lower point amplitudes replacing the particle of the type going soft by entirely another type of particle (such as a graviton being replaced by a scalar). This raises the question of whether these amplitudes respect at least the chiral $\mathfrak{bms}_4$ algebra or not. Recall that the chiral $\mathfrak{bms}_4$ symmetries are sufficient to show that the leading and subleading soft theorems hold in Einstein-type theories. We show that even with modification of the subleading soft theorem these celestial amplitudes continue to respect the chiral $\mathfrak{bms}_4$ symmetries.

To demonstrate that this phenomenon of theories with local symmetries but with non-standard kinetic terms (propagators) still give rise to interesting realisations of the asymptotic symmetry algebras, albeit with different representations than in theories with standard kinetic terms, we examine another theory, namely the $DF^2$ theory of Johansson et al \cite{Johansson:2017srf,Johansson:2018ues}. Here too we show that, even though the leading soft gluon theorem gets modified, it does so in a remarkable way to keep the symmetry algebra to be still the current algebra version of the gauge group. Again, curiously enough, we find that particle changing operators appear at the leading soft expansion where, upon a gluon becoming soft a scalar participating in the lower point amplitude turns into a gluon.  

The rest of the paper is organised as follows. In section \ref{cga}, we discuss the tree-level scattering amplitudes in the BW theory, particularly focusing on the 6- and 5-point MHV amplitudes required for our OPE analysis. By Mellin transforming these amplitudes, in section \ref{OPE_from_amp} we write them as correlation functions on the celestial sphere and extract the OPE between different primary operators. The section \ref{summary} involves a summary of the OPEs in the celestial CFT dual of the BW theory and their implications to the bulk. In section \ref{soft_fac}, we explicitly show, by working out the soft expansion of a generic $(n+1)$-point MHV amplitude in detail, that the subleading soft graviton theorem is modified. In section \ref{algebra}, we show that though the subleading soft graviton theorem is corrected, the chiral $\mathfrak{sl}(2,\mathbb{R})$ current algebra symmetry remains unchanged. We end the paper with a discussion and future directions in section \ref{discussion}. Appendix \ref{review} briefly reviews the modified Mellin transform for massless scattering amplitudes. In appendix \ref{delta_param}, we provide the parameterisation for 5- and 6-point momentum conserving delta functions useful for OPE decomposition of scattering amplitudes. In appendix \ref{ope_cal}, we give some details of the higher order OPE computation. In appendix \ref{con_bms4}, we construct the chiral conformal $\mathfrak{bms}_4$ algebra which is a conformal extension of chiral $\mathfrak{bms}_4$ algebra. Finally, in appendix \ref{dfsquare} we sketch our analysis of leading soft gluon theorem of $(DF)^2$ theory.

\section{Conformal gravity amplitudes}
\label{cga}

In this section, we will briefly summarise the essential details of the BW theory and its tree-level MHV scattering amplitudes of bosonic particles of our interest.

The simplest example of conformally invariant gravitational theories is obtained by considering fluctuations of the Weyl invariant theory with Lagrangian given by the square of the Weyl tensor around the Minkowski spacetime. This is a four-derivative theory that consists of a physical spin-2 graviton and associated spin-2 and spin-1 massless ghosts. In this theory, the tree-level amplitudes of physical gravitons vanish \cite{Maldacena:2011mk,Adamo:2013tja,Adamo:2016ple}, and hence we will not consider it in this work. However, the pure Weyl$^2$ theory can be generalised in various ways. One such example is a bosonic extension of the theory where one non-minimally couples a complex scalar to (the self-dual and the anti-self-dual parts of) the Weyl tensor, keeping the Weyl invariance unbroken (see \cite{Johansson:2017srf} for details). The complex scalar $\Phi$ is made up of a dilaton $\phi(x)$ and a pseudo-scalar axion field, $a(x)$, $\Phi(x) = \phi(x)+i a(x)$.  This bosonic theory can be considered as a sub-sector of the Berkovits-Witten non-minimal $\mathcal{N}=4$ conformal supergravity theory. That is, we consider the tree-level amplitudes of the Berkovits-Witten theory, given by the top and bottom components of the $\mathcal{N}=4$ supermultiplet. For the sake of the reader's convenience, we state the compact formula for the tree-level superamplitudes of this theory, which is given by \cite{Johansson:2018ues},
\begin{equation} \label{sup_amp}
    M_n^{\textnormal{BW} \, \textnormal{CSG}}(\mathcal{H}_1^+, \cdots, \mathcal{H}_k^+, \mathcal{H}_{k+1}^-, \cdots, \mathcal{H}_n^- ) = (-1)^n i \delta^{8}(Q) \prod_{i=1}^k \sum_{j=1,j\neq i}^n \frac{[i,j]\left< j,q\right>^2}{\left< i,j \right> \left< i,q \right>^2}
\end{equation}
where $q$ is a reference spinor and $\delta^{8}(Q) = \delta^{8}(\sum_i \lambda_i^\alpha \eta_i^I)$ is the usual supermomentum conserving delta function in terms of on-shell spinors $\lambda_i^\alpha$ and Grassmann vaiables $\eta^I_i$. Here, $I,J,\ldots $ are fundamental indices of $ \textnormal{SU}(4) $ R-symmetry group, and $\mathcal{H}^{\pm}$ are the $\mathcal{N} =4$ conformal supermultiplets given by,
\begin{equation}\label{grav_multiplet}
    \begin{split}
        \mathcal{H}^+ &= h^{++} + \eta^I \psi^+_I + \frac{1}{2} \eta^I \eta^J A^+_{IJ} + \frac{1}{3!} \epsilon_{IJKL} \eta^I \eta^J \eta^K \Lambda^L_+ + \eta^1 \eta^2 \eta^3 \eta^4 \bar \Phi \\
         \mathcal{H}^- &= \Phi + \eta^I \Lambda^-_I + \frac{1}{2} \eta^I \eta^J A^-_{IJ} + \frac{1}{3!} \epsilon_{IJKL} \eta^I \eta^J \eta^K \psi^L_- + \eta^1 \eta^2 \eta^3 \eta^4 h^{--}
    \end{split}
\end{equation}
These are the same on-shell graviton supermultiplets of $\mathcal{N} = 4$ Einstein supergravity. The additional ghost states that are present in the conformal supergravity can also be considered, but we will be interested in the scattering of physical states only. 
Without discussing further about the general conformal supergravity amplitudes, we will, from now on concentrate on the MHV amplitudes involving only $(h^{++}, h^{--}, \Phi)$ particles.

We will use 6-point MHV amplitudes for the purpose of OPE decomposition. The reason for working with the 6-point amplitudes is that the lower-point celestial amplitudes are distributional in nature, and hence some of the terms in the OPE decomposition may vanish due to this constraint. We could have chosen any other higher point amplitudes as well to extract the OPE between the above-mentioned operators; however, it turns out that working with the six-point amplitudes is sufficient as higher point ones provide no further information in this regard.

\subsection{6-point MHV amplitudes }

We will be interested in extracting the celestial graviton-graviton and graviton-scalar OPEs from the appropriate scattering amplitudes of the BW theory. The scattering amplitudes for different constituent particles in the supermultiplet \eqref{grav_multiplet} can be obtained by taking appropriate derivatives of \eqref{sup_amp} with respect to the Grassmann variables. For additional details on how to do this, see \cite{Elvang:2013cua}. So let us first start with the 6-point MHV amplitude with all the external states as gravitons (we call this amplitude the pure graviton MHV amplitude) as this will help us to obtain graviton-graviton celestial OPE.

\subsubsection*{6-point pure graviton MHV amplitude}
 We obtain the 6-point pure graviton MHV amplitude as,
\begin{equation}\label{6pt_mom_amp}
\begin{gathered}
M_6(1^{--},2^{--},3^{++},4^{++},5^{++},6^{++})= i \left< 1,2 \right>^4\left( \frac{\left< 1,2\right>^2 [2,3]}{\left< 1,3 \right>^2 \left< 2,3 \right>} + \frac{\left< 1,4\right>^2 [3,4]}{\left< 1,3 \right>^2 \left< 3,4 \right>} + \frac{\left< 1,5\right>^2 [3,5]}{\left< 1,3 \right>^2 \left< 3,5 \right>} \right. \\
\left. + \frac{\left< 1,6\right>^2 [3,6]}{\left< 1,3 \right>^2 \left< 3,6 \right>} \right) \left( \frac{\left< 1,2\right>^2 [2,4]}{\left< 1,4 \right>^2 \left< 2,4 \right>} + \frac{\left< 1,3\right>^2 [3,4]}{\left< 1,4 \right>^2 \left< 3,4 \right>} + \frac{\left< 1,5\right>^2 [4,5]}{\left< 1,4 \right>^2 \left< 4,5 \right>} + \frac{\left< 1,6\right>^2 [4,6]}{\left< 1,4 \right>^2 \left< 4,6 \right>} \right) \\
\times \left( \frac{\left< 1,2\right>^2 [2,5]}{\left< 1,5 \right>^2 \left< 2,5 \right>} + \frac{\left< 1,3\right>^2 [3,5]}{\left< 1,5 \right>^2 \left< 3,5 \right>} + \frac{\left< 1,4\right>^2 [4,5]}{\left< 1,5 \right>^2 \left< 4,5 \right>} + \frac{\left< 1,6\right>^2 [5,6]}{\left< 1,5 \right>^2 \left< 5,6 \right>} \right) \\
\times \left( \frac{\left< 1,2\right>^2 [2,6]}{\left< 1,6 \right>^2 \left< 2,6 \right>} + \frac{\left< 1,3\right>^2 [3,6]}{\left< 1,6 \right>^2 \left< 3,6 \right>} + \frac{\left< 1,4\right>^2 [4,6]}{\left< 1,6 \right>^2 \left< 4,6 \right>} + \frac{\left< 1,5\right>^2 [5,6]}{\left< 1,6 \right>^2 \left< 5,6 \right>} \right) \, .
\end{gathered}
\end{equation}
One can work either with (1,3) signature with complexified momenta or with  (2,2) signature and real momenta, and this choice would have no bearing on either the analysis or the results. Here we choose to use (1,3) signature with mostly minus signs. In our convention, the momentum of the $i$-th massless particle $p_i^\mu$, satisfying the onshell condition $p_i^2=0$, is parametrised as,
\begin{eqnarray}
 \nonumber   p_i^\mu &=& \epsilon_i \omega_i q_i^\mu(z_i, \bar z_i) \\
    &=& \epsilon_i \omega_i (1+z_i \bar z_i , z_i +\bar z_i, -i( z_i -\bar z_i), 1-z_i \bar z_i  ) \label{null_mom}
\end{eqnarray}
where $\epsilon_i = \pm 1 $ for the outgoing/incoming particles. The positive real number $\omega_i$ is the energy of the $i$-th particle, and $(z_i, \bar z_i)$ are the coordinates on the celestial sphere at null infinity which represents the direction of motion of the $i$-th particle. The Lorentz group in $ (1,3) $ signature is given by $ SO^{+}(1,3) \simeq \frac{SL(2,\mathbb{C})}{\mathbb{Z}_2} $ and acts as the group of conformal transformations on the celestial sphere as:
\begin{equation}  \label{lt}
\begin{gathered}
 z \to \frac{a z + b}{cz+d} ~, \qquad \bar z \to \frac{\bar a \bar z + \bar b}{\bar c \bar z + \bar d} ~, \qquad ad-bc=1 \, . 
\end{gathered}
\end{equation} 
We treat ($z_i,\bar z_i$) as two independent variables. We also use the following parameterisation for the spinor helicity brackets
\begin{equation}\label{spin_brack}
\left< i,j \right> = 2 \epsilon_i \epsilon_j \sqrt{\omega_i \omega_j} z_{ij}, \ [ i,j ] = 2 \sqrt{\omega_i \omega_j} \bar z_{ij}
\end{equation}
where $z_{ij}=z_i-z_j$ and $\bar z_{ij} = \bar z_{i} - \bar z_j $. Using the parameterisation \eqref{spin_brack}, we can write the amplitude \eqref{6pt_mom_amp} in the $ (\omega, z, \bar z) $ space as follows:
\begin{equation}
\begin{gathered}
M_6(1^{--},2^{--},3^{++},4^{++},5^{++},6^{++})= 2^4 i \, z_{12}^4 (\omega_1\omega_2)^2\left( \epsilon_2 \epsilon_3 \frac{\omega_2}{\omega_3}\frac{ z_{12}^2 \bar z_{23}}{z_{13}^2 z_{23}} + \epsilon_3 \epsilon_4 \frac{\omega_4}{\omega_3}\frac{z_{14}^2 \bar z_{34}}{z_{13}^2 z_{34}} \right.\\
\left. + \epsilon_3 \epsilon_5 \frac{\omega_5}{\omega_3}\frac{z_{15}^2 \bar z_{35}}{ z_{13}^2 z_{35}} + \epsilon_3 \epsilon_6 \frac{\omega_6}{\omega_3}\frac{z_{16}^2 \bar z_{36}}{ z_{13}^2 z_{36}} \right) \left( \epsilon_2 \epsilon_4 \frac{\omega_2}{\omega_4}\frac{z_{12}^2 \bar z_{24}}{z_{14}^2 z_{24}} + \epsilon_3 \epsilon_4 \frac{\omega_3}{\omega_4}\frac{z_{13}^2 \bar z_{34}}{z_{14}^2 z_{34}} + \epsilon_4 \epsilon_5 \frac{\omega_5}{\omega_4}\frac{z_{15}^2 \bar z_{45}}{z_{14}^2 z_{45}} \right.\\
\left. + \epsilon_4 \epsilon_6 \frac{\omega_6}{\omega_4}\frac{z_{16}^2 \bar z_{46}}{z_{14}^2 z_{46}}  \right) \left( \epsilon_2 \epsilon_5 \frac{\omega_2}{\omega_5}\frac{ z_{12}^2 \bar z_{25}}{ z_{15}^2  z_{25} } + \epsilon_3 \epsilon_5 \frac{\omega_3}{\omega_5}\frac{ z_{13}^2 \bar z_{35}}{ z_{15}^2 z_{35}} + \epsilon_4 \epsilon_5 \frac{\omega_4}{\omega_5}\frac{ z_{14}^2 \bar z_{45}}{ z_{15}^2 z_{45} } + \epsilon_5 \epsilon_6 \frac{\omega_6}{\omega_5}\frac{ z_{16}^2 \bar z_{56}}{ z_{15}^2 z_{56}} \right) \\
\times \left( \epsilon_2 \epsilon_6 \frac{\omega_2}{\omega_6}\frac{ z_{12}^2 \bar z_{26}}{ z_{16}^2 z_{26}} + \epsilon_3 \epsilon_6 \frac{\omega_3}{\omega_6}\frac{ z_{13}^2 \bar z_{36}}{ z_{16}^2 z_{36} } + \epsilon_4 \epsilon_6 \frac{\omega_4}{\omega_6}\frac{ z_{14}^2 \bar z_{46}}{ z_{16}^2 z_{46}} + \epsilon_5 \epsilon_6 \frac{\omega_5}{\omega_6}\frac{ z_{15}^2 \bar z_{56}}{ z_{16}^2 z_{56}} \right) \, .
\end{gathered}\label{6pt_omzbz}
\end{equation}
We will use this form of the 6-point pure graviton amplitude for the Mellin transformation in the later part of this section. 

\subsubsection*{6-point scalar-graviton amplitude}

The 6-point amplitude with one external scalar is given by,
\begin{equation}
\begin{gathered}
M_6(1^{--},2^{--},3^{++},4^{++},5^{++},6_\Phi)= i \left< 1,2 \right>^4\left( \frac{\left< 1,2\right>^2 [2,3]}{\left< 1,3 \right>^2 \left< 2,3 \right>} + \frac{\left< 1,4\right>^2 [3,4]}{\left< 1,3 \right>^2 \left< 3,4 \right>} + \frac{\left< 1,5\right>^2 [3,5]}{\left< 1,3 \right>^2 \left< 3,5 \right>} \right.\\
\left. + \frac{\left< 1,6\right>^2 [3,6]}{\left< 1,3 \right>^2 \left< 3,6 \right>}  \right) \left( \frac{\left< 1,2\right>^2 [2,4]}{\left< 1,4 \right>^2 \left< 2,4 \right>} + \frac{\left< 1,3\right>^2 [3,4]}{\left< 1,4 \right>^2 \left< 3,4 \right>} + \frac{\left< 1,5\right>^2 [4,5]}{\left< 1,4 \right>^2 \left< 4,5 \right>} + \frac{\left< 1,6\right>^2 [4,6]}{\left< 1,4 \right>^2 \left< 4,6 \right>} \right) \\
\times \left( \frac{\left< 1,2\right>^2 [2,5]}{\left< 1,5 \right>^2 \left< 2,5 \right>} + \frac{\left< 1,3\right>^2 [3,5]}{\left< 1,5 \right>^2 \left< 3,5 \right>} + \frac{\left< 1,4\right>^2 [4,5]}{\left< 1,5 \right>^2 \left< 4,5 \right>} + \frac{\left< 1,6\right>^2 [5,6]}{\left< 1,5 \right>^2 \left< 5,6 \right>} \right)  \, .
\end{gathered}
\end{equation}
In terms of $ (\omega_i, z_i, \bar z_i) $ this becomes,
\begin{equation} \label{6pt_scalar_grav}
\begin{gathered}
M_6(1^{--},2^{--},3^{++},4^{++},5^{++},6_\Phi)= 2^4 i \, z_{12}^4 (\omega_1\omega_2)^2\left( \epsilon_2 \epsilon_3 \frac{\omega_2}{\omega_3}\frac{ z_{12}^2 \bar z_{23}}{z_{13}^2 z_{23}} + \epsilon_3 \epsilon_4 \frac{\omega_4}{\omega_3}\frac{z_{14}^2 \bar z_{34}}{z_{13}^2 z_{34}}  \right.\\
\left. + \epsilon_3 \epsilon_5 \frac{\omega_5}{\omega_3}\frac{z_{15}^2 \bar z_{35}}{ z_{13}^2 z_{35}} + \epsilon_3 \epsilon_6 \frac{\omega_6}{\omega_3}\frac{z_{16}^2 \bar z_{36}}{ z_{13}^2 z_{36}} \right) \left( \epsilon_2 \epsilon_4 \frac{\omega_2}{\omega_4}\frac{z_{12}^2 \bar z_{24}}{z_{14}^2 z_{24}} + \epsilon_3 \epsilon_4 \frac{\omega_3}{\omega_4}\frac{z_{13}^2 \bar z_{34}}{z_{14}^2 z_{34}} + \epsilon_4 \epsilon_5 \frac{\omega_5}{\omega_4}\frac{z_{15}^2 \bar z_{45}}{z_{14}^2 z_{45}} \right.\\
\left. + \epsilon_4 \epsilon_6 \frac{\omega_6}{\omega_4}\frac{z_{16}^2 \bar z_{46}}{z_{14}^2 z_{46}} \right) \left( \epsilon_2 \epsilon_5 \frac{\omega_2}{\omega_5}\frac{ z_{12}^2 \bar z_{25}}{ z_{15}^2  z_{25} } + \epsilon_3 \epsilon_5 \frac{\omega_3}{\omega_5}\frac{ z_{13}^2 \bar z_{35}}{ z_{15}^2 z_{35}} + \epsilon_4 \epsilon_5 \frac{\omega_4}{\omega_5}\frac{ z_{14}^2 \bar z_{45}}{ z_{15}^2 z_{45} } + \epsilon_5 \epsilon_6 \frac{\omega_6}{\omega_5}\frac{ z_{16}^2 \bar z_{56}}{ z_{15}^2 z_{56}} \right) \, .
\end{gathered}
\end{equation}
Before Mellin transforming the 6-point amplitudes and writing them as correlation functions on the celestial sphere, let us also write down the 5-point amplitudes in momentum space that will be required for the OPE analysis.

\subsection{5-point MHV amplitudes}
\label{5-point_amp}
We will be interested in expanding the 6-point amplitudes around the collinear/OPE limit of two of their external particle momenta and write them in terms of lower point amplitudes. Therefore, we need the expressions for the relevant 5-point amplitudes as well.
\subsubsection*{The 5-point pure graviton MHV amplitude}

The 5-point pure graviton amplitude is given by,
\begin{equation}\label{5pt_mom}
\begin{gathered}
M_5(1^{--},2^{--},3^{++},4^{++},5^{++})=-i\left< 1,2 \right>^4 \left( \frac{\left< 1,2 \right>^2 [2,3]}{\left< 1,3 \right>^2\left< 2,3 \right>} + \frac{\left< 1,4 \right>^2 [3,4]}{\left< 1,3 \right>^2\left< 3,4 \right>} \right.\\
\left. + \frac{\left< 1,5 \right>^2 [3,5]}{\left< 1,3 \right>^2\left< 3,5 \right>}\right) \left( \frac{\left< 1,2 \right>^2 [2,5]}{\left< 1,5 \right>^2\left< 2,5 \right>} + \frac{\left< 1,3 \right>^2 [3,5]}{\left< 1,5 \right>^2\left< 3,5 \right>} + \frac{\left< 1,4 \right>^2 [4,5]}{\left< 1,5 \right>^2\left< 4,5 \right>}\right) \\
\times\left( \frac{\left< 1,2 \right>^2 [2,4]}{\left< 1,4 \right>^2\left< 2,4 \right>} + \frac{\left< 1,3 \right>^2 [3,4]}{\left< 1,4 \right>^2\left< 3,4 \right>} + \frac{\left< 1,5 \right>^2 [4,5]}{\left< 1,4 \right>^2\left< 4,5 \right>}\right) \, .
\end{gathered}
\end{equation}
Since we are interested in taking the OPE limit $ 5 \to 6 $ we will label the 5-point amplitude as $ M_5(1^{--},2^{--},3^{++},4^{++},6^{++})$. In terms of $ (\omega_i, z_i, \bar z_i) $ variables, the amplitude \eqref{5pt_mom} then becomes,
\begin{equation}\label{5pt_pure_grav}
\begin{gathered}
M_5(1^{--},2^{--},3^{++},4^{++},6^{++})=-2^4 i(\omega_1 \omega_2)^2 z_{12}^4 \left( \epsilon_2 \epsilon_3 \frac{\omega_2}{\omega_3}\frac{z_{12}^2 \bar z_{23}}{z_{13}^2 z_{23} } + \epsilon_3 \epsilon_4 \frac{\omega_4}{\omega_3}\frac{z_{14}^2 \bar z_{34}}{z_{13}^2 z_{34}} \right.\\
\left. + \epsilon_3 \epsilon_6 \frac{\omega_6}{\omega_3} \frac{z_{16}^2 \bar z_{36} }{ z_{13}^2 z_{36}}\right) \left( \epsilon_2 \epsilon_6 \frac{\omega_2}{\omega_6}\frac{z_{12}^2 \bar z_{26}}{z_{16}^2 z_{26} } + \epsilon_3 \epsilon_6 \frac{\omega_3}{\omega_6}\frac{z_{13}^2 \bar z_{36}}{z_{16}^2 z_{36}} + \epsilon_4 \epsilon_6 \frac{\omega_4}{\omega_6} \frac{z_{14}^2 \bar z_{46} }{ z_{16}^2 z_{46}}\right) \\
\times \left( \epsilon_2 \epsilon_4 \frac{\omega_2}{\omega_4}\frac{z_{12}^2 \bar z_{24}}{z_{14}^2 z_{24} } + \epsilon_3 \epsilon_4 \frac{\omega_3}{\omega_4}\frac{z_{13}^2 \bar z_{34}}{z_{14}^2 z_{34}} + \epsilon_4 \epsilon_6 \frac{\omega_6}{\omega_4} \frac{z_{16}^2 \bar z_{46} }{ z_{14}^2 z_{46}}\right) \, .
\end{gathered}
\end{equation}

\subsubsection*{5-point scalar-graviton amplitude}

We now write the 4-graviton and one scalar amplitude, where the last particle is the holomorphic complex scalar. This amplitude is given by,
\begin{equation}
\begin{gathered}
M_5(1^{--},2^{--},3^{++},4^{++},6_\Phi)=-i\left< 1,2 \right>^4 \left( \frac{\left< 1,2 \right>^2 [2,3]}{\left< 1,3 \right>^2\left< 2,3 \right>} + \frac{\left< 1,4 \right>^2 [3,4]}{\left< 1,3 \right>^2\left< 3,4 \right>} + \frac{\left< 1,6 \right>^2 [3,6]}{\left< 1,3 \right>^2\left< 3,6 \right>}\right)\\
\times \left( \frac{\left< 1,2 \right>^2 [2,4]}{\left< 1,4 \right>^2\left< 2,4 \right>} + \frac{\left< 1,3 \right>^2 [3,4]}{\left< 1,4 \right>^2\left< 3,4 \right>} + \frac{\left< 1,6 \right>^2 [4,6]}{\left< 1,4 \right>^2\left< 4,6 \right>}\right) \, .
\end{gathered}
\end{equation}
In terms of $ (\omega_i, z_i, \bar z_i) $ variables this reads,
\begin{equation}\label{5pt_scalar_grav}
\begin{gathered}
M_5(1^{--},2^{--},3^{++},4^{++},6_\Phi)=- 2^4 i(\omega_1 \omega_2)^2 z_{12}^4 \left( \epsilon_2 \epsilon_3 \frac{\omega_2}{\omega_3}\frac{z_{12}^2 \bar z_{23}}{z_{13}^2 z_{23} } + \epsilon_3 \epsilon_4 \frac{\omega_4}{\omega_3}\frac{z_{14}^2 \bar z_{34}}{z_{13}^2 z_{34}} \right.\\
\left. + \epsilon_3 \epsilon_6 \frac{\omega_6}{\omega_3} \frac{z_{16}^2 \bar z_{36} }{ z_{13}^2 z_{36}}\right) \left( \epsilon_2 \epsilon_4 \frac{\omega_2}{\omega_4}\frac{z_{12}^2 \bar z_{24}}{z_{14}^2 z_{24} } + \epsilon_3 \epsilon_4 \frac{\omega_3}{\omega_4}\frac{z_{13}^2 \bar z_{34}}{z_{14}^2 z_{34}} + \epsilon_4 \epsilon_6 \frac{\omega_6}{\omega_4} \frac{z_{16}^2 \bar z_{46} }{ z_{14}^2 z_{46}}\right) \, .
\end{gathered}
\end{equation}
Now that we have written down all the necessary momentum space amplitudes, let us briefly describe the method we will use for the extraction of OPE from them.  We will follow the method developed by \cite{Banerjee:2020zlg}. In the current context, the method involves starting with the 6-point amplitudes of gravitons and scalars above, and Mellin transforming away the energies $\omega_i$ for the conformal dimensions $\Delta_i$ for each external particle. This gives the corresponding 6-point celestial amplitudes. Then one expands the result in the OPE limit $z_{56} \rightarrow 0$, $\bar z_{56} \rightarrow 0 $, and identifies the coefficients of the expansion again in terms of the 5-point celestial amplitudes of gravitons and scalars. Finally, we reinterpret the answer as the OPE of two primary operators corresponding to the 5-th and 6-th particles of appropriate helicities ($\sigma_i$) in terms of the celestial CFT primary operators of gravitons and other particles. This gives very specific singularity structures and OPE coefficients in terms of $\Delta_i$ and $\sigma_i$. One then needs to figure out which symmetries of the putative celestial CFT would lead to precisely such OPE expansions. 

\subsection{6-point MHV celestial amplitudes}

The modified Mellin transform \cite{Banerjee:2018gce} of the 6-point amplitude is given by,
\begin{equation}\label{6pt_gen_mellin}
\begin{gathered}
\mathcal{M}_6\left( 1^{--}_{\Delta_1},2^{--}_{\Delta_2},3^{++}_{\Delta_3},4^{++}_{\Delta_4},5^{++}_{\Delta_5}, 6^{++}_{\Delta_6}/6^\Phi \right) \\
= \left< G^{--}_{\Delta_1} (1) G^{--}_{\Delta_2} (2) G^{++}_{\Delta_3} (3) G^{++}_{\Delta_4} (4) G^{++}_{\Delta_5} (5) \left( G^{++}_{\Delta_6} (6)/\Phi_{\Delta_6}(6)\right) \right>\\
=\left( \prod_{i=1}^6 \int_{0}^\infty d\omega_i \, \omega_i^{\Delta_i-1} \right) e^{-i\sum_{k=1}^6 \epsilon_k \omega_k u_k} M_6(1^{--},2^{--},3^{++},4^{++},5^{++}, 6^{++}/6_\Phi) \\
\times \delta^{(4)}\left( \sum_{i=1}^6 \epsilon_i \omega_i q^\mu_i \right)
\end{gathered}
\end{equation}
where $ G^{\sigma_i}_{\Delta_i}(i) = G^{\sigma_i}_{\Delta_i}(u_i,z_i,\bar z_i) $ is the $i$-th graviton primary operator with helicity  $\sigma_i $ and conformal dimension $\Delta_i$ living in $(u,z,\bar z)$ space, i.e. at null infinity, corresponding to the $i$-th external graviton in the $S$-matrix element. Similarly $ \Phi_{\Delta_i}(i) = \Phi_{\Delta_i}(u_i,z_i,\bar z_i) $ is the scalar primary operator. In the amplitude \eqref{6pt_gen_mellin} the $6$-th particle can either be a graviton or a scalar. We have also restored the momentum-conserving delta function.

The integral in \eqref{6pt_gen_mellin} becomes highly oscillatory in the limit $\omega \to \infty$. To regulate this behaviour, one introduces a small imaginary part to each $u_i$ variable via the shift $ u_i \to u_i + i\delta_i$, where $\delta_i \to 0^{\pm}$ with the sign determined by $\epsilon_i$. The standard celestial amplitude \cite{Pasterski:2017kqt}, does not have the exponential $u$-factor in the Mellin transformation of \eqref{6pt_gen_mellin}, but requires a regulator for it to be well-defined. It transforms as a 2$d$ conformal correlator on the celestial sphere. As explained in \cite{Banerjee:2019prz}, the standard celestial amplitude can be recovered from the modified one as follows. Time translation invariance ensures that the modified celestial amplitude depends only on the differences $u_{ij} = u_i - u_j$. Setting all $u_i$ equal (i.e., $u_i = u \ \forall ~ i$) reduces the modified celestial amplitude to the standard form. Therefore, we work with the modified celestial amplitude throughout our analysis and impose the condition $u_i=u$ only when extracting OPE from the correlators. This procedure allows us to recover the standard celestial OPE between operators on the celestial sphere. For brevity, we suppress the regulator dependence in our expressions for the modified celestial amplitudes.

\subsubsection*{6-point pure graviton celestial amplitude}

We are interested in the celestial OPE between the primary operators inserted at the points $(z_5, \bar z_5)$ and $( z_6, \bar z_6)$ on the celestial sphere. The parametrisation of the 6-point delta function needed for our OPE analysis is discussed in appendix \ref{6pt_delta}. Using that parametrisation, we can perform four of the energy integrals over $(\omega_1,\ldots,\omega_4)$ in \eqref{6pt_gen_mellin}. Then, using \eqref{6pt_delta_param} and \eqref{6pt_omzbz} in \eqref{6pt_gen_mellin} and taking 6-th particle as a graviton, we get the following result,
\begin{equation}
\begin{gathered}
\mathcal{M}_6\left( 1^{--}_{\Delta_1},2^{--}_{\Delta_2},3^{++}_{\Delta_3},4^{++}_{\Delta_4},5^{++}_{\Delta_5}, 6^{++}_{\Delta_6} \right) =  4 i \frac{ z_{12}^4}{ z_{14}z_{23} \bar z_{14} \bar z_{23}(r_{13,42} - \bar r_{13,42})} \int_0^\infty d\omega_5 \, \omega_5^{\Delta_5-1} \\
\times \int_0^\infty d\omega_6 \, \omega_6^{\Delta_6-1} (\omega_1^* \omega_2^*)^2 \left(\prod_{i=1}^4 {(\omega_i^*)}^{\Delta_i-1}\right)e^{-i\sum_{k=1}^4 \epsilon_k \omega_k^* u_k -i\epsilon_5\omega_5 u_5 -i\epsilon_6 \omega_6 u_6}\\
\times  \left( \epsilon_2 \epsilon_3 \frac{\omega_2^*}{\omega_3^*}\frac{ z_{12}^2 \bar z_{23}}{z_{13}^2 z_{23}} + \epsilon_3 \epsilon_4 \frac{\omega_4^*}{\omega_3^*}\frac{z_{14}^2 \bar z_{34}}{z_{13}^2 z_{34}} + \epsilon_3 \epsilon_5 \frac{\omega_5}{\omega_3^*}\frac{z_{15}^2 \bar z_{35}}{ z_{13}^2 z_{35}} + \epsilon_3 \epsilon_6 \frac{\omega_6}{\omega_3^*}\frac{z_{16}^2 \bar z_{36}}{ z_{13}^2 z_{36}} \right) \\
\times \left( \epsilon_2 \epsilon_4 \frac{\omega_2^*}{\omega_4^*}\frac{z_{12}^2 \bar z_{24}}{z_{14}^2 z_{24}} + \epsilon_3 \epsilon_4 \frac{\omega_3^*}{\omega_4^*}\frac{z_{13}^2 \bar z_{34}}{z_{14}^2 z_{34}} + \epsilon_4 \epsilon_5 \frac{\omega_5}{\omega_4^*}\frac{z_{15}^2 \bar z_{45}}{z_{14}^2 z_{45}} + \epsilon_4 \epsilon_6 \frac{\omega_6}{\omega_4^*}\frac{z_{16}^2 \bar z_{46}}{z_{14}^2 z_{46}} \right) \\
\times \left( \epsilon_2 \epsilon_5 \frac{\omega_2^*}{\omega_5}\frac{ z_{12}^2 \bar z_{25}}{ z_{15}^2  z_{25} } + \epsilon_3 \epsilon_5 \frac{\omega_3^*}{\omega_5}\frac{ z_{13}^2 \bar z_{35}}{ z_{15}^2 z_{35}} + \epsilon_4 \epsilon_5 \frac{\omega_4^*}{\omega_5}\frac{ z_{14}^2 \bar z_{45}}{ z_{15}^2 z_{45} } + \epsilon_5 \epsilon_6 \frac{\omega_6}{\omega_5}\frac{ z_{16}^2 \bar z_{56}}{ z_{15}^2 z_{56}} \right) \\
\times \left( \epsilon_2 \epsilon_6 \frac{\omega_2^*}{\omega_6}\frac{ z_{12}^2 \bar z_{26}}{ z_{16}^2 z_{26}} + \epsilon_3 \epsilon_6 \frac{\omega_3^*}{\omega_6}\frac{ z_{13}^2 \bar z_{36}}{ z_{16}^2 z_{36} } + \epsilon_4 \epsilon_6 \frac{\omega_4^*}{\omega_6}\frac{ z_{14}^2 \bar z_{46}}{ z_{16}^2 z_{46}} + \epsilon_5 \epsilon_6 \frac{\omega_5}{\omega_6}\frac{ z_{15}^2 \bar z_{56}}{ z_{16}^2 z_{56}} \right)
\end{gathered}
\label{bope}
\end{equation}
where $\omega_i^*$'s are given by,
\begin{equation} \label{omega_stars22}
\begin{split}
\omega_1^* &= \epsilon_1 \epsilon_6 \omega_6 \sigma_{1,1} + \epsilon_1 \epsilon_5 \omega_5 \sigma_{1,2}\\
\omega_2^* &= \epsilon_2 \epsilon_6 \omega_6 \sigma_{2,1} + \epsilon_2 \epsilon_5 \omega_5 \sigma_{2,2}\\
\omega_3^* &= \epsilon_3 \epsilon_6 \omega_6 \sigma_{3,1} + \epsilon_3 \epsilon_5 \omega_5 \sigma_{3,2}\\
\omega_4^* &= \epsilon_4 \epsilon_6 \omega_6 \sigma_{4,1} + \epsilon_4 \epsilon_5 \omega_5 \sigma_{4,2}
\end{split}
\end{equation}
and $r_{ij,kl}, \ \sigma_{i,j}$'s are given in appendix \ref{delta_param}. Let us now make a change of variables,
\begin{equation}
\omega_5 = \omega_P t, \ \omega_6 = \omega_P(1-\epsilon_5 \epsilon_6 t) \, .
\end{equation}
Then we have 
\begin{equation}
\omega_i^* = \epsilon_i \epsilon_6 \Sigma_i \omega_P, \ 
\Sigma_i = \sigma_{i,1} - \epsilon_5\epsilon_6(\sigma_{i,1}-\sigma_{i,2})t, \ i=1,\ldots,4
\end{equation}
Then \eqref{bope} becomes
\begin{equation}
\begin{gathered}
\mathcal{M}_6\left( 1^{--}_{\Delta_1},2^{--}_{\Delta_2},3^{++}_{\Delta_3},4^{++}_{\Delta_4},5^{++}_{\Delta_5}, 6^{++}_{\Delta_6} \right) \\
= 4 i \frac{z_{12}^4}{z_{14}z_{23} \bar z_{14} \bar z_{23}(r_{13,42} - \bar r_{13,42})} \int_{0}^1 dt \, t^{\Delta_5-1} (1-\epsilon_5 \epsilon_6 t)^{\Delta_6-1} (\Sigma_1 \Sigma_2)^2 \\
\times \prod_{k=1}^4 \Theta(\epsilon_6 \epsilon_k \Sigma_{k}) \left(\prod_{i=1}^4 {(\epsilon_i \epsilon_6 \Sigma_i)}^{\Delta_i-1}\right)\int_0^\infty d\omega_P \, \omega_P^{\Delta-1}  e^{-i \omega_P ( \mathcal{U} + u_{56}) }\\
\times \left( \frac{\Sigma_2}{\Sigma_3}\frac{ z_{12}^2 \bar z_{23}}{z_{13}^2 z_{23}} + \frac{\Sigma_4}{\Sigma_3}\frac{z_{14}^2 \bar z_{34}}{z_{13}^2 z_{34}} + \epsilon_5 \epsilon_6 \frac{t}{\Sigma_3}\frac{z_{15}^2 \bar z_{35}}{ z_{13}^2 z_{35}} +  \frac{(1-\epsilon_5\epsilon_6 t)}{\Sigma_3}\frac{z_{16}^2 \bar z_{36}}{ z_{13}^2 z_{36}} \right) \\
\times \left( \frac{\Sigma_2}{\Sigma_4}\frac{z_{12}^2 \bar z_{24}}{z_{14}^2 z_{24}} + \frac{\Sigma_3}{\Sigma_4}\frac{z_{13}^2 \bar z_{34}}{z_{14}^2 z_{34}} + \frac{\epsilon_5 \epsilon_6 t}{\Sigma_4}\frac{z_{15}^2 \bar z_{45}}{z_{14}^2 z_{45}} + \frac{(1-\epsilon_5 \epsilon_6 t)}{\Sigma_4}\frac{z_{16}^2 \bar z_{46}}{z_{14}^2 z_{46}} \right) \\
\times \left( \epsilon_5 \epsilon_6 \frac{\Sigma_2}{t}\frac{ z_{12}^2 \bar z_{25}}{ z_{15}^2  z_{25} } + \epsilon_5 \epsilon_6 \frac{\Sigma_3}{t}\frac{ z_{13}^2 \bar z_{35}}{ z_{15}^2 z_{35}} + \epsilon_5 \epsilon_6 \frac{\Sigma_4}{t}\frac{ z_{14}^2 \bar z_{45}}{ z_{15}^2 z_{45} } + \epsilon_5 \epsilon_6 \frac{(1-\epsilon_5 \epsilon_6 t)}{t}\frac{ z_{16}^2 \bar z_{56}}{ z_{15}^2 z_{56}} \right) \\
\times \left(  \frac{\Sigma_2}{(1-\epsilon_5\epsilon_6 t)}\frac{ z_{12}^2 \bar z_{26}}{ z_{16}^2 z_{26}} +  \frac{\Sigma_3}{(1-\epsilon_5 \epsilon_6 t)}\frac{ z_{13}^2 \bar z_{36}}{ z_{16}^2 z_{36} } + \frac{\Sigma_4}{(1-\epsilon_5 \epsilon_6 t)}\frac{ z_{14}^2 \bar z_{46}}{ z_{16}^2 z_{46}} + \epsilon_5 \epsilon_6 \frac{t}{(1-\epsilon_5 \epsilon_6 t)}\frac{ z_{15}^2 \bar z_{56}}{ z_{16}^2 z_{56}} \right)
\end{gathered}
\label{bmmope}
\end{equation}
where, \begin{equation}
\mathcal{U} = \epsilon_6 \sum_{k=1}^4 \sigma_{k,1} u_{k6} +  \epsilon_5 t z_{56} \sum_{k=1}^4 \partial_6 \sigma_{k,1} u_{k6} + \epsilon_5 t \bar z_{56} \sum_{k=1}^4 \bar \partial_6 \sigma_{k,1} u_{k6} + \epsilon_5 t z_{56} \sum_{k=1}^4 \partial_6 \bar \partial_6 \sigma_{k,1} u_{k6} \bar z_{56} \, .
\end{equation}
Equation \eqref{bmmope} will be used for the OPE expansion between two positive helicity gravitons $G^{++}_{\Delta_5}(5)$ and $G^{++}_{\Delta_6}(6)$. We can also set $u_{56}=0$, which does not affect our OPE analysis. Next, we Mellin transform the scalar-graviton 6-point amplitude.

\subsubsection*{6-point scalar-graviton celestial amplitude}

To obtain the celestial amplitude for the 6-point scalar-graviton amplitude \eqref{6pt_scalar_grav}, we follow the same procedure as described above. The result is,
\begin{equation}
\begin{gathered}
\mathcal{M}_6\left( 1^{--}_{\Delta_1},2^{--}_{\Delta_2},3^{++}_{\Delta_3},4^{++}_{\Delta_4},5^{++}_{\Delta_5}, 6^{\Phi}_{\Delta_6} \right) \\
= 4 i \frac{z_{12}^4}{z_{14}z_{23} \bar z_{14} \bar z_{23}(r_{13,42} - \bar r_{13,42})} \int_{0}^1 dt \, t^{\Delta_5-1} (1-\epsilon_5 \epsilon_6 t)^{\Delta_6-1} (\Sigma_1 \Sigma_2)^2  \\
\times \prod_{k=1}^4 \Theta(\epsilon_6 \epsilon_k \Sigma_{k}) \left(\prod_{i=1}^4 {(\epsilon_i \epsilon_6 \Sigma_i)}^{\Delta_i-1}\right)\int_0^\infty d\omega_P \, \omega_P^{\Delta-1}  e^{-i \omega_P \mathcal{U} }\\
\times \left( \frac{\Sigma_2}{\Sigma_3}\frac{ z_{12}^2 \bar z_{23}}{z_{13}^2 z_{23}} + \frac{\Sigma_4}{\Sigma_3}\frac{z_{14}^2 \bar z_{34}}{z_{13}^2 z_{34}} + \epsilon_5 \epsilon_6 \frac{t}{\Sigma_3}\frac{z_{15}^2 \bar z_{35}}{ z_{13}^2 z_{35}} +  \frac{(1-\epsilon_5\epsilon_6 t)}{\Sigma_3}\frac{z_{16}^2 \bar z_{36}}{ z_{13}^2 z_{36}} \right) \\
\times \left( \frac{\Sigma_2}{\Sigma_4}\frac{z_{12}^2 \bar z_{24}}{z_{14}^2 z_{24}} + \frac{\Sigma_3}{\Sigma_4}\frac{z_{13}^2 \bar z_{34}}{z_{14}^2 z_{34}} + \frac{\epsilon_5 \epsilon_6 t}{\Sigma_4}\frac{z_{15}^2 \bar z_{45}}{z_{14}^2 z_{45}} + \frac{(1-\epsilon_5 \epsilon_6 t)}{\Sigma_4}\frac{z_{16}^2 \bar z_{46}}{z_{14}^2 z_{46}} \right) \\
\times \left( \epsilon_5 \epsilon_6 \frac{\Sigma_2}{t}\frac{ z_{12}^2 \bar z_{25}}{ z_{15}^2  z_{25} } + \epsilon_5 \epsilon_6 \frac{\Sigma_3}{t}\frac{ z_{13}^2 \bar z_{35}}{ z_{15}^2 z_{35}} + \epsilon_5 \epsilon_6 \frac{\Sigma_4}{t}\frac{ z_{14}^2 \bar z_{45}}{ z_{15}^2 z_{45} } + \epsilon_5 \epsilon_6 \frac{(1-\epsilon_5 \epsilon_6 t)}{t}\frac{ z_{16}^2 \bar z_{56}}{ z_{15}^2 z_{56}} \right)  \, .
\end{gathered}
\label{bmscope}
\end{equation}
This amplitude will help us to extract the OPE between a positive helicity graviton $G^{++}_{\Delta_5}(5)$ and a scalar $\Phi_{\Delta_6}(6)$. We also need all the 5-point celestial amplitudes that will arise in the OPE expansions of the 6-point amplitudes derived so far.

\subsection{5-point MHV celestial amplitudes}

In this section, we Mellin transform the 5-point amplitudes \eqref{5pt_pure_grav} and \eqref{5pt_scalar_grav}.

\subsubsection*{5-point  pure graviton celestial amplitude}
The modified Mellin transform of the 5-point momentum space amplitude is given by,
\begin{equation}
\begin{gathered}
\mathcal{M}_5\left( 1^{--}_{\Delta_1},2^{--}_{\Delta_2},3^{++}_{\Delta_3},4^{++}_{\Delta_4},6^{++}_{\Delta_6} \right) = \prod_{i=1,i\neq 5}^6 \int_{0}^\infty d\omega_i \, \omega_i^{\Delta_i-1} e^{-i\sum_{k=1,k\neq 5}^6 \epsilon_k \omega_k u_k} \\
\times M_5(1^{--},2^{--},3^{++},4^{++},6^{++})\delta^{(4)}\left( \sum_{i=1,i\neq 5}^6 \epsilon_i \omega_i q^\mu_i \right) \, .
\end{gathered}
\end{equation}
We have discussed the parameterisation of the 5-point momentum-conserving delta function in section \ref{5pt_delta}. Using that parametrisation, we can perform four of the $ \omega_i$ integrals in the above equation, and the remaining one gives the gamma function. The result is as follows
\begin{equation} \label{5pt_pure_grav_mellin}
\begin{gathered}
\mathcal{M}_5\left( 1^{--}_{\Delta_1},2^{--}_{\Delta_2},3^{++}_{\Delta_3},4^{++}_{\Delta_4},6^{++}_{\Delta_6} \right) = -4 i\frac{z_{12}^4}{(z_{13}z_{24} \bar{z}_{14} \bar z_{23} - \bar z_{13} \bar z_{24} {z}_{14} z_{23})} \prod_{k=1}^4 \Theta(\epsilon_6 \epsilon_k \sigma_{k,1}) \\
\times  \prod_{k=1}^4 \left( \epsilon_6 \epsilon_k \sigma_{k,1} \right)^{\Delta_k-1} \frac{\Gamma(\Delta')}{\left( i \mathcal{U}_1 \right)^{\Delta'}} \sigma_{1,1}^2 \sigma_{2,1}^2 \mathcal{T}^1_0 \mathcal{T}^2_0 \mathcal{T}^3_0
\end{gathered}
\end{equation}

where 
\begin{equation}
\begin{gathered}
\Delta' = \sum_{k=1,k\neq 5}^6 \Delta_k \\
\mathcal{U}_1 = \epsilon_6 \sum_{k=1}^4 \sigma_{k,1} u_{k6}
\end{gathered}
\end{equation}

\begin{equation}\label{taus}
\begin{split} 
\mathcal{T}^1_{0} &= \frac{\sigma_{2,1}}{\sigma_{3,1}}\frac{ z_{12}^2 \bar z_{23}}{z_{13}^2 z_{23}} + \frac{\sigma_{4,1}}{\sigma_{3,1}}\frac{z_{14}^2 \bar z_{34}}{z_{13}^2 z_{34}}  + \frac{1}{\sigma_{3,1}}\frac{z_{16}^2 \bar z_{36}}{ z_{13}^2 z_{36}}\\
\mathcal{T}^2_{0}  &= \frac{\sigma_{2,1}}{\sigma_{4,1}}\frac{z_{12}^2 \bar z_{24}}{z_{14}^2 z_{24}} + \frac{\sigma_{3,1}}{\sigma_{4,1}}\frac{z_{13}^2 \bar z_{34}}{z_{14}^2 z_{34}} + \frac{1}{\sigma_{4,1}}\frac{z_{16}^2 \bar z_{46}}{z_{14}^2 z_{46}}\\
\mathcal{T}^3_{0} &= \sigma_{2,1}\frac{ z_{12}^2 \bar z_{26}}{ z_{16}^2  z_{26} } +  \sigma_{3,1} \frac{ z_{13}^2 \bar z_{36}}{ z_{16}^2 z_{36}} + \sigma_{4,1} \frac{ z_{14}^2 \bar z_{46}}{ z_{16}^2 z_{46} } 
\end{split}
\end{equation}

\subsubsection*{5-point scalar-graviton celestial amplitude}

We follow the same procedure as before for the 5-point scalar-graviton amplitude as well and get the following result,
\begin{equation}\label{5pt_scalar_grav_mellin}
\begin{gathered}
\mathcal{M}_5\left( 1^{--}_{\Delta_1},2^{--}_{\Delta_2},3^{++}_{\Delta_3},4^{++}_{\Delta_4},6^\Phi_{\Delta_6} \right) = -4 i\frac{z_{12}^4}{(z_{13}z_{24} \bar{z}_{14} \bar z_{23} - \bar z_{13} \bar z_{24} {z}_{14} z_{23})} \prod_{k=1}^4 \Theta(\epsilon_6 \epsilon_k \sigma_{k,1}) \\
\times \sigma_{1,1}^2 \sigma_{2,1}^2 \prod_{k=1}^4 \left( \epsilon_6 \epsilon_k \sigma_{k,1} \right)^{\Delta_k-1} \frac{\Gamma(\Delta')}{\left( i \mathcal{U}_1 \right)^{\Delta'}} \mathcal{T}^1_0 \mathcal{T}^2_0 \, .
\end{gathered}
\end{equation}
Now, we are in a position to extract the OPEs from the amplitudes discussed above. From now on, we take the 5-th and 6-th particles to be outgoing, that is, we set $\epsilon_5 = \epsilon_6 = +1 $ and the rest will be unspecified.

\section{Celestial OPE from 6-point MHV amplitudes}
\label{OPE_from_amp}
We now discuss the OPE decomposition of the 6-point amplitudes. We have two 6-point MHV amplitudes: one with all external states as gravitons and another one where one external particle is the holomorphic scalar. The first one will give us the OPE between two graviton primaries, whereas the second one will give us the OPE between a graviton and a scalar primary operators. 

We expand both the 6-point amplitudes around $z_{56}=0, \bar z_{56}=0 $ while keeping the other $z_{ij}, \bar z_{ij}$ fixed and non-zero. Our amplitudes contain $\Theta$-functions of different $z_i,\bar z_i$ coordinates. As we expand these amplitudes around $z_{56}=0, \bar z_{56}=0$, we will get delta functions as derivatives of $\Theta$-functions with arguments $z_{ij}, \bar z_{ij}, \, i,j=1,2,3,4,6$. However, as none of the operators insertion points in the celestial amplitudes are coincident, except the pair whose OPE is being considered, we can neglect these contact terms. The following formulae will be useful for our OPE expansions that can be obtained from the expressions of $\sigma_{ij}$'s given in sections \ref{5pt_delta} and \ref{6pt_delta}:  
\begin{equation}
\begin{gathered}
\sigma_{i,2} = \sigma_{i,1} + z_{56} \frac{\partial \sigma_{i,1}}{\partial z_6} + \bar z_{56} \frac{\partial \sigma_{i,1}}{  \partial \bar z_6} + z_{56} \bar z_{56} \frac{\partial^2 \sigma_{i,1}}{ \partial z_6  \partial \bar z_6},\\
\Sigma_{i} = \sigma_{i,1} + t\left[ z_{56} \frac{\partial \sigma_{i,1}}{\partial z_6} + \bar z_{56} \frac{\partial \sigma_{i,1}}{  \partial \bar z_6} + z_{56} \bar z_{56} \frac{\partial^2 \sigma_{i,1}}{ \partial z_6  \partial \bar z_6} \right].
\end{gathered}
\end{equation}
Let us start with the OPE between the graviton operators.

\subsection{OPE between two positive helicity outgoing gravitons}

We start with  equation \eqref{bmmope}, and expand the right-hand side around $z_{56}=0, \bar z_{56}=0$. After the expansion, one can perform the $t$ and $\omega_P$ integrals. The $t$-integral will produce the beta functions, whereas the $\omega_P$ integral gives us the gamma functions below. 

\subsubsection*{The first two terms}

The first two terms in the OPE expansion of \eqref{bmmope} are given by,
\begin{equation}
\begin{gathered}
\mathcal{M}_6\left( 1^{--}_{\Delta_1},2^{--}_{\Delta_2},3^{++}_{\Delta_3},4^{++}_{\Delta_4},5^{++}_{\Delta_5}, 6^{++}_{\Delta_6} \right) = 4 i \frac{z_{12}^4}{z_{14}z_{23} \bar z_{14} \bar z_{23}(r_{13,42} - \bar r_{13,42})}  (\sigma_{1,1} \sigma_{2,1})^2  \\
\times  \prod_{k=1}^4 \Theta( \epsilon_k \sigma_{k,1}) \left(\prod_{i=1}^4 {(\epsilon_i \sigma_{i,1})}^{\Delta_i-1}\right) \frac{\Gamma(\Delta)}{(i\mathcal{U}_1)^{\Delta}} \left[ B(\Delta_5-1,\Delta_6-1)\left( \frac{\bar z_{56}}{z_{56}} \right) \mathcal{T}^1_0 \mathcal{T}^2_0 \mathcal{T}^3_0 \right.\\
 \left. + B(\Delta_5,\Delta_6)\left( \frac{\bar z_{56}}{z_{56}} \right)^2  \mathcal{T}^1_0 \mathcal{T}^2_0 + \cdots\right]
\end{gathered}
\label{f2ope}
\end{equation}
where $\mathcal{T}_0^i$'s are given by \eqref{taus}. By comparison with the 5-point amplitudes \eqref{5pt_pure_grav_mellin} and \eqref{5pt_scalar_grav_mellin}, we can write the above equation as follows:
\begin{equation}
    \begin{gathered}
 \mathcal{M}_6\left( 1^{--}_{\Delta_1},2^{--}_{\Delta_2},3^{++}_{\Delta_3},4^{++}_{\Delta_4},5^{++}_{\Delta_5}, 6^{++}_{\Delta_6} \right) =   - \frac{\bar z_{56}}{z_{56}}B(\Delta_5-1,\Delta_6-1)  \mathcal{M}_5\left( 1^{--}_{\Delta_1},2^{--}_{\Delta_2},3^{++}_{\Delta_3},4^{++}_{\Delta_4},\right.\\
 \left. 6^{++}_{\Delta_5+\Delta_6} \right) - \left( \frac{\bar z_{56}}{z_{56}}\right)^2 B(\Delta_5,\Delta_6) \mathcal{M}_5\left( 1^{--}_{\Delta_1},2^{--}_{\Delta_2},3^{++}_{\Delta_3},4^{++}_{\Delta_4} , 6^{\Phi}_{\Delta_5+\Delta_6} \right) + \cdots
    \end{gathered}
\end{equation}
In terms of celestial correlators, the above equation can be written as,
\begin{equation}
    \begin{gathered}
 \left< G^{--}_{\Delta_1}(1)G^{--}_{\Delta_2}(2) G^{++}_{\Delta_3}(3) G^{++}_{\Delta_4}(4) G^{++}_{\Delta_5}(5) G^{++}_{\Delta_6}(6) \right> \\
 =   - \frac{\bar z_{56}}{z_{56}}B(\Delta_5-1,\Delta_6-1)  \left< G^{--}_{\Delta_1}(1) G^{--}_{\Delta_2}(2) G^{++}_{\Delta_3}(3) G^{++}_{\Delta_4}(4)G^{++}_{\Delta_5+\Delta_6}(6) \right> \\
 - \left( \frac{\bar z_{56}}{z_{56}}\right)^2 B(\Delta_5,\Delta_6) \left< G^{--}_{\Delta_1}(1) G^{--}_{\Delta_2}(2)  G^{++}_{\Delta_3}(3) G^{++}_{\Delta_4}(4) \Phi_{\Delta_5+\Delta_6}(6) \right> + \cdots
    \end{gathered}
\end{equation}
This equation implies that the first two holomorphic singular terms in the OPE are
\begin{equation}
\label{GG-OPE}
\begin{gathered}
G^{++}_{\Delta_5}(z_5,\bar z_5) G^{++}_{\Delta_6}(z_6,\bar z_6) = - \frac{\bar z_{56}}{z_{56}}B(\Delta_5-1,\Delta_6-1)G^{++}_{\Delta_5+\Delta_6}(z_6,\bar z_6) \\
- \left( \frac{\bar z_{56}}{z_{56}}\right)^2 B(\Delta_5,\Delta_6)\Phi_{\Delta_5+\Delta_6}(z_6,\bar z_6) + \cdots
\end{gathered}
\end{equation}
This OPE is one of the important results we were after. We will discuss its implications in the next section. For now, let us compute one more higher-order term.

\subsubsection*{The $ \mathcal{O}\left( \frac{\bar z_{56}^2}{z_{56}} \right) $ term}
We now extract the $ \mathcal{O}\left(\frac{\bar z_{56}^2}{z_{56}}\right) $ term. From \eqref{bmmope} and appendix \ref{ope_cal}, we find,
\begin{equation}\label{higher_order_term}
\begin{gathered}
\mathcal{M}_6\left( 1^{--}_{\Delta_1},2^{--}_{\Delta_2},3^{++}_{\Delta_3},4^{++}_{\Delta_4},5^{++}_{\Delta_5}, 6^{++}_{\Delta_6} \right) = 4 i \frac{z_{12}^4}{z_{14}z_{23} \bar z_{14} \bar z_{23}(r_{13,42} - \bar r_{13,42})} \prod_{k=1}^4 \Theta(\epsilon_k \sigma_{k,1})\\
\times \int_{0}^1 dt \, t^{\Delta_5-1} (1-t)^{\Delta_6-1} \left[ (\sigma_{1,1}\sigma_{2,1})^2 + t z_{56} \partial_6 \left( \sigma_{1,1}\sigma_{2,1}\right)^2 + t \bar z_{56} \bar \partial_6 \left( \sigma_{1,1}\sigma_{2,1}\right)^2 \right]\\
\times \left[ \left(\prod_{i=1}^4 {(\epsilon_i \sigma_{i,1})}^{\Delta_i-1}\right) + t z_{56} \partial_6 \left(\prod_{i=1}^4 {(\epsilon_i \epsilon_6 \sigma_{i,1})}^{\Delta_i-1}\right) + t \bar z_{56} \bar \partial_6 \left(\prod_{i=1}^4 {(\epsilon_i \epsilon_6 \sigma_{i,1})}^{\Delta_i-1}\right)\right] \\
\times  \frac{\Gamma(\Delta)}{(i\mathcal{U}_1)^{\Delta}}\left[ 1 - t z_{56}\Delta \frac{\mathcal{U}_2 }{\mathcal{U}_1} - t \bar z_{56} \Delta \frac{ \mathcal{U}_3 }{\mathcal{U}_1}\right] \\
\times \left[ \left( \frac{\bar z_{56}}{z_{56}} \right)^2 \mathcal{T}^1_0 \mathcal{T}^2_0 + \frac{1}{t(1-t)} \left( \frac{\bar z_{56}}{z_{56}} \right) \mathcal{T}^1_0 \mathcal{T}^2_0 \mathcal{T}^3_0 + \frac{1}{t(1-t)}\frac{\bar z_{56}^2}{z_{56}} \left[ \mathcal{T}^1_0 \mathcal{T}^2_0 \lbrace t\mathcal{T}^3_{\bar z} + (1-t)\mathcal{T}^4_{\bar z} \rbrace \right.\right.\\
\left.\left. + \mathcal{T}^3_0 \lbrace \mathcal{T}^1_0 \mathcal{T}^2_{\bar z} + \mathcal{T}^2_0  \mathcal{T}^1_{\bar z} \rbrace + t(1-t)\lbrace \mathcal{T}^1_0 \mathcal{T}^2_{ z} + \mathcal{T}^2_0  \mathcal{T}^1_{ z} \rbrace \right]  \right] + \cdots
\end{gathered}
\end{equation}
where $\mathcal{T}$s are given in the appendix \ref{ope_cal}. From \eqref{higher_order_term} we can now write the $ \mathcal{O}\left(\frac{\bar z_{56}^2}{z_{56}}\right) $ term in the OPE (\ref{GG-OPE}). Let us first note the relations (See appendix \ref{ope_cal} for details.),
\begin{equation}
\begin{gathered}
\mathcal{T}^1_{ z} = t \partial_6 \mathcal{T}^1_0, \ \mathcal{T}^2_{z} = t \partial_6 \mathcal{T}^2_0,\\
\mathcal{T}^1_{\bar z} = t\bar \partial_6 \mathcal{T}^1_0, \ \mathcal{T}^2_{\bar z} = t\bar \partial_6 \mathcal{T}^2_0, \ t\mathcal{T}^3_{\bar z} + (1-t)\mathcal{T}^4_{\bar z} = t\bar \partial_6 \mathcal{T}^3_0, \\
\mathcal{U}_2 = \partial_6 \mathcal{U}_1, \ \mathcal{U}_3 = \bar \partial_6 \mathcal{U}_1.
\end{gathered}
\end{equation}
A straightforward but lengthy computation leads us to the following result, 
\begin{equation}
\begin{gathered}
\mathcal{M}_6\left( 1^{--}_{\Delta_1},2^{--}_{\Delta_2},3^{++}_{\Delta_3},4^{++}_{\Delta_4},5^{++}_{\Delta_5}, 6^{++}_{\Delta_6} \right)\Big{|}_{\mathcal{O}\left(\frac{\bar z_{56}^2}{z_{56}}\right)} = -\frac{\bar z_{56}^2}{z_{56}} \left[ B(\Delta_5,\Delta_6-1) \bar \partial_6 \mathcal{M}_5\left( 1^{--}_{\Delta_1},2^{--}_{\Delta_2}, \right. \right. \\
\left. \left. 3^{++}_{\Delta_3},4^{++}_{\Delta_4},6^{++}_{\Delta_5+\Delta_6}  \right) + B(\Delta_5+1,\Delta_6) \partial_6  \mathcal{M}_5\left( 1^{--}_{\Delta_1},2^{--}_{\Delta_2},3^{++}_{\Delta_3},4^{++}_{\Delta_4},6^{\Phi}_{\Delta_5+\Delta_6} \right)  \right] \, .
\end{gathered}
\label{bmmopeexp2}
\end{equation}
This translates to the following in the OPE:
\begin{equation}
\begin{gathered}
    G^{++}_{\Delta_5}(z_5,\bar z_5) G^{++}_{\Delta_6}(z_6,\bar z_6) \Big{|}_{\mathcal{O}\left(\frac{\bar z_{56}^2}{z_{56}}\right)} = - \frac{\bar z_{56}^2}{z_{56}} \left[ B(\Delta_5,\Delta_6-1) \bar \partial_6 G^{++}_{\Delta_5+\Delta_6}(z_6,\bar z_6) \right.\\
    \left. +  B(\Delta_5+1,\Delta_6) \partial_6 \Phi_{\Delta_5+\Delta_6}(z_6, \bar z_6) \right] \, .
\end{gathered}
\end{equation}
This result will be relevant in reading out the subleading conformal soft terms later. Computing more higher-order terms is beyond the scope of this paper. However, they are important in analysing the null states which give rise to the differential equations for the scattering amplitudes under consideration \cite{Banerjee:2020zlg}. We leave these questions for future investigations. We now move on to computing OPE between a positive helicity outgoing graviton and an outgoing scalar operators.
 
\subsection{OPE between a positive helicity graviton and a scalar}

Expanding RHS of \eqref{bmscope} around $ z_{56}=0, \bar z_{56}=0 $ and keeping the first two holomorphic singular terms, we find,
\begin{equation}
\begin{gathered}
\mathcal{M}_6\left( 1^{--}_{\Delta_1},2^{--}_{\Delta_2},3^{++}_{\Delta_3},4^{++}_{\Delta_4},5^{++}_{\Delta_5}, 6^{\Phi}_{\Delta_6} \right) = - \frac{\bar z_{56}}{z_{56}}B(\Delta_5-1,\Delta_6+1) \mathcal{M}_5\left( 1^{--}_{\Delta_1},2^{--}_{\Delta_2},3^{++}_{\Delta_3}, \right. \\
\left.  4^{++}_{\Delta_4},6^{\Phi}_{\Delta_5+\Delta_6} \right)  - \frac{\bar z_{56}^2}{z_{56}}B(\Delta_5,\Delta_6+1) \bar \partial_6 \mathcal{M}_5\left( 1^{--}_{\Delta_1},2^{--}_{\Delta_2},3^{++}_{\Delta_3},4^{++}_{\Delta_4},6^{\Phi}_{\Delta_5+\Delta_6} \right) + \cdots
\end{gathered}
\label{bmscopeexp1}
\end{equation}
At the level of OPE, we obtain from \eqref{bmscopeexp1}, 

\begin{equation}
\begin{gathered}
G^{++}_{\Delta_5}(z_5, \bar z_5) \Phi_{\Delta_6}(z_6, \bar z_6 ) = - \frac{\bar z_{56}}{z_{56}}B(\Delta_5-1,\Delta_6+1) \Phi_{\Delta_5+\Delta_6}(z_6, \bar z_6)\\
- \frac{\bar z_{56}^2}{z_{56}}B(\Delta_5,\Delta_6+1) \bar \partial_{6} \Phi_{\Delta_5+\Delta_6}(z_6, \bar z_6) + \cdots
\end{gathered}
\end{equation}
This completes our extraction of the relevant OPEs in the putative celestial dual of the BW theory. 
\section{Summary and implications of OPEs }
\label{summary}

Let us summarise the results we obtained so far and discuss their implications. In the celestial CFT dual of BW theory, the tree-level OPE between two positive helicity outgoing graviton primary operators with conformal dimensions $\Delta_1$ and $\Delta_2$, inserted at the points $(z,\bar z)$ and $(w, \bar w)$ on the celestial sphere is given by,
\begin{equation}\label{gg_OPE_summary}
\begin{gathered}
G^{++}_{\Delta_1}(z,\bar z) G^{++}_{\Delta_2}(w,\bar w) = - \frac{(\bar z - \bar w)}{(z-w)}B(\Delta_1-1,\Delta_2-1)G^{++}_{\Delta_1+\Delta_2}(w,\bar w) \\
 - \frac{(\bar z - \bar w)^2}{(z-w)}B(\Delta_1,\Delta_2-1) \partial_{\bar w} G^{++}_{\Delta_1+\Delta_2}(w,\bar w) - \frac{(\bar z- \bar w)^2}{(z-w)^2} B(\Delta_1,\Delta_2) \Phi_{\Delta_1+\Delta_2}(w,\bar w) \\
 - \frac{(\bar z - \bar w)^2}{(z-w)} B(\Delta_1+1,\Delta_2) \partial_{w}\Phi_{\Delta_1+\Delta_2}(w,\bar w) + \cdots \\
 \end{gathered}
\end{equation}
The tree-level OPE between a positive helicity outgoing graviton primary operator and an outgoing scalar primary operator is given by,
\begin{equation}\label{gphi_OPE_summary}
\begin{gathered}
G^{++}_{\Delta_1}(z, \bar z) \Phi_{\Delta_2}(w, \bar w ) = - \frac{(\bar z - \bar w)}{(z-w)}B(\Delta_1-1,\Delta_2+1) \Phi_{\Delta_1+\Delta_2}(w, \bar w)\\
- \frac{(\bar z - \bar w)^2}{(z-w)}B(\Delta_1,\Delta_2+1) \partial_{\bar w} \Phi_{\Delta_1+\Delta_2}(w, \bar w) + \cdots
\end{gathered}
\end{equation}

\subsection{Implications on the bulk theory}

Suppose there is a hypothetical 2$d$ celestial CFT dual of a gravitational theory with spin-2 and scalar primary operators, and the OPEs among them are given by \eqref{gg_OPE_summary}, \eqref{gphi_OPE_summary}. Given these OPEs what can one say about the bulk theory? We try to answer this question by analysing the OPE between different conformal soft operators (currents) and hard primary operators. The leading conformal soft graviton operator for a positive helicity graviton is defined as \cite{Donnay:2018neh,Pate:2019mfs,Fan:2019emx,Nandan:2019jas,Adamo:2019ipt,Puhm:2019zbl,Guevara:2019ypd,Pano:2021ewd},
\begin{equation}
    H^1(z,\bar z) = \lim_{\Delta \to 1} (\Delta-1)G^{++}_{\Delta}(z,\bar z) \, .
\end{equation}
Taking this limit in equation \eqref{gg_OPE_summary} and \eqref{gphi_OPE_summary} we find
\begin{equation}\label{ls_OPE}
\begin{gathered}
H^1(z,\bar z) G^{++}_{\Delta}(w,\bar w) \sim - \frac{(\bar z - \bar w)}{(z-w)}G^{++}_{\Delta + 1}(w,\bar w), \\
H^1(z,\bar z) \Phi_{\Delta}(w, \bar w ) \sim - \frac{(\bar z - \bar w)}{(z-w)} \Phi_{\Delta + 1}(w, \bar w) \, .
 \end{gathered}
\end{equation}
This is the same OPE between the leading conformal soft graviton operator and a hard primary operator that follows from the leading conformal soft theorems in two derivative theories of gravity \cite{Banerjee:2020zlg,Pate:2019lpp,Banerjee:2020kaa,Banerjee:2021cly,Himwich:2021dau,Banerjee:2021dlm,Guevara:2021abz,Strominger:2021lvk,Banerjee:2023zip,Banerjee:2020vnt,Ebert:2020nqf,Banerjee:2023rni,Adamo:2022wjo,Ren:2023trv,Hu:2021lrx,Bhardwaj:2022anh,Krishna:2023ukw}. Let us proceed and compute the OPE between the subleading conformally soft graviton operator and a hard primary operator. The subleading conformally soft graviton operator is defined by,
\begin{equation}
    H^0(z,\bar z) = \lim_{\Delta \to 0} \Delta G^{++}_{\Delta}(z,\bar z).
\end{equation}
Taking this limit in equation \eqref{gg_OPE_summary} and \eqref{gphi_OPE_summary} we get
\begin{equation}\label{sl_OPE}
\begin{gathered}
H^0(z,\bar z) G^{++}_{\Delta}(w,\bar w) \sim \frac{(\bar z - \bar w)}{(z-w)} (\Delta-2)G^{++}_{\Delta}(w,\bar w) - \frac{(\bar z - \bar w)^2}{(z-w)} \partial_{\bar w} G^{++}_{\Delta}(w,\bar w) \\
 - \frac{(\bar z- \bar w)^2}{(z-w)^2} \Phi_{\Delta}(w,\bar w) \, , \\
 H^0(z,\bar z) \Phi_{\Delta}(w,\bar w) \sim \frac{(\bar z - \bar w)}{(z-w)} \Delta \Phi_{\Delta}(w,\bar w) - \frac{(\bar z - \bar w)^2}{(z-w)} \partial_{\bar w} \Phi_{\Delta}(w,\bar w).
 \end{gathered}
\end{equation}
Now, it is known \cite{Banerjee:2020zlg, Guevara:2021abz,Strominger:2021lvk} that, if we consider conformal soft graviton theorem for a positive helicity graviton of any Einstein-type theory in the bulk, then the OPE between the positive helicity subleading conformally soft graviton operator and any hard primary is given by \eqref{sl_OPE} with just the simple pole terms. Thus, we see that the OPE we have considered for a 2$d$ celestial CFT dictates that the subleading soft graviton theorem in the bulk must have changed due to the presence of the extra term, namely, the double pole term of the first equation in \eqref{sl_OPE}. In the next section, we will indeed show, by directly analysing the momentum space amplitudes, that the subleading soft graviton theorem is modified for the BW-theory amplitudes. The modification is due to one of the hard graviton primary operators getting changed to a scalar primary operator. This kind of particle-changing phenomenon has been seen in effective field theories also; however, they do not modify the subleading soft graviton theorem, but only the higher order ones \cite{Elvang:2016qvq,Laddha:2017vfh}.

It has been shown that for Einstein-type theories of gravity, the subleading soft graviton theorem is universal \cite{Laddha:2017kpo}. So our OPE analysis suggests that if we start with an OPE such as  \eqref{gg_OPE_summary}, then the dual bulk gravity theory cannot be Einstein-type. Therefore, the OPE structure in the celestial CFT can differentiate Einstein-type theories from others in the bulk. 
However, surprisingly, as we will show in section \ref{algebra}, the chiral $\mathfrak{bms}_4$ symmetry algebra remains unchanged, even though the subleading soft graviton theorem has been modified, albeit in a well-controlled fashion. In other words, the subleading conformally soft graviton theorem can be interpreted as the Ward identity of the $\mathfrak{sl}(2,\mathbb{R})$ current algebra, but the representation is different.

\section{Momentum space soft expansions}
\label{soft_fac}

In this section, we take a generic $(n+1)$-point momentum space amplitude in the MHV configuration and derive its leading and subleading soft expansions. Following \cite{Cachazo:2014kd}, we will work with stripped amplitudes only. Let us consider an $(n+1)$-point amplitude with two negative helicity gravitons, $(r-3)$ scalars and $(n-r-2)$ positive helicity gravitons. We denote the amplitude by $M_{n+1}(1^{--},2^{--}, 3_\Phi,\ldots,r_{\Phi}, (r+1)^{++},\ldots,(n+1)^{++})$. From \eqref{sup_amp}, we can write the explicit form of this amplitude as,
\begin{eqnarray} \label{n+1_point_amp}
    \begin{gathered}
     M_{n+1}(1^{--},2^{--}, 3_\Phi,\ldots,r_{\Phi}, (r+1)^{++},\ldots,(n+1)^{++}) \\
     = (-1)^{n+1}  i \left< 1,2 \right>^4 \prod_{i=r+1}^{n+1} \left(\sum_{j=1, j\neq i}^{n+1} \frac{[i,j]\left<j,1\right>^2}{\left< i,j \right>\left< i,1 \right>^2}\right) = (-1)^{n+1}  i \left< 1,2 \right>^4 \prod_{i=r+1}^{n+1} A_{n+1}(i)
    \end{gathered}
\end{eqnarray}
where we have chosen the reference spinor to be 1 and
\begin{equation}\label{A_n}
    A_{n}(i) = \sum_{j=1, j\neq i}^{n} \frac{[i,j]\left<j,1\right>^2}{\left< i,j \right>\left< i,1 \right>^2} \,\, \, .
\end{equation}
Note that the product in \eqref{n+1_point_amp} runs over positive helicity gravitons only. Similarly, the $n$-point amplitude with one less positive helicity graviton is given by,
\begin{eqnarray} \label{n_point_amp}
    \begin{gathered}
     M_{n}(1^{--},2^{--}, 3_\Phi,\ldots,r_{\Phi}, (r+1)^{++},\ldots,n^{++}) \\
     = (-1)^{n}  i \left< 1,2 \right>^4 \prod_{i=r+1}^{n} \left(\sum_{j=1, j\neq i}^{n} \frac{[i,j]\left<j,1\right>^2}{\left< i,j \right>\left< i,1 \right>^2}\right) = (-1)^{n}  i \left< 1,2 \right>^4 \prod_{i=r+1}^{n} A_{n}(i) \, .
    \end{gathered}
\end{eqnarray}
We will also require an $n$-point amplitude where one positive helicity graviton in the above amplitude has been replaced by a scalar. This is given by
\begin{eqnarray} \label{n_point_amp_phi}
    \begin{gathered}
    M_{n}(1^{--},2^{--}, 3_\Phi,\ldots,r_{\Phi}, (r+1)^{++},\ldots, a_\phi,\ldots,n^{++}) \\
     = (-1)^{n}  i \left< 1,2 \right>^4 \prod_{i=r+1, i\neq a}^{n} \left(\sum_{j=1, j\neq i}^{n} \frac{[i,j]\left<j,1\right>^2}{\left< i,j \right>\left< i,1 \right>^2}\right) = (-1)^{n}  i \left< 1,2 \right>^4 \prod_{i=r+1, i \neq a}^{n} A_{n}(i) \, .
    \end{gathered}
\end{eqnarray}
Using momentum conservation, we can replace $ \tilde{\lambda}_{1\dot{\alpha}}$ and  $\tilde{\lambda}_{2\dot{\alpha}}$  in $(n+1)$- and $n$-point amplitudes. However, due to our choice of reference spinor, the amplitudes do not depend on $ \tilde{\lambda}_{1\dot{\alpha}}$. On the support of the $n$-point delta function, $\tilde{\lambda}_{2\dot{\alpha}}$ is given by,
\begin{eqnarray}\label{lambdat2}
    \tilde{\lambda}_{2\dot{\alpha}} = - \sum_{i=3}^n \frac{\left<1, i\right>}{\left<1, 2\right>}  \tilde{\lambda}_{i\dot{\alpha}}
\end{eqnarray}
Substituting $\tilde{\lambda}_{2\dot{\alpha}}$ from \eqref{lambdat2},  into \eqref{A_n} and performing some straightforward algebra, we obtain,
\begin{equation}\label{A_n_final}
    A_{n}(i) = \sum_{j=3, j\neq i}^{n} \frac{[i,j]\left<1,j\right> \left<2,j\right>}{\left< i,j \right>\left< 1,i \right> \left< 2,i \right>} \, .
\end{equation}
We will consider the $(n+1)$-th graviton in the amplitude \eqref{n+1_point_amp} to be soft. The momentum of this graviton can be written as $p_{n+1, \alpha \dot{\alpha}}=\lambda_{n+1,\alpha} \tilde{\lambda}_{n+1,\dot{\alpha}} $. As discussed in \cite{Cachazo:2014kd}, the soft limit, $p_{n+1} \to 0 $, can be taken by sending the holomorphic spinor to 0, that is, $\lambda_{n+1} \to 0 $, keeping the anti-holomorphic spinor fixed and generic. So we scale the holomorphic spinor as $\lambda_{n+1} \to \epsilon \lambda_{n+1} $ and send $\epsilon \to 0$. As we can see from \eqref{A_n_final}, $A_{n+1}(n+1)$ contains three $\lambda_{n+1}$ in the denominator but none in the numerator. Hence, scaling $\lambda_{n+1} $ by $\epsilon \lambda_{n+1} $ in $A_{n+1}(n+1)$, we get
\begin{equation} \label{Anp1}
  A_{n}(n+1) = \frac{1}{\epsilon^3}\sum_{j=3}^{n} \frac{[n+1,j]\left<1,j\right> \left<2,j\right>}{\left< n+1,j \right>\left< 1, n+1 \right> \left< 2, n+1 \right>} ~~ . 
\end{equation}
A similar calculation for $A_{n+1}(i), i \neq n+1 $ gives the following result:
\begin{equation} \label{ai}
\begin{split}
 A_{n+1}(i)  &=  \sum_{j=3, j\neq i}^{n} \frac{[i,j]\left<1,j\right> \left<2,j\right>}{\left< i,j \right>\left< 1,i \right> \left< 2,i \right>} + \epsilon \frac{[i,n+1]\left<1,n+1\right> \left<2,n+1\right>}{\left< i, n+1 \right>\left< 1,i \right> \left< 2,i \right>}\\
 &= A_n(i) + \epsilon \frac{[i,n+1]\left<1,n+1\right> \left<2,n+1\right>}{\left< i, n+1 \right>\left< 1,i \right> \left< 2,i \right>} ~~.
 \end{split}
\end{equation}
We now use these results to derive leading and subleading soft terms for the $(n+1)$-point amplitude \eqref{n+1_point_amp}.

\subsection{Leading soft factor}

Substituting \eqref{Anp1} and \eqref{ai} in \eqref{n+1_point_amp} and keeping the leading term in  $\epsilon \to 0$ (equivalent to $p_{n+1} \to 0$) limit gives the following result:
\begin{equation}
\label{lsf-momsp}
\begin{split}
    \lim_{p_{n+1}\to 0}  M_{n+1}(1^{--},2^{--}, 3_\Phi,\ldots,r_{\Phi}, (r+1)^{++},\ldots, (n+1)^{++})\big|_{\textnormal{leading}} \\
    =  (-1)^{n+1}  i \left< 1,2 \right>^4 \sum_{j=3}^{n} \frac{[n+1,j]\left<1,j\right> \left<2,j\right>}{\left< n+1,j \right>\left< 1, n+1 \right> \left< 2, n+1 \right>}  \prod_{i=r+1}^{n} A_{n}(i) \\
    = - S^{(0)}  M_{n}(1^{--},2^{--}, 3_\Phi,\ldots,r_{\Phi}, (r+1)^{++},\ldots,n^{++})
\end{split}
\end{equation}
where $S^{(0)} = \sum_{j=3}^{n} \frac{[n+1,j]\left<1,j\right> \left<2,j\right>}{\left< n+1,j \right>\left< 1, n+1 \right> \left< 2, n+1 \right>}$ is the same as the universal leading soft factor for Einstein-type theories of gravity. The overall minus sign in \eqref{lsf-momsp} is there because the amplitude alternates sign with the number of external particles. Thus, we see, at least in one example of four derivative theories of gravity, that the leading soft factorisation of the amplitude in the MHV configuration is still universal and is the same as that of Einstein-type theories.

\subsection{Subleading soft factor}
\label{ssf}

The subleading term in the soft expansion of \eqref{n+1_point_amp} is given by,
\begin{equation}
    \begin{gathered}
    \lim_{p_{n+1}\to 0} M_{n+1}(1^{--},2^{--}, 3_\Phi,\ldots,r_{\Phi}, r+1^{++},\ldots,n+1^{++})\big|_{\textnormal{subleading}} \\
     = (-1)^{n+1}  i \left< 12 \right>^4 \sum_{a=3}^n \frac{[n+1, a]}{\left< n+1,a \right>}\left< 1,a \right>\left< 2,a \right> \sum_{j=r+1}^n \frac{[n+1, j]}{\left< n+1, j \right>}\frac{1}{\left< 1,j \right>\left< 2,j \right>}\prod_{i=r+1, i\neq j}^{n} A_{n}(i)
    \end{gathered}
\end{equation}  
We can divide the RHS of the above equation into two pieces depending on whether $a=j$ or $a\neq j$. By doing this, we obtain,
\begin{equation} \label{sub_fac}
    \begin{gathered}
       \lim_{p_{n+1}\to 0} M_{n+1}(1^{--},2^{--}, 3_\Phi,\ldots,r_{\Phi}, r+1^{++},\ldots,n+1^{++})\big|_{\textnormal{Subleading}} \\
     = (-1)^{n+1}  i \left< 12 \right>^4 \sum_{j=r+1}^n \sum_{a=3,a\neq j}^n \frac{[n+1, a][n+1, j]\left< 1,a \right>\left< 2,a \right>}{\left< n+1,a \right>\left< n+1, j \right>\left< 1,j \right>\left< 2,j \right>}\prod_{i=r+1, i\neq j}^{n} A_{n}(i) \\
     + (-1)^{n+1}  i \left< 12 \right>^4 \sum_{a=r+1}^n \frac{[n+1,a]^2}{\left< n+1,a\right>^2} \prod_{i=r+1, i\neq a}^{n} A_{n}(i) \, .
      \end{gathered}
\end{equation}
Now, it is not hard to see that the second term of the RHS of the above equation is proportional to an $n$-point amplitude \eqref{n_point_amp_phi} where one of the positive helicity gravitons  has been replaced by a scalar (recall that we started with an $(n+1)$-point amplitude where we had positive helicity gravitons from $r+1$ to $n+1$ and we took $(n+1)$-th graviton to be soft). More precisely,
\begin{equation} \label{partcl_chang}
    \begin{gathered}
    (-1)^{n+1}  i \left< 12 \right>^4 \sum_{a=r+1}^n \frac{[n+1,a]^2}{\left< n+1,a\right>^2} \prod_{i=r+1, i\neq a}^{n} A_{n}(i) \\
    = - \sum_{a=r+1}^n \frac{[n+1,a]^2}{\left< n+1,a\right>^2}  M_{n}(1^{--},2^{--}, 3_\Phi,\ldots,r_{\Phi}, (r+1)^{++},\ldots, a_\Phi,\ldots,n^{++}) \, .
    \end{gathered}
\end{equation}
Let us now concentrate on the first term of \eqref{sub_fac}. Recall that the subleading soft operator for two derivative theories of gravity is given by
\begin{equation}
    \begin{gathered}
        S^{(1)} = \frac{1}{2} \sum_{a=1}^n \frac{[n+1,a]}{\left< n+1,a\right>}\left( \frac{\left< x,a \right>}{\left< x,n+1\right>} + \frac{\left< y,a \right>}{\left< y,n+1\right>}\right) \tilde{\lambda}_{n+1}^{\dot{\alpha}} \frac{\partial}{\partial \tilde{\lambda}_{a}^{\dot{\alpha}}}
    \end{gathered}
\end{equation}
where $x,y$ are two reference spinors. We choose $x=1,y=2$. Applying this operator on the $n$-point amplitude \eqref{n_point_amp} we find,
\begin{equation}\label{SM_n}
    \begin{gathered}
     S^{(1)}  M_{n}(1^{--},2^{--}, 3_\Phi,\ldots,r_{\Phi}, r+1^{++},\ldots,n^{++})\\
     = (-1)^n i \left<12\right>^4 \sum_{j=r+1}^n \left\lbrace \left( S^{(1)} A_n(j) \right) \prod_{i=r+1,i \neq j}^n A_n(i) \right\rbrace.
    \end{gathered}
\end{equation}
Now, another straightforward algebra gives,
\begin{equation}\label{SA_n}
    \begin{gathered}
        S^{(1)} A_n(j) = \sum_{a=3, a\neq j}^n \frac{[n+1,a][n+1,j]\left< 1,a \right> \left<2,a \right>}{\left< n+1,a \right> \left< n+1,j \right> \left< 1,j \right> \left< 2,j \right>} ~~ .
    \end{gathered}
\end{equation}
In deriving the above equation, we used the Shouten identity 
$$\left<i,j\right>\left<k,l\right>+\left<i,k\right>\left<l,j\right>=\left<i,l\right>\left<k,j\right>$$ in the intermediate steps. Substituting \eqref{SA_n} in \eqref{SM_n} we get
\begin{equation}\label{SM_n_fin}
    \begin{gathered}
     S^{(1)}  M_{n}(1^{--},2^{--}, 3_\Phi,\ldots,r_{\Phi}, (r+1)^{++},\ldots,n^{++})\\
     = (-1)^n i \left<12\right>^4 \sum_{j=r+1}^n  \left( \sum_{a=3, a\neq j}^n \frac{[n+1,a][n+1,j]\left< 1,a \right> \left<2,a \right>}{\left< n+1,a \right> \left< n+1,j \right> \left< 1,j \right> \left< 2,j \right>} \right) \prod_{i=r+1,i \neq j}^n A_n(i). 
    \end{gathered}
\end{equation}
Using  \eqref{partcl_chang} and \eqref{SM_n_fin} in \eqref{sub_fac}, we then finally obtain
\begin{eqnarray}
    \begin{gathered}
        \lim_{p_{n+1}\to 0} M_{n+1}(1^{--},2^{--}, 3_\Phi,\ldots,r_{\Phi}, (r+1)^{++},\ldots, (n+1)^{++})\big|_{\textnormal{subleading}} \\
     = - S^{(1)}  M_{n}(1^{--},2^{--}, 3_\Phi,\ldots,r_{\Phi}, (r+1)^{++},\ldots,n^{++})\\
     - \sum_{a=r+1}^n \frac{[n+1,a]^2}{\left< n+1,a\right>^2}  M_{n}(1^{--},2^{--}, 3_\Phi,\ldots,r_{\Phi}, (r+1)^{++},\ldots, a_\Phi,\ldots,n^{++})
    \end{gathered}
\end{eqnarray}
Thus, as discussed before, we indeed see that the subleading soft graviton theorem is modified in the BW theory. However, the interesting fact is that the new term in the subleading soft factor is again quadratic in the anti-holomorphic coordinate of the subleading soft graviton operator since $ [n+1,a]^2 \sim (\bar z_{n+1}-\bar z_a)^2 $. So we still have three currents from the modified subleading soft factor, and as we will show in the next section, the mode algebra of these currents is still the good old $\mathfrak{sl}(2,\mathbb{R}) $ current algebra.

\section{Symmetry algebra}
\label{algebra}

In this section, we compute the symmetry algebra that follows from the OPE given by results \eqref{ls_OPE} and \eqref{sl_OPE}. The leading conformally soft positive helicity graviton operator admits a truncated mode expansion in the anti-holomorphic variable \cite{Banerjee:2020zlg}, given by
\begin{eqnarray}
    \label{lead_exp}
    H^1(z,\bar z) = H^1_{\frac{1}{2}}(z) + \bar z H^1_{-\frac{1}{2}}(z)
\end{eqnarray}
where $H^1_{\frac{1}{2}}(z)$ and $H^1_{-\frac{1}{2}}(z)$ are two holomorphic supertranslation currents. Then, the OPEs between these currents and other primary operators follow from \eqref{ls_OPE} and \eqref{sl_OPE}, given by,  
\begin{equation}\label{ls_currents_OPE}
\begin{gathered}
H^1_{-\frac{1}{2}}(z) G^{++}_{\Delta}(w,\bar w) \sim - \frac{1}{(z-w)}G^{++}_{\Delta + 1}(w,\bar w), \ H^1_{\frac{1}{2}}(z) G^{++}_{\Delta}(w,\bar w) \sim  \frac{\bar w}{(z-w)}G^{++}_{\Delta + 1}(w,\bar w) ,\\
H^1_{-\frac{1}{2}}(z) \Phi_{\Delta}(w, \bar w ) \sim - \frac{1}{(z-w)} \Phi_{\Delta + 1}(w, \bar w), \ H^1_{\frac{1}{2}}(z) \Phi_{\Delta}(w, \bar w ) \sim  \frac{\bar w}{(z-w)} \Phi_{\Delta + 1}(w, \bar w) \, .
 \end{gathered}
\end{equation}
The holomorphic modes of these supertranslation currents $H^1_{m,\pm \frac{1}{2}}$ satisfy the abelian algebra, 
\begin{eqnarray}
    [H^1_{m,\pm \frac{1}{2}}, H^1_{n,\pm \frac{1}{2}} ] = 0 \, .
\label{leading_comm}
\end{eqnarray}
As we discussed in the previous section,  the subleading conformal soft graviton theorem gets modified in the BW theory. However, its quadratic dependence on the anti-holomorphic coordinate of the subleading conformally soft graviton operator remains the same as the subleading conformal soft graviton theorem of Einstein-type theories. Hence, we can again decompose the subleading conformal soft graviton operator as follows \cite{Banerjee:2020zlg}:
\begin{equation}\label{sl2_currents}
 H^0(z,\bar z) = H^0_1(z) + \bar z H^0_0(z) + \bar z^2 H^0_{-1}(z)
\end{equation}
where $H^0_{a}(z), \, a = 0, \pm 1,$ are three holomorphic currents. Here we have used the standard notation for subleading soft graviton currents \cite{Strominger:2021lvk}.

Now, using \eqref{sl_OPE}, we can write the OPEs between the above currents and any of the hard primary operators. They are given by
\begin{equation}
\begin{gathered}
\label{sl_currents_OPE}
H^0_1(z) G^{++}_{\Delta}(w,\bar w) \sim  - \frac{(\Delta-2)\bar w}{(z-w)}G^{++}_\Delta(w, \bar w) - \frac{\bar w^2}{(z-w)} \partial_{\bar w}G^{++}_\Delta(w, \bar w) \\
- \frac{\bar w^2}{(z-w)^2} \Phi_\Delta(w, \bar w) \, , \\
H^0_0(z) G^{++}_{\Delta}(w,\bar w) \sim \frac{(\Delta-2)}{(z-w)}G^{++}_\Delta(w, \bar w) + \frac{2\bar w}{(z-w)}\partial_{\bar w} G^{++}_\Delta(w, \bar w) \\
+ \frac{2\bar w}{(z-w)^2} \Phi_\Delta(w,\bar w) \, ,  \\
H^0_{-1}(z) G^{++}_{\Delta}(w,\bar w) \sim -\frac{1}{(z-w)}\partial_{\bar w} G^{++}_\Delta(w, \bar w) - \frac{1}{(z-w)^2} \Phi_\Delta(w,\bar w) \, , \\
H^0_1(z) \Phi_{\Delta}(w,\bar w) \sim -\frac{\Delta \, \bar w}{(z-w)}\Phi_\Delta(w,\bar w) - \frac{\bar w^2}{(z-w)}\partial_{\bar w} \Phi_\Delta(w,\bar w)  \, ,\\
H^0_0(z) \Phi_{\Delta}(w,\bar w) \sim \frac{\Delta}{(z-w)}\Phi_\Delta(w,\bar w) + \frac{2\bar w}{(z-w)}\partial_{\bar w} \Phi_\Delta(w,\bar w) , \\
H^0_{-1}(z) \Phi_{\Delta}(w,\bar w) \sim -\frac{1}{z-w} \partial_{\bar w} \Phi_\Delta(w,\bar w)  \, .
\end{gathered}
\end{equation}
%
\begin{comment}
The holomorphic mode decomposition of any operator with weight $ (h,\bar h) $ is given by \cite{DiFrancesco:1997nk},
\begin{equation}
\Phi_l(\bar z) = \oint \frac{dz}{2\pi i} z^{l+h-1} \Phi_{h,\bar h}(z, \bar z)
\end{equation}
\end{comment}
For the currents $ H^0_{a} $, with holomorphic weight $ h=1 $, the holomorphic mode decomposition is \cite{DiFrancesco:1997nk}
\begin{equation}
H^0_{n,a} = \oint \frac{dz}{2\pi i} z^{n} H^0_{a}(z)
\end{equation}
Using the OPE \eqref{sl_currents_OPE}, one can compute the following action of the holomorphic modes on the graviton and scalar primaries,
\begin{equation}
\begin{split}
\label{commutators}
[H^{0}_{n,1},G^{++}_{\Delta}(z,\bar z)] &= - z^n[(\Delta-2)\bar z + \bar z^2 \partial_{\bar z} ]G^{++}_{\Delta}(z,\bar z) - n z^{n-1} \bar z^2 \Phi_{\Delta}(z,\bar z) , \\
[H^{0}_{n,0},G^{++}_{\Delta}(z,\bar z)] &= z^n[(\Delta-2) + 2 \bar z \partial_{\bar z}] G^{++}_{\Delta}(z,\bar z) + 2 n z^{n-1} \bar z \Phi_{\Delta}(z,\bar z) , \\
[H^{0}_{n,-1},G^{++}_{\Delta}(z, \bar z)] &= -z^n \partial_{\bar z} G^{++}_{\Delta}(z,\bar z) - n z^{n-1} \Phi_{\Delta}(z, \bar z) , \\
[H^{0}_{n,1},\Phi_{\Delta}(z, \bar z)] &= - z^n[ \Delta \bar z + \bar z^2 \partial_{\bar z} ] \Phi_{\Delta}(z,\bar z) , \\
[H^{0}_{n,0}, \Phi_{\Delta}(z, \bar z)] &= z^n[\Delta + 2 \bar z \partial_{\bar z}] \Phi_{\Delta}(z,\bar z) , \\
[H^{0}_{n,-1}, \Phi_{\Delta}(z, \bar z)] &= -z^n \partial_{\bar z} \Phi_{\Delta}(z, \bar z) \, .
\end{split}
\end{equation}
 We, now impose the Jacobi identity, 
 \begin{equation}\label{jacobi_id}
[A,[B,C]]+[B,[C,A]]+[C,[A,B]] = 0
\end{equation}
for two $H^0_{n,a}$ and one of $\{G^{++}_\Delta(z, \bar z), \Phi_\Delta(z, \bar z)\}$, and use the above commutators \eqref{commutators}, to compute the algebra between different modes of the three currents $H^0_{m,a}$. We find that the algebra, modulo central terms, is given by,
\begin{equation}
\begin{split}
[H^0_{m,1},H^0_{n,-1}]=H^0_{m+n,0}, [H^0_{m,1},H^0_{n,0}]=2H^0_{m+n,1}, [H^0_{m,0},H^0_{n,-1}]=2H^0_{m+n,-1}
\end{split}
\label{current-alg}
\end{equation}
This is simply the $\mathfrak{sl}(2,\mathbb{R})$ algebra, first discussed in \cite{Banerjee:2020zlg}, by analysing the subleading soft positive helicity graviton theorem in the MHV sector of the Einstein gravity and later realised as the asymptotic symmetry algebra of asymptotically locally flat spacetimes in \cite{Gupta:2021cwo}. So, we conclude that, though the subleading soft graviton theorem has changed, the $\mathfrak{sl}(2,\mathbb{R})$ current algebra symmetries remains the same. In other words, the subleading soft graviton theorem can be thought of as the Ward identities for the three $\mathfrak{sl}(2,\mathbb{R})$ currents but with a different realisation. One can also check that the commutators between the modes $H^{1}_{m,\pm\frac{1}{2}}$  of the supertranslation generators, and  the modes  $ \lbrace H^0_{n,\pm 1}, H^0_{n,0} \rbrace  $ of the $\mathfrak{sl}(2,\mathbb{R})$ generators are the same as the chiral $\mathfrak{bms}_4$.

The action of the modes $H^0_{n,a}$ of the $\mathfrak{sl}(2, {\mathbb R})$ currents in \eqref{commutators} provides an interesting representation, mixing the two primaries $\{G^{++}_\Delta(z, \bar z), \Phi_\Delta(z, \bar z)\}$. Note, however, that for the zero-mode $\mathfrak{sl}(2, {\mathbb R})$ subalgebra generators $H^0_{0,a}$ the $\Phi_\Delta(z, \bar z)$ dependent terms on the RHS of the first three equations drops out. Therefore, the upper triangular nature of this representation is only for the non-zero modes $\{H^0_{n,a}, n\ne 0\}$ of the currents. It will be intersting to understand such representations and their role in the current context better.

\section{Discussion}
\label{discussion}

Operator product expansions play an important role in celestial CFTs, with the singularity structure of the OPE encoding information about the bulk interactions and propagators. In Einstein-type theories, for instance, the OPE between an outgoing positive helicity graviton and any other primary operator always exhibits a holomorphic simple pole singularity as shown in  \cite{Pate:2019lpp}. As we have shown in this paper, the conformally invariant theory of gravity, specifically the BW theory, also gives the same singularity structure in the OPE between an outgoing positive helicity graviton primary  and a scalar primary operator (equation \eqref{gphi_OPE_summary}). However, the OPE between two positive helicity outgoing gravitons displays a double pole singularity multiplied by a scalar primary operators, apart from the usual simple pole holomorphic singularity (equation \eqref{gg_OPE_summary}). Thus, we need to scan over all the OPE relations among the primary operators in the boundary theory to better characterise the bulk dynamics. 

We have also shown that, in the BW theory, the OPE of the leading conformally soft graviton current for a positive helicity graviton with any other primary operator shows no difference from that of the Einstein-type theories, while that of the subleading conformally soft graviton current is different. This modification manifests as a correction to the subleading soft graviton theorem, which we confirmed through soft expansion analysis of scattering amplitudes of the BW theory in momentum space. In particular, by considering a generic $(n+1)$-point tree-level MHV scattering amplitude, we have shown that the leading soft term remains the same as that expected in Einstein-type theories, whereas the subleading term gets corrected. Interestingly, however, the chiral $\mathfrak{sl}(2,\mathbb{R})$ current algebra that follows from the subleading positive helicity soft graviton theorem remains the same. This raises an important question: can we classify all gravitational theories whose dual celestial primary operators transforms under non-trivial representations, such as the one we encountered here, of the chiral $\mathfrak{bms}_4$ algebra? Attempts in this direction were pursued in \cite{Banerjee:2023zip}, however, without taking into account representations of the kind that arose here.

In the context of the non-abelian gauge theory with a kinetic term of the type $(DF)^2$ considered in Appendix \ref{dfsquare}, we have found that the leading soft gluon theorem itself is modified, and yet leaving the algebra responsible for the factorisation unchanged. That is, the algebra is still the same Kac-Moody algebra one obtains from the positive helicity leading soft gluon theorem in Yang-Mills type theories.

In the case of BW theory it is not clear to us why the leading soft terms are the same as those expected from Einstein-type theories. There is some folk-lore (see for example \cite{Schwartz:2014sze}) that the amplitudes in any diffeomorphism invariant theory of gravity are expected to have this universal leading soft behaviour. A simple re-run of these arguments, even though do predict a universal term at the leading order, do not seem to necessarily rule out corrections to it at the same order. The fact that there are no such corrections in the BW theory might be due to some other hidden symmetries of the theory. We comment on one such possibility below.

The BW theory we considered is known to include gravitational interactions that respect both diffeomorphism and Weyl symmetries. One expects, on general grounds, that the scattering amplitudes of this theory ( for degrees of freedom around the Minkowski spacetime) to respect not just the Poincaré symmetries but the full conformal symmetries. It is therefore natural to ask, just as the enhancement of Poincaré symmetries in Einstein-type theories to the (appropriate extension/variation of the) famous $\mathfrak{bms}_4$ symmetries, if the relevant symmetries in the context of BW theory would be a conformal variant of the chiral $\mathfrak{bms}_4$. There does exist a chiral ${\cal W}$-algebra extension of the chiral $\mathfrak{bms}_4$ which can be referred to as the chiral conformal $\mathfrak{bms}_4$ (see the appendix \ref{con_bms4} for details) that admits the chiral $\mathfrak{bms}_4$ as a proper subalgebra.\footnote{In \cite{Haco:2017ekf}, a non-chiral extension of the $\mathfrak{bms}_4$ algebra to a conformal version has been discussed, which, unlike our extension, is a linear algebra.}  Therefore, it becomes interesting to ask if there is a hidden symmetry algebra of the MHV scattering amplitudes of the BW theory that is bigger than the chiral $\mathfrak{bms}_4$ and if it coincides with this chiral conformal $\mathfrak{bms}_4$ or not. We hope to report on some progress in this direction in the near future.

\section*{Acknowledgements} 

We would like to thank Alok Laddha for his very helpful discussions and suggestions throughout the course of this work. We also want to thank Akavoor Manu and Arkajyoti Manna for various helpful conversations.

\appendix

\section{Brief review of celestial amplitudes for massless scattering}
\label{review}
The Celestial amplitude for massless particles in four dimensions is defined as the Mellin transformation of the $S$-matrix element, $A_n\big(\{\omega_i,z_i,\bar z_i, \sigma_i\}$, given by \cite{Pasterski:2017kqt}
\begin{equation}\label{mellin}
\mathcal M_n\big(\{z_i, \bar z_i, h_i, \bar h_i\}\big) = \prod_{i=1}^{n} \int_{0}^{\infty} d\omega_i \ \omega_i^{\Delta_i -1} A_n\big(\{\omega_i,z_i,\bar z_i, \sigma_i\}\big)\,,
\end{equation} 
where $\sigma_i$ denotes the helicity of the $i$-th particle and the on-shell momenta are parametrised by \eqref{null_mom}. The conformal weights $(h_i,\bar h_i)$ are given by
\begin{equation}
h_i = \frac{\Delta_i + \sigma_i}{2}, \quad \bar h_i = \frac{\Delta_i - \sigma_i}{2}.
\end{equation}
Under the Lorentz transformation \eqref{lt}, the celestial amplitude $\mathcal M_n$ transforms as
\begin{equation}
\mathcal M_n\big(\{z_i, \bar z_i, h_i, \bar h_i\}\big) = \prod_{i=1}^{n} \frac{1}{(cz_i + d)^{2h_i}} \frac{1}{(\bar c \bar z_i + \bar d)^{2\bar h_i}} \mathcal M_n\bigg(\frac{az_i+b}{cz_i+d} \ ,\frac{\bar a \bar z_i + \bar b}{\bar c \bar z_i + \bar d} \ , h_i,\bar h_i\bigg)\,.
\end{equation}
This is the familiar transformation law of the correlation function of primary operators of weight $(h_i,\bar h_i)$ in a $2d$ CFT under the global conformal group.

In Einstein gravity, the celestial amplitude as defined in \eqref{mellin} typically diverges. This divergence can be regulated by defining a modified celestial amplitude  \cite{Banerjee:2018gce, Banerjee:2019prz}, 
\begin{equation}\label{mellinmod}
\mathcal M_n\big(\{u_i,z_i, \bar z_i, h_i, \bar h_i\}\big) = \prod_{i=1}^{n} \int_{0}^{\infty} d\omega_i \ \omega_i^{\Delta_i -1} e^{-i\sum_{i=1}^n \epsilon_i \omega_i u_i} A_n\big(\{\omega_i,z_i,\bar z_i, \sigma_i\}\big)\,,
\end{equation}
where $u_i$ can be thought of as a time coordinate. Under global conformal transformations, the modified celestial amplitude $\mathcal M_n$ transforms as,
\begin{equation}
\begin{gathered}
\mathcal M_n\big(\{u_i,z_i, \bar z_i, h_i, \bar h_i\}\big) \\ = \prod_{i=1}^{n} \frac{1}{(cz_i + d)^{2h_i}} \frac{1}{(\bar c \bar z_i + \bar d)^{2\bar h_i}} \mathcal M_n\bigg(\frac{u_i}{|cz_i + d|^2} \ , \frac{az_i+b}{cz_i+d} \ ,\frac{\bar a \bar z_i + \bar b}{\bar c \bar z_i + \bar d} \ , h_i,\bar h_i\bigg)\,.
\end{gathered}
\end{equation}
Under global spacetime translation, $u \rightarrow u + A + B\,z + \bar B\,\bar z + C z\bar z$, the modified celestial amplitude is invariant, i.e, 
\begin{equation}
\mathcal M_n\big(\{u_i + A + Bz_i + \bar B\bar z_i + C z_i\bar z_i ,z_i, \bar z_i, h_i, \bar h_i\}\big) = \mathcal M_n\big(\{u_i,z_i, \bar z_i, h_i, \bar h_i\}\big)\,.
\end{equation}
Now, in order to make manifest the conformal nature of the dual theory living on the celestial sphere, it is useful to write the (modified) celestial amplitude as a correlation function of conformal primary operators. So let us define a generic conformal primary operator as, 
\begin{equation}
\label{confprim}
\phi^{\epsilon}_{h,\bar h}(z,\bar z) = \int_{0}^{\infty} d\omega \  \omega^{\Delta-1} a(\epsilon\,\omega, z, \bar z, \sigma)\,,
\end{equation}
where $\epsilon=\pm 1$ corresponds to an annihilation/creation operator of a massless particle of helicity $\sigma$. Under global conformal transformations, it transforms as a primary operator of weights $(h,\bar h)$,
\begin{equation}
\phi'^{\epsilon}_{h,\bar h}(z,\bar z) = \frac{1}{(cz + d)^{2h}} \frac{1}{(\bar c \bar z + \bar d)^{2\bar h}} \mathcal \phi^{\epsilon}_{h,\bar h}\bigg(\frac{az+b}{cz+d} \ ,\frac{\bar a \bar z + \bar b}{\bar c \bar z + \bar d}\bigg)
\end{equation}
Similarly, in the presence of the time coordinate $u$, one has,
\begin{equation}
\label{confprimu}
\phi^{\epsilon}_{h,\bar h}(u,z,\bar z) = \int_{0}^{\infty} d\omega \ \omega^{\Delta-1} e^{-i \epsilon \omega u} a(\epsilon\omega, z, \bar z, \sigma)\,.
\end{equation}
Under global conformal transformations 
\begin{equation}
\phi'^{\epsilon}_{h,\bar h}(u,z,\bar z) = \frac{1}{(cz + d)^{2h}} \frac{1}{(\bar c \bar z + \bar d)^{2\bar h}} \mathcal \phi^{\epsilon}_{h,\bar h}\bigg(\frac{u}{|cz+d|^2},\frac{az+b}{cz+d} \ ,\frac{\bar a \bar z + \bar b}{\bar c \bar z + \bar d}\bigg)
\end{equation}
In terms of \eqref{confprim},  the celestial amplitude can be written as the correlation function of conformal primary operators
\begin{equation}
\mathcal M_n = \bigg\langle{\prod_{i=1}^n \phi^{\epsilon_i}_{h_i,\bar h_i}(z_i,\bar z_i)}\bigg\rangle\,.
\end{equation}
Similarly, using \eqref{confprimu}, the modified celestial amplitude can be written as,
\begin{equation}
\mathcal M_n = \bigg\langle{\prod_{i=1}^n \phi^{\epsilon_i}_{h_i,\bar h_i}(u_i,z_i,\bar z_i)}\bigg\rangle\,.
\end{equation}

\section{Parameterisation of the delta functions}
\label{delta_param}
Here, we work out the parameterisation of the momentum-conserving delta function.

\subsection{5-point delta function} 
\label{5pt_delta}
For 5-particle scattering, the momentum conservation in terms of spinor helicity brackets can be written as

\begin{equation}\label{mom_cons_5}
\sum_{i=1,i\neq 5}^6 \left< qi \right>[ir] = 0.
\end{equation}
First by choosing $ q=3, r=4 $ and then $ q=4, r=3 $ we get the following two equations,
\begin{equation}
\begin{gathered}
\epsilon_1 \omega_1 z_{13} \bar z_{14} + \epsilon_2 \omega_2 z_{23} \bar{z}_{24} + \epsilon_6 \omega_6 z_{36} \bar z_{46} = 0, \\
\epsilon_1 \omega_1 z_{14} \bar z_{13} + \epsilon_2 \omega_2 z_{24} \bar{z}_{23} + \epsilon_6 \omega_6 z_{46} \bar z_{36} = 0.
\end{gathered}
\end{equation}
These two equations can simultaneously be solved for $ \omega_1, \omega_2 $, and we get,
\begin{eqnarray}
\omega_1 &= \epsilon_1 \epsilon_6 \omega_6 \sigma_{1,1}\,,\\[2pt]
\omega_2 &= \epsilon_2 \epsilon_6 \omega_6 \sigma_{2,1}\,,
\end{eqnarray}
where
\begin{equation} \label{omega_stars0}
\begin{gathered}
\sigma_{1,1} = -\frac{z_{46}\bar z_{46}}{z_{14}\bar z_{14}} \frac{r_{24,36} - \bar r_{24,36}}{r_{13,42} - \bar r_{13,42}}\,,\\[4pt]
\sigma_{2,1} = -\frac{z_{36}\bar z_{36}}{z_{23}\bar z_{23}} \frac{r_{13,46} - \bar r_{13,46}}{r_{13,42} - \bar r_{13,42}}\,,\\[4pt]
r_{ij,kl} = \frac{z_{ij}z_{kl}}{z_{ik}z_{jl}}\,,\quad \bar r_{ij,kl} = \frac{\bar z_{ij}\bar z_{kl}}{\bar z_{ik} \bar z_{jl}}\,.
\end{gathered}
\end{equation}
The Jacobian is $ \epsilon_1 \epsilon_2 (z_{13}z_{24} \bar{z}_{14} \bar z_{23} - \bar z_{13} \bar z_{24} {z}_{14} z_{23}) $. Next we choose $ q=1, r=2 $ and then $ q=2, r=1 $ in \eqref{mom_cons_5} and follow the same procedure as above to get,
\begin{equation}
\begin{gathered}
\omega_3 = \epsilon_3 \epsilon_6 \omega_6 \sigma_{3,1}\,,\\
\omega_4 = \epsilon_4 \epsilon_6 \omega_6 \sigma_{4,1}\,,\\
\sigma_{3,1}= - \frac{ z_{26} \bar{z}_{26}}{ z_{23} \bar z_{23}} \frac{r_{16,42}-\bar r_{16,42}}{ r_{13,42} - \bar r_{13,42}}\,,\\[4pt]
\sigma_{4,1}= - \frac{ z_{16} \bar{z}_{16}}{ z_{14} \bar z_{14}} \frac{r_{13,62}-\bar r_{13,62}}{ r_{13,42} - \bar r_{13,42}}\,.
\end{gathered}\label{omega_stars1}
\end{equation}
Hence, we can write the 5-point delta function as,
\begin{equation}
\begin{gathered}
\delta^{(4)}\left( \sum_{i=1, i\neq 5}^6 \epsilon_i \omega_i q^\mu_i \right) =\frac{1}{4} \frac{1}{(z_{13}z_{24} \bar{z}_{14} \bar z_{23} - \bar z_{13} \bar z_{24} {z}_{14} z_{23})} \prod_{i=1}^4 \delta(\omega_i - \epsilon_6 \omega_6 \epsilon_i \sigma_{i,1}).
\end{gathered}
\end{equation}

\subsection{6-point delta function}
\label{6pt_delta}

The parameterisation for the 6-point delta function is:
\begin{equation}\label{6pt_delta_param}
\begin{gathered}
\delta^{(4)}\left( \sum_{i=1}^6 \epsilon_i \omega_i q_i^\mu \right) =\frac{1}{4} \frac{1}{z_{14}z_{23} \bar z_{14} \bar z_{23}(r_{13,42} - \bar r_{13,42})}\prod_{i=1}^4 \delta(\omega_i - \omega_i^*)
\end{gathered}
\end{equation}
where
\begin{equation} \label{omega_stars}
\begin{split}
\omega_1^* &= \epsilon_1 \epsilon_6 \omega_6 \sigma_{1,1} + \epsilon_1 \epsilon_5 \omega_5 \sigma_{1,2}\\
\omega_2^* &= \epsilon_2 \epsilon_6 \omega_6 \sigma_{2,1} + \epsilon_2 \epsilon_5 \omega_5 \sigma_{2,2}\\
\omega_3^* &= \epsilon_3 \epsilon_6 \omega_6 \sigma_{3,1} + \epsilon_3 \epsilon_5 \omega_5 \sigma_{3,2}\\
\omega_4^* &= \epsilon_4 \epsilon_6 \omega_6 \sigma_{4,1} + \epsilon_4 \epsilon_5 \omega_5 \sigma_{4,2}\\
\sigma_{1,2} &= -\frac{z_{45}\bar z_{45}}{z_{14}\bar z_{14}} \frac{r_{24,35}-\bar{r}_{24,35}}{r_{13,42}-\bar r_{13,42}}\\
\sigma_{2,2} &= -\frac{z_{35}\bar z_{35}}{z_{23}\bar z_{23}} \frac{r_{13,45}-\bar{r}_{13,45}}{r_{13,42}-\bar r_{13,42}}\\
\sigma_{3,2} &= -\frac{z_{25} \bar z_{25}}{z_{23} \bar z_{23}} \frac{r_{15,42}-\bar r_{15,42}}{r_{13,42} - \bar r_{13,42}}\\
\sigma_{4,2} &= -\frac{z_{15} \bar z_{15}}{z_{14} \bar z_{14}} \frac{r_{13,52}-\bar r_{13,52}}{r_{13,42} - \bar r_{13,42}}
\end{split}
\end{equation}

%---------------------------------
%
\section{OPE computation}
\label{ope_cal}
%-----------------------------
Here, we write some of the explicit calculations going into the computation of OPEs for completeness. Let us first start with the following expansion,
\begin{equation}
\begin{gathered}
\frac{\Sigma_i}{\Sigma_j} = \frac{\sigma_{i,1}}{\sigma_{j,1}} + z_{56} \frac{t}{\sigma_{j,1}^2}(\sigma_{j,1}\partial_6 \sigma_{i,1} - \sigma_{i,1} \partial_6 \sigma_{j,1}) + \bar z_{56} \frac{t}{\sigma_{j,1}^2}(\sigma_{j,1}\bar \partial_6 \sigma_{i,1} - \sigma_{i,1}\bar \partial_6 \sigma_{j,1})\\
+ z^2_{56} \frac{t^2 }{\sigma_{j,1}^3}(\sigma_{i,1}\partial_6 \sigma_{j,1} - \sigma_{j,1} \partial_6 \sigma_{i,1}) \partial_6 \sigma_{j,1} + z_{56} \bar z_{56}  \frac{t}{\sigma_{3,1}^3} \left[ 2t \sigma_{i,1} (\partial_6 \sigma_{j,1} \bar \partial_6 \sigma_{j,1})  \right.\\
\left. - t \sigma_{j,1} (\bar \partial_6 \sigma_{j,1} \partial_6 \sigma_{i,1}) - t \sigma_{j,1} (\bar \partial_6 \sigma_{i,1} \partial_6 \sigma_{j,1}) -\sigma_{i,1}\sigma_{j,1} \partial_6 \bar \partial_6 \sigma_{j,1} + \sigma_{j,1}^2 \partial_6 \bar \partial_6 \sigma_{i,1})\right] + \cdots\,.
\end{gathered}
\end{equation}

\noindent We also require the following expansions:
\begin{equation}
\begin{gathered}
\frac{t}{\Sigma_{3}}\left( \frac{z_{15}^2 \bar{z}_{35}}{z_{13}^2 z_{35}} - \frac{z_{16}^2 \bar{z}_{36}}{z_{13}^2 z_{36}}\right) = \frac{t}{\sigma_{3,1}} \left[ z_{56}  \frac{z_{16} \bar{z}_{36}}{z_{13} z_{36}}\left( \frac{1}{z_{36}} - \frac{1}{z_{13}}\right) - \bar z_{56} \frac{z_{16}^2 }{z_{13}^2 z_{36}} - z_{56} \bar z_{56} \frac{z_{16}}{z_{13}z_{36}}\left( \frac{1}{z_{36}} - \frac{1}{z_{13}}\right)  \right.\\
\left.  + z_{56}^2 \frac{ \bar{z}_{36}}{ z_{36}^3} - \frac{t}{\sigma_{3,1}}  \frac{z_{16} \bar{z}_{36}}{z_{13} z_{36}}\left( \frac{1}{z_{36}} - \frac{1}{z_{13}}\right) (z_{56}^2\partial_6 \sigma_{3,1} + z_{56}\bar z_{56} \bar \partial_6 \sigma_{3,1}) + \frac{t}{\sigma_{3,1}} z_{56} \bar z_{56} \frac{z_{16}^2 }{z_{13}^2 z_{36}} \partial_6 \sigma_{3,1} \right] + \cdots\,,
\end{gathered}
\end{equation}
and 
\begin{equation}
\begin{gathered}
\frac{t}{\Sigma_{4}}\left( \frac{z_{15}^2 \bar{z}_{45}}{z_{14}^2 z_{45}} - \frac{z_{16}^2 \bar{z}_{46}}{z_{14}^2 z_{46}}\right) = \frac{t}{\sigma_{4,1}} \left[ z_{56}  \frac{z_{16} \bar{z}_{46}}{z_{14} z_{46}}\left( \frac{1}{z_{46}} - \frac{1}{z_{14}}\right) - \bar z_{56} \frac{z_{16}^2 }{z_{14}^2 z_{46}} \right] + \cdots\,.
\end{gathered}
\end{equation}
Using the above equations, we can now expand the following expression:
\begin{equation}
\begin{gathered}
\left( \frac{\Sigma_2}{\Sigma_3}\frac{ z_{12}^2 \bar z_{23}}{z_{13}^2 z_{23}} + \frac{\Sigma_4}{\Sigma_3}\frac{z_{14}^2 \bar z_{34}}{z_{13}^2 z_{34}} + \frac{t}{\Sigma_3}\frac{z_{15}^2 \bar z_{35}}{ z_{13}^2 z_{35}} +  \frac{(1- t)}{\Sigma_3}\frac{z_{16}^2 \bar z_{36}}{ z_{13}^2 z_{36}} \right) 
\times \Big( \frac{\Sigma_2}{\Sigma_4}\frac{z_{12}^2 \bar z_{24}}{z_{14}^2 z_{24}} + \frac{\Sigma_3}{\Sigma_4}\frac{z_{13}^2 \bar z_{34}}{z_{14}^2 z_{34}} + \frac{t}{\Sigma_4}\frac{z_{15}^2 \bar z_{45}}{z_{14}^2 z_{45}} \\+ \frac{(1- t)}{\Sigma_4}\frac{z_{16}^2 \bar z_{46}}{z_{14}^2 z_{46}} \Big) 
\times \frac{1}{t}\left( \Sigma_2\frac{ z_{12}^2 \bar z_{25}}{ z_{15}^2  z_{25} } +  \Sigma_3 \frac{ z_{13}^2 \bar z_{35}}{ z_{15}^2 z_{35}} + \Sigma_4 \frac{ z_{14}^2 \bar z_{45}}{ z_{15}^2 z_{45} } + (1- t) \frac{ z_{16}^2 \bar z_{56}}{ z_{15}^2 z_{56}} \right) \\
\times \frac{1}{(1-t)}\left(  \Sigma_2 \frac{ z_{12}^2 \bar z_{26}}{ z_{16}^2 z_{26}} +  \Sigma_3 \frac{ z_{13}^2 \bar z_{36}}{ z_{16}^2 z_{36} } + \Sigma_4 \frac{ z_{14}^2 \bar z_{46}}{ z_{16}^2 z_{46}} +  t \frac{ z_{15}^2 \bar z_{56}}{ z_{16}^2 z_{56}} \right) \\
= \frac{1}{t(1-t)}\left( \mathcal{T}^1_0 + z_{56} \mathcal{T}^1_{z} + \bar z_{56} \mathcal{T}^1_{\bar z} + z_{56}^2 \mathcal{T}^1_{z^2} + z_{56}\bar z_{56} \mathcal{T}^1_{z\bar z} \right) \\
\times \left( \mathcal{T}^2_{0}   + z_{56} \mathcal{T}^2_{z} + \bar z_{56} \mathcal{T}^2_{\bar z} + z_{56}^2 \mathcal{T}^2_{z^2} + z_{56}\bar z_{56} \mathcal{T}^2_{z\bar z} \right) \\
\times \left((1- t) \frac{ \bar z_{56}}{ z_{56}} + \mathcal{T}^3_0 + z_{56} \mathcal{T}^3_{z} + \bar z_{56} \mathcal{T}^3_{\bar z} + z_{56}^2 \mathcal{T}^3_{z^2} + z_{56}\bar z_{56} \mathcal{T}^3_{z\bar z} \right) \\
\times \left(  t \frac{ \bar z_{56}}{ z_{56}} + \mathcal{T}^4_0 + z_{56} \mathcal{T}^4_{z} + \bar z_{56} \mathcal{T}^4_{\bar z} + z_{56}^2 \mathcal{T}^4_{z^2} + z_{56}\bar z_{56} \mathcal{T}^4_{z\bar z} \right)\\
= \left( \frac{\bar z_{56}}{z_{56}} \right)^2 \mathcal{T}^1_0 \mathcal{T}^2_0 + \frac{1}{t(1-t)} \left( \frac{\bar z_{56}}{z_{56}} \right) \mathcal{T}^1_0 \mathcal{T}^2_0 \mathcal{T}^3_0 + \frac{1}{t(1-t)}\frac{\bar z_{56}^2}{z_{56}} \left[ \mathcal{T}^1_0 \mathcal{T}^2_0 \lbrace t\mathcal{T}^3_{\bar z} + (1-t)\mathcal{T}^4_{\bar z} \rbrace \right.\\
\left. + \mathcal{T}^3_0 \lbrace \mathcal{T}^1_0 \mathcal{T}^2_{\bar z} + \mathcal{T}^2_0  \mathcal{T}^1_{\bar z} \rbrace + t(1-t)\lbrace \mathcal{T}^1_0 \mathcal{T}^2_{ z} + \mathcal{T}^2_0  \mathcal{T}^1_{ z} \rbrace \right] + \cdots\,,
\end{gathered}
\end{equation}
\newpage
\noindent where 
\begin{equation}
\begin{split}
\mathcal{T}^1_{0} &= \frac{\sigma_{2,1}}{\sigma_{3,1}}\frac{ z_{12}^2 \bar z_{23}}{z_{13}^2 z_{23}} + \frac{\sigma_{4,1}}{\sigma_{3,1}}\frac{z_{14}^2 \bar z_{34}}{z_{13}^2 z_{34}}  + \frac{1}{\sigma_{3,1}}\frac{z_{16}^2 \bar z_{36}}{ z_{13}^2 z_{36}}\,,\\
\mathcal{T}^1_{z} &= \frac{t}{\sigma_{3,1}} \left[ \frac{1}{\sigma_{3,1}} \frac{z_{12}^2 \bar z_{23}}{z_{13}^2 z_{23}} (\sigma_{3,1} \partial_6 \sigma_{2,1} - \sigma_{2,1} \partial_6 \sigma_{3,1}) + \frac{1}{\sigma_{3,1}} \frac{z_{14}^2 \bar z_{34}}{z_{13}^2 z_{34}} (\sigma_{3,1} \partial_6 \sigma_{4,1} - \sigma_{4,1} \partial_6 \sigma_{3,1}) \right.\\
&  \hspace{8cm}\left. + \frac{z_{16} \bar z_{36}}{z_{13} z_{36}} \left( \frac{1}{z_{36}} - \frac{1}{z_{13}} \right) - \frac{z_{16}^2 \bar z_{36}}{ z_{13}^2 z_{36}} \frac{\partial_6 \sigma_{3,1}}{\sigma_{3,1}} \right]\,, \\
\mathcal{T}^1_{\bar z} &= \frac{t}{\sigma_{3,1}} \left[ \frac{1}{\sigma_{3,1}} \frac{z_{12}^2 \bar z_{23}}{z_{13}^2 z_{23}} (\sigma_{3,1} \bar \partial_6 \sigma_{2,1} - \sigma_{2,1} \bar \partial_6 \sigma_{3,1}) + \frac{1}{\sigma_{3,1}} \frac{z_{14}^2 \bar z_{34}}{z_{13}^2 z_{34}} (\sigma_{3,1} \bar \partial_6 \sigma_{4,1} - \sigma_{4,1} \bar \partial_6 \sigma_{3,1}) \right.\\
& \left. \hspace{5cm} - \frac{z_{16}^2}{z_{13}^2 z_{36}}  - \frac{z_{16}^2 \bar z_{36}}{ z_{13}^2 z_{36}} \frac{\bar \partial_6 \sigma_{3,1}}{\sigma_{3,1}} \right]\,,
\end{split}
\end{equation}

\begin{equation}
\begin{gathered}
\mathcal{T}^2_{0}  = \frac{\sigma_{2,1}}{\sigma_{4,1}}\frac{z_{12}^2 \bar z_{24}}{z_{14}^2 z_{24}} + \frac{\sigma_{3,1}}{\sigma_{4,1}}\frac{z_{13}^2 \bar z_{34}}{z_{14}^2 z_{34}} + \frac{1}{\sigma_{4,1}}\frac{z_{16}^2 \bar z_{46}}{z_{14}^2 z_{46}}\,,\\
\mathcal{T}^2_{z} = \frac{t}{\sigma_{4,1}^2} \frac{ z_{12}^2 \bar z_{24}}{z_{14}^2 z_{24}}(\sigma_{4,1}\partial_6 \sigma_{2,1} - \sigma_{2,1}\partial_6 \sigma_{4,1}) + \frac{t}{\sigma_{4,1}^2} \frac{z_{13}^2 \bar z_{34}}{z_{14}^2 z_{34}} (\sigma_{4,1}\partial_6 \sigma_{3,1} - \sigma_{3,1}\partial_6 \sigma_{4,1})\\
- \frac{t}{\sigma_{4,1}^2} \frac{z_{16}^2 \bar z_{46}}{ z_{14}^2 z_{46}}\partial_6 \sigma_{4,1} + \frac{t}{\sigma_{4,1}} \frac{z_{16} \bar{z}_{46}}{z_{14} z_{46}}\left( \frac{1}{z_{46}} - \frac{1}{z_{14}}\right)\,, \\
\mathcal{T}^2_{\bar z} = \frac{t}{\sigma_{4,1}^2} \frac{ z_{12}^2 \bar z_{24}}{z_{14}^2 z_{24}} (\sigma_{4,1}\bar \partial_6 \sigma_{2,1} - \sigma_{2,1}\bar \partial_6 \sigma_{4,1}) + \frac{t}{\sigma_{4,1}^2} \frac{z_{13}^2 \bar z_{34}}{z_{14}^2 z_{34}} (\sigma_{4,1} \bar \partial_6 \sigma_{3,1} - \sigma_{3,1}\bar \partial_6 \sigma_{4,1}) \\
- \frac{t}{\sigma_{4,1}^2} \frac{z_{16}^2 \bar z_{46}}{ z_{14}^2 z_{46}}\bar \partial_6 \sigma_{4,1} - \frac{t}{\sigma_{4,1}} \frac{z_{16}^2 }{z_{14}^2 z_{46}}\,,\\
\end{gathered}
\end{equation}

\begin{equation}
\begin{gathered}
\mathcal{T}^3_{0} = \sigma_{2,1}\frac{ z_{12}^2 \bar z_{26}}{ z_{16}^2  z_{26} } +  \sigma_{3,1} \frac{ z_{13}^2 \bar z_{36}}{ z_{16}^2 z_{36}} + \sigma_{4,1} \frac{ z_{14}^2 \bar z_{46}}{ z_{16}^2 z_{46} }\,,\\
\mathcal{T}^3_{z} = \frac{z_{12}^2 \bar z_{26}}{z_{16}^2 z_{26}}\left[\left( \frac{1}{z_{26}}+\frac{2}{z_{16}}\right)\sigma_{2,1} + t\partial_6 \sigma_{2,1} \right] + \frac{z_{13}^2 \bar z_{36}}{z_{16}^2 z_{36}}\left[\left( \frac{1}{z_{36}}+\frac{2}{z_{16}}\right)\sigma_{3,1} + t \partial_6 \sigma_{3,1} \right]\\
+ \frac{z_{14}^2 \bar z_{46}}{z_{16}^2 z_{46}}\left[\left( \frac{1}{z_{46}}+\frac{2}{z_{16}}\right)\sigma_{4,1} + t\partial_6 \sigma_{4,1} \right]\,,\\
\mathcal{T}^3_{\bar z} = t\frac{z_{12}^2 \bar z_{26}}{z_{16}^2 z_{26}}\bar \partial_6 \sigma_{2,1} - \frac{z_{12}^2}{z_{16}^2 z_{26}}\sigma_{2,1} + t\frac{z_{13}^2 \bar z_{36}}{z_{16}^2 z_{36}}\bar \partial_6 \sigma_{3,1} - \frac{z_{13}^2}{z_{16}^2 z_{36}}\sigma_{3,1}\\
+ t\frac{z_{14}^2 \bar z_{46}}{z_{16}^2 z_{46}}\bar \partial_6 \sigma_{4,1} - \frac{z_{14}^2}{z_{16}^2 z_{46}}\sigma_{4,1} + 2\frac{(1-t)}{z_{16}}\,.
\end{gathered}
\end{equation}
We have used these expressions in section \ref{OPE_from_amp}.

\section{Chiral conformal $\mathfrak{bms}_4$ algebra}
\label{con_bms4}

The chiral conformal $\mathfrak{bms}_4$ algebra that we seek here can be viewed as a conformal extension of the chiral $\mathfrak{bms}_4$ algebra. Its operator content consists of a chiral $\mathfrak{sl}(2,\mathbb{R})$ current algebra generated by currents $J_a(z)$ with $a = 0, 1, 2$, a spin 1 current $D(z)$, four spin-$\tfrac{3}{2}$ chiral primary operators $G_i^{\pm}(z)$ with $i = 1, 2$, and a chiral stress tensor $T(z)$ of spin 2. 

We identify two supertranslation currents from \eqref{lead_exp}, defined as
\begin{align}
    H^{1}_{\frac{1}{2}}(z)=G^{+}_{1}(z),~  H^{1}_{-\frac{1}{2}}(z)=-G^{+}_{2}(z),
\end{align}
which serve as the spin-$\tfrac{3}{2}$ generators in the chiral algebra. Similarly, we identify the $\mathfrak{sl}(2,\mathbb{R})$ currents defined in \eqref{sl2_currents} as
\begin{align}
    H^{0}_1(z)=-J_1(z),~ H^{0}_0(z)=2J_0(z),~H^{0}_{-1}(z)=-J_{-1}(z).
\end{align}

We now propose an ansatz for OPEs among the chiral operators introduced above. The general structure of these OPEs is given by
\begin{equation}
\begin{gathered}
 T(z)T(w)=\frac{c/2}{(z-w)^4}+\frac{2 T(w)}{(z-w)^2}+\frac{\partial_w T(w)}{z-w},\\
     J_a(z)J_b(w)=\frac{-\frac{k}{2} \eta_{ab}}{(z-w)^2}+\frac{f^{c}_{ab}~J_c(w)}{z-w},\\
     D(z)D(w)=\frac{1}{(z-w)^2}, \quad
    D(z)J_a(w)=0,\\
    J_a(z)G_i^{+}(w)=\frac{(\lambda_a)^j_{~~i}G_j^{+}(w)}{z-w},\quad
    J_a(z)G_i^{-}(w)=\frac{(\lambda_a)^{j}_{~~i}G_j^{-}(w)}{z-w},\\ 
    D(z)G_i^{\pm}(w)=\pm\frac{q~G_i^{\pm}(w)}{z-w},\\
    G_i^{+}(z)G_{j}^{+}(w)=0, \quad
    G_{i}^{-}(z)G_{j}^{-}(w)=0,\\
    T(z)G_i^{\pm}(w)=\frac{\tfrac{3}{2} G_i^{\pm}(w)}{(z-w)^2}+\frac{\partial_w G_i^{\pm}(w)}{z-w},\\
     T(z)J_a(w)=\frac{J_a(w)}{(z-w)^2}+\frac{\partial_w J_a(w)}{z-w},\\
      T(z)D(w)=\frac{D(w)}{(z-w)^2}+\frac{\partial_w D(w)}{z-w}.
\end{gathered}
\end{equation}
The mixed OPE between $G_i^{+}$ and $G_j^{-}$ takes the following general form, dictated by conformal invariance:
\begin{equation}
\begin{gathered}
    G_{i}^{+}(z)G_{j}^{-}(w)=\epsilon_{ij}\left(\frac{d_1}{(z-w)^3}+\frac{d_2~T(w)}{(z-w)}+\frac{d_3 ~\Xi(w)}{z-w}+\textcolor{black}{\frac{d_6~ \Lambda(w)}{z-w}+\frac{d_5 \partial_w D(w)}{z-w}}\right)\\
    +\left(\frac{2d_4 (\lambda^a)_{ij}J_a(w)}{(z-w)^2}+\frac{d_4 (\lambda^a)_{ij} \partial_w J_a(w)}{z-w}+\textcolor{black}{\frac{d_7~\Sigma(w)}{z-w}}\right)
    +\textcolor{black}{\frac{2 d_5 D(w)}{(z-w)^2}\epsilon_{ij}},
\label{opes}
\end{gathered}
\end{equation}
where, \(\Xi(z)\), \(\Lambda(z)\), and \(\Sigma(z)\) are the quasi-primary operators defined as
\begin{equation}
\begin{gathered}
    \Xi(z):=\eta^{ab}(J_aJ_b)(z),\quad \Lambda(z):=(DD)(z),\quad \Sigma(z):=(\lambda^a)_{ij}(DJ_a)(z).
\end{gathered}
\end{equation}
Here, \( (\lambda_a)^s_{~s'}\) and \(\eta_{ab}\) are defined as in Refs.~\cite{Gupta:2023fmp, Gupta:2022mdt}:
\begin{equation}
\begin{aligned}
    (\lambda_a)^s_{~s'} &= \frac{1}{2}\,(a - 2s')\,\delta^s_{a+s'}, \quad
    \eta_{ab} = (3a^2 - 1)\,\delta_{a+b,0}.
\end{aligned}
\end{equation}
We have used parentheses \((AB)(z)\) to denote normal ordering between two operators.

To determine the coefficients \(q\)\,, Virasoro central charge \(c\) and \(d_1\) through \(d_7\)  in  \eqref{opes}, we impose the associativity condition on these OPEs. The resulting algebraic constraints can be efficiently solved using \texttt{Mathematica}~\cite{Thielemans:1991uw}, yielding
\begin{equation}
\begin{aligned}
    d_1 &= -\tfrac{1}{2}d_3(1+k), \quad
    d_2 =  \tfrac{1}{2}d_3(3+k), \quad
    d_4 =  d_3(1+k), \quad
    d_5 = -\tfrac{1}{2}d_3 k\sqrt{1+k}, \\[4pt]
    d_6 &= -\tfrac{3}{4}d_3(1+k), \quad
    d_7 =  2d_3\sqrt{1+k}, \quad
    q   =  \frac{1}{\sqrt{1+k}}, \quad
    c   =  \frac{3 + 3k - 6k^2}{3 + k},
\end{aligned}
\end{equation}
with \(k\neq-1\) \& \(k\neq -3\).
When the operators \(D(z)\) and \(G^{-}_{i}(z)\) are omitted, the remaining set of operators \(\{T(z), J^{a}(z), G^{+}_{i}(z)\}\) closes to form the chiral \(\mathfrak{bms}_4\) subalgebra. This chiral conformal \(\mathfrak{bms}_4\) algebra constitutes one of the four possible chiral extensions of the \(\mathfrak{so}(2,4)\) algebra.\footnote{Closely related constructions were investigated by Creutzig \textit{et al.}~\cite{Creutzig:2020ffn}, who classified all possible chiral extensions of the \(\mathfrak{sl}(4,\mathbb{R})\) algebra.}

\begin{comment}
\begin{enumerate}
    \item \textbf{Virasoro algbera}
    \begin{align}
        [L_m,L_n]&=(m-n)L_{m+n}+\frac{c}{12}m(m^2-1)\delta_{m+n,0}
    \end{align}
    \item \textbf{Virasoro Primaries}
    \begin{align}
        [L_m,H^{0}_{n,a}]&=-n H^{0}_{m+n,a},~~ [L_m,G^{\pm}_{s,r}]=\frac{1}{2}(m-2r)G^{\pm}_{s,m+r},~~[L_m,D_n]=-nD_{m+n}
    \end{align}
    \item \textbf{$\mathfrak{sl}_2(\mathbb{R})$ current algebra}:
    \begin{align}
        [H^{0}_{m,a},H^{0}_{n,a}]&=-\frac{1}{2}k~ m~\eta_{ab}\delta_{m+n,0}+f^{~~c}_{ab}H^{0}_{m+n,c}
    \end{align}
    \item  \textbf{Current algebra primaries}
    \begin{align}
        [H^{0}_{n,a},G^{\pm}_{s,r}]=G^{\pm}_{s',n+r}(\lambda_a)^{s'}_{~~s}
    \end{align}
    \item \textbf{U(1) current algebra \& primaries}
    \begin{align}
        [D_m,D_n]&=m\delta_{m+n},~~[D_m,G^{\pm}_{s,r}]=\pm q G^{\pm}_{s,r+m}
    \end{align}
    \item \textbf{Commutators of} $[G^{+}_{s,r},G^{+}_{s',r'}]$ \& $[G^{-}_{s,r},G^{-}_{s',r'}]$
    \begin{align}
        [G^{+}_{s,r},G^{+}_{s',r'}]&=0\nonumber\\
         [G^{-}_{s,r},G^{-}_{s',r'}]&=0
    \end{align}
    \item \textbf{Commutator of} $[G^{+}_{s,r},G^{-}_{s',r'}]$
    \begin{align}
        [G^{+}_{s,r},G^{-}_{s',r'}]&=\epsilon_{ss'}\Big[d_1 \left(r^2-\frac{1}{4}\right)\delta_{r+r',0}+d_2 L_{r+r'}+d_3((H^0)^2)_{r+r'}+d_5~(r-r')D_{r+r'}+d_6 (D^2)_{r+r'}\Big]\nonumber\\
        &+d_4~(r-r')D_{a,r+r'} (\lambda^a)_{ss'}
         +d_7 (DH^{0}_a)_{r+r'}(\lambda^a)_{ss'}
    \end{align}
\end{enumerate}
\end{comment}

%---------------------------------------------------------------
%
\section{The leading soft gluon theorem in $(DF)^2$ theory}
\label{dfsquare}
%-------------------------------------------------------------------
We have seen in section \ref{ssf} that the subleading soft graviton theorem in conformal gravity theory gets corrected, leaving the algebra responsible for the subleading soft factorisation unchanged. However, the realisation of ${\mathfrak{sl}(2,\mathbb{R}})$ current is quite different. In \cite{Johansson:2017srf, Johansson:2018ues}, the scattering amplitudes of the conformal gravity theory that we considered here were computed using the amplitudes of a four-derivative $SU(N)$ gauge theory, called $(DF)^2$ theory. In this appendix, we show that the leading soft (positive-helicity) gluon theorem in this theory gets corrected, in a way such that the algebra which was responsible for the leading soft gluon theorem in $SU(N)$ Yang-Mills type gauge theory, remains unchanged, though its representation is different. The particle content of the $(DF)^2$ theory is a spin 1 gluon in the adjoint representation of $SU(N)$ and scalars in some auxiliary representation whose generators are given by $T_{\mathfrak{R}}^a$. The Lagrangian of the theory is given by
\begin{equation}\label{dfLag}
    \begin{gathered}
        \mathcal{L} = \frac{1}{2}(D_\mu F^{a \mu\nu})^2-\frac{1}{3}g F^3 + \frac{1}{2}(D_\mu \phi^\alpha)^2 + \frac{1}{2}g C^{\alpha a b}\phi^\alpha F^a_{\mu\nu}F^{b\mu\nu}+\frac{1}{3!}g d^{\alpha\beta\gamma}\phi^\alpha \phi^\beta \phi^\gamma\,,
    \end{gathered}
\end{equation}
where we have 
\begin{equation}
\begin{gathered}
D_{\rho}F^{a}_{\mu\nu}=\partial_\rho F^{a}_{\mu\nu}+g f^{abc}A^{b}_\rho F^{c}_{\mu\nu}\,,\\[4pt]
 F^{a}_{\mu\nu}=\partial_{\mu}A^{a}_{\nu}-\partial_{\nu}A^{a}_{\mu}+g f^{abc}A^{b}_{\mu}A^{c}_\nu\,,\\[4pt]
 F^3=f^{abc}F_{\nu}^{a ~\mu}F^{b~\nu}_{\rho}F^{c~\rho}_{\mu}\,.
\end{gathered}
\end{equation}
In the Feynman-like gauge, one has the following gluon and scalar propagator
\vspace{5pt}

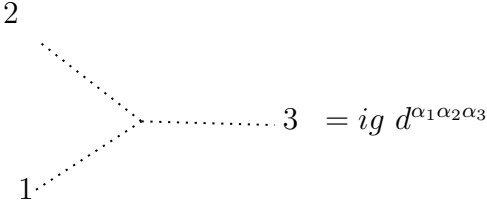
\tikzset{
  gluon/.style  = {decorate,
                   decoration={coil, aspect=0.5, segment length=5pt,
                               amplitude=4pt, pre length=2pt, post length=2pt}},
  scalar/.style = {dashed, dash pattern=on 3pt off 2.5pt},
  vtx/.style    = {circle, fill=black, inner sep=1.5pt},
}

\textbf{Propagators:}

\begin{center}
% --- gluon propagator ---
\begin{tikzpicture}
  \draw[gluon] (0,0) -- (2.2,0);
  \node[left]        at (0,0)   {$\begin{array}{c}\mu\\ a\end{array}$};
  \node[right]       at (2.2,0) {$\begin{array}{c}\nu\\ b\end{array}$};
  \node[anchor=west] at (2.9,0) {$=\,i\,\dfrac{\eta^{\mu\nu}\,\delta^{ab}}{p^{4}}$};
\end{tikzpicture}

\vspace{14pt}

% --- scalar propagator ---
\begin{tikzpicture}
  \draw[scalar] (0,0) -- (2.2,0);
  \node[above left]  at (0,0)   {$\alpha$};
  \node[above right] at (2.2,0) {$\beta$};
  \node[above]       at (1.1,0) {$p$};
  \node[anchor=west] at (2.9,0) {$=\,i\,\dfrac{\delta^{\alpha\beta}}{p^{2}}$};
\end{tikzpicture}
\end{center}

\bigskip
\textbf{3-point vertices:}
From the Lagrangian \eqref{dfLag}, we can write down all possible 3-point
vertices, with all momenta taken to be incoming.

\bigskip
\noindent\underline{gluon-gluon-gluon vertex:}

\begin{center}
\begin{tikzpicture}
  \coordinate (v)  at (0,0);
  \coordinate (l1) at (135:1.6);
  \coordinate (l2) at (225:1.6);
  \coordinate (l3) at (0:1.7);
  \draw[gluon] (v) -- (l1);
  \draw[gluon] (v) -- (l2);
  \draw[gluon] (v) -- (l3);
  \node[vtx] at (v) {};
  \node[above left] at (l1) {$1$};
  \node[below left] at (l2) {$2$};
  \node[right]      at (l3) {$3$};
\end{tikzpicture}
\end{center}

\begin{align}
={}&2g\,f^{a_1a_2a_3}\big(p_1^{\mu_3}p_2^{\mu_1}p_3^{\mu_2}-p_1^{\mu_2}p_2^{\mu_3}p_3^{\mu_1}\big)
+2g\,f^{a_1a_2a_3}\Big[\eta^{\mu_1\mu_2}\Big((p_1\cdot p_3)p_2^{\mu_3}-(p_2\cdot p_3)p_1^{\mu_3}\Big)\nonumber\\
&+\eta^{\mu_2\mu_3}\Big((p_1\cdot p_2)p_3^{\mu_1}-(p_1\cdot p_3)p_2^{\mu_1}\Big)
+\eta^{\mu_1\mu_3}\Big((p_2\cdot p_3)p_1^{\mu_2}-(p_1\cdot p_2)p_3^{\mu_2}\Big)\Big]\nonumber\\
&+g\Big(f^{a_1a_2a_3}p_1^2\Big[(p_2^{\mu_2}+2p_3^{\mu_2})\eta^{\mu_1\mu_3}-p_3^{\mu_1}\eta^{\mu_2\mu_3}\Big]+\text{perm.}(1,2,3)\Big)\nonumber\\
&+g\Big(f^{a_1a_2a_3}(p_1\cdot p_3)p_1^{\mu_1}\eta^{\mu_2\mu_3}+\text{perm.}(1,2,3)\Big)\nonumber\\
&-g\Big(f^{a_1a_2a_3}p_1^{\mu_1}p_1^{\mu_3}(p_2^{\mu_2}+2p_3^{\mu_2})+\text{perm.}(1,2,3)\Big)\nonumber
\end{align}

\noindent\underline{gluon-gluon-scalar vertex:}

\begin{center}
\begin{tikzpicture}
  \coordinate (v)  at (0,0);
  \coordinate (l2) at (135:1.6);
  \coordinate (l1) at (225:1.6);
  \coordinate (l3) at (0:1.7);
  \draw[gluon]  (v) -- (l2);
  \draw[gluon]  (v) -- (l1);
  \draw[scalar] (v) -- (l3);
  \node[vtx] at (v) {};
  \node[above left] at (l2) {$2$};
  \node[below left] at (l1) {$1$};
  \node[right]      at (l3) {$3$};
  \node[anchor=west] at (2.3,0)
       {$=-2ig\,C^{\alpha_3 a_1 a_2}\big(p_1\cdot p_2\,\eta^{\mu_1\mu_2}-p_1^{\mu_2}p_2^{\mu_1}\big)$};
\end{tikzpicture}
\end{center}

\noindent\underline{scalar-scalar-gluon vertex:}

\begin{center}
\begin{tikzpicture}
  \coordinate (v)  at (0,0);
  \coordinate (l2) at (135:1.6);
  \coordinate (l1) at (225:1.6);
  \coordinate (l3) at (0:1.7);
  \draw[scalar] (v) -- (l2);
  \draw[scalar] (v) -- (l1);
  \draw[gluon]  (v) -- (l3);
  \node[vtx] at (v) {};
  \node[above left] at (l2) {$2$};
  \node[below left] at (l1) {$1$};
  \node[right]      at (l3) {$3$};
  \node[anchor=west] at (2.3,0)
       {$=ig\,\big(T^{a_3}_{\mathfrak{R}}\big)^{\alpha_1\alpha_2}(p_1-p_2)^{\mu_3}$};
\end{tikzpicture}
\end{center}

\noindent\underline{scalar-scalar-scalar vertex:}

\begin{center}
\begin{tikzpicture}
  \coordinate (v)  at (0,0);
  \coordinate (l2) at (135:1.6);
  \coordinate (l1) at (225:1.6);
  \coordinate (l3) at (0:1.7);
  \draw[scalar] (v) -- (l2);
  \draw[scalar] (v) -- (l1);
  \draw[scalar] (v) -- (l3);
  \node[vtx] at (v) {};
  \node[above left] at (l2) {$2$};
  \node[below left] at (l1) {$1$};
  \node[right]      at (l3) {$3$};
  \node[anchor=west] at (2.3,0) {$=ig\,d^{\alpha_1\alpha_2\alpha_3}$};
\end{tikzpicture}
\end{center}

\subsection*{The 4-point  tree amplitudes}
\begin{figure}
\begin{tikzpicture}[x=0.55pt,y=0.55pt,yscale=-1,xscale=1]
%uncomment if require: \path (0,330); %set diagram left start at 0, and has height of 330

%Flowchart: Connector [id:dp5117358376537894] 
\draw  [fill={rgb, 255:red, 155; green, 155; blue, 155 }  ,fill opacity=1 ] (173.62,159.47) .. controls (173.62,157) and (174.89,155) .. (176.45,155) .. controls (178.01,155) and (179.28,157) .. (179.28,159.47) .. controls (179.28,161.94) and (178.01,163.94) .. (176.45,163.94) .. controls (174.89,163.94) and (173.62,161.94) .. (173.62,159.47) -- cycle ;
%Shape: Spring [id:dp5384579836233768] 
\draw   (114.96,77.98) .. controls (117.01,75.18) and (120.24,74.16) .. (123.4,78.89) .. controls (129.71,88.34) and (123.12,101.88) .. (119.96,97.15) .. controls (116.8,92.42) and (123.4,78.89) .. (129.71,88.34) .. controls (136.03,97.8) and (129.44,111.33) .. (126.28,106.6) .. controls (123.12,101.88) and (129.71,88.34) .. (136.03,97.8) .. controls (142.35,107.25) and (135.76,120.78) .. (132.6,116.06) .. controls (129.44,111.33) and (136.03,97.8) .. (142.35,107.25) .. controls (148.67,116.7) and (142.08,130.24) .. (138.92,125.51) .. controls (135.76,120.78) and (142.35,107.25) .. (148.67,116.7) .. controls (154.99,126.16) and (148.39,139.69) .. (145.23,134.96) .. controls (142.08,130.24) and (148.67,116.7) .. (154.99,126.16) .. controls (161.31,135.61) and (154.71,149.14) .. (151.55,144.42) .. controls (148.39,139.69) and (154.99,126.16) .. (161.31,135.61) .. controls (167.63,145.06) and (161.03,158.6) .. (157.87,153.87) .. controls (154.71,149.14) and (161.31,135.61) .. (167.63,145.06) .. controls (173.95,154.52) and (167.35,168.05) .. (164.19,163.32) .. controls (161.03,158.6) and (167.63,145.06) .. (173.95,154.52) .. controls (175.14,156.3) and (175.87,158.24) .. (176.26,160.16) ;
%Straight Lines [id:da6329996042018575] 
\draw  [dash pattern={on 0.84pt off 2.51pt}]  (125.27,236.73) -- (160.5,183.56) -- (176.45,159.47) ;
%Shape: Spring [id:dp14414020941590588] 
\draw   (175.7,158.48) .. controls (176.22,155.14) and (178.3,151.8) .. (182.45,151.8) .. controls (190.76,151.8) and (190.76,165.15) .. (186.61,165.15) .. controls (182.45,165.15) and (182.45,151.8) .. (190.76,151.8) .. controls (199.07,151.8) and (199.07,165.15) .. (194.91,165.15) .. controls (190.76,165.15) and (190.76,151.8) .. (199.07,151.8) .. controls (207.38,151.8) and (207.38,165.15) .. (203.22,165.15) .. controls (199.07,165.15) and (199.07,151.8) .. (207.38,151.8) .. controls (215.69,151.8) and (215.69,165.15) .. (211.53,165.15) .. controls (207.38,165.15) and (207.38,151.8) .. (215.69,151.8) .. controls (223.99,151.8) and (223.99,165.15) .. (219.84,165.15) .. controls (215.69,165.15) and (215.69,151.8) .. (223.99,151.8) .. controls (232.3,151.8) and (232.3,165.15) .. (228.15,165.15) .. controls (223.99,165.15) and (223.99,151.8) .. (232.3,151.8) .. controls (240.61,151.8) and (240.61,165.15) .. (236.46,165.15) .. controls (232.3,165.15) and (232.3,151.8) .. (240.61,151.8) .. controls (242.95,151.8) and (244.63,152.86) .. (245.75,154.38) ;
%Flowchart: Connector [id:dp06336465531454394] 
\draw  [fill={rgb, 255:red, 155; green, 155; blue, 155 }  ,fill opacity=1 ] (244.36,154.84) .. controls (244.36,151.82) and (245.76,149.38) .. (247.48,149.38) .. controls (249.2,149.38) and (250.59,151.82) .. (250.59,154.84) .. controls (250.59,157.85) and (249.2,160.3) .. (247.48,160.3) .. controls (245.76,160.3) and (244.36,157.85) .. (244.36,154.84) -- cycle ;
%Straight Lines [id:da32923217231103363] 
\draw  [dash pattern={on 0.84pt off 2.51pt}]  (247.48,154.84) -- (297.68,224.33) ;
%Straight Lines [id:da5105882614881602] 
\draw  [dash pattern={on 0.84pt off 2.51pt}]  (247.48,154.84) -- (303.91,94.52) ;
%Straight Lines [id:da5654334766543777] 
\draw    (132.08,80.99) -- (150.96,106.22) ;
\draw [shift={(152.16,107.82)}, rotate = 233.19] [color={rgb, 255:red, 0; green, 0; blue, 0 }  ][line width=0.75]    (10.93,-3.29) .. controls (6.95,-1.4) and (3.31,-0.3) .. (0,0) .. controls (3.31,0.3) and (6.95,1.4) .. (10.93,3.29)   ;
%Straight Lines [id:da28266845808758256] 
\draw    (139.7,225.37) -- (152.47,205.36) ;
\draw [shift={(153.55,203.67)}, rotate = 122.55] [color={rgb, 255:red, 0; green, 0; blue, 0 }  ][line width=0.75]    (10.93,-3.29) .. controls (6.95,-1.4) and (3.31,-0.3) .. (0,0) .. controls (3.31,0.3) and (6.95,1.4) .. (10.93,3.29)   ;
%Straight Lines [id:da49381422288610655] 
\draw    (281.64,221.87) -- (268.81,200.39) ;
\draw [shift={(267.79,198.68)}, rotate = 59.16] [color={rgb, 255:red, 0; green, 0; blue, 0 }  ][line width=0.75]    (10.93,-3.29) .. controls (6.95,-1.4) and (3.31,-0.3) .. (0,0) .. controls (3.31,0.3) and (6.95,1.4) .. (10.93,3.29)   ;
%Straight Lines [id:da3215115013551638] 
\draw    (285.1,101.62) -- (269.87,117.31) ;
\draw [shift={(268.48,118.74)}, rotate = 314.13] [color={rgb, 255:red, 0; green, 0; blue, 0 }  ][line width=0.75]    (10.93,-3.29) .. controls (6.95,-1.4) and (3.31,-0.3) .. (0,0) .. controls (3.31,0.3) and (6.95,1.4) .. (10.93,3.29)   ;
\draw   (335.19,154.71) -- (345.85,154.71)(340.52,147.87) -- (340.52,161.56) ;
%Flowchart: Connector [id:dp5087938881242372] 
\draw  [fill={rgb, 255:red, 155; green, 155; blue, 155 }  ,fill opacity=1 ] (475.71,130.94) .. controls (477.77,130.92) and (479.45,132.02) .. (479.48,133.39) .. controls (479.5,134.76) and (477.85,135.88) .. (475.79,135.9) .. controls (473.73,135.91) and (472.05,134.82) .. (472.03,133.45) .. controls (472.01,132.08) and (473.66,130.95) .. (475.71,130.94) -- cycle ;
%Shape: Spring [id:dp038266112381810546] 
\draw   (414.84,68.75) .. controls (416.64,66.98) and (419.61,66.54) .. (422.73,70.11) .. controls (428.97,77.26) and (423.35,86.16) .. (420.23,82.58) .. controls (417.1,79.01) and (422.73,70.11) .. (428.97,77.26) .. controls (435.22,84.4) and (429.59,93.3) .. (426.47,89.73) .. controls (423.35,86.16) and (428.97,77.26) .. (435.22,84.4) .. controls (441.46,91.55) and (435.84,100.45) .. (432.72,96.88) .. controls (429.59,93.3) and (435.22,84.4) .. (441.46,91.55) .. controls (447.71,98.7) and (442.08,107.6) .. (438.96,104.02) .. controls (435.84,100.45) and (441.46,91.55) .. (447.71,98.7) .. controls (453.95,105.84) and (448.33,114.74) .. (445.21,111.17) .. controls (442.08,107.6) and (447.71,98.7) .. (453.95,105.84) .. controls (460.2,112.99) and (454.58,121.89) .. (451.45,118.32) .. controls (448.33,114.74) and (453.95,105.84) .. (460.2,112.99) .. controls (466.45,120.14) and (460.82,129.04) .. (457.7,125.46) .. controls (454.58,121.89) and (460.2,112.99) .. (466.45,120.14) .. controls (472.69,127.28) and (467.07,136.18) .. (463.94,132.61) .. controls (460.82,129.04) and (466.45,120.14) .. (472.69,127.28) .. controls (474.52,129.37) and (475.33,131.61) .. (475.48,133.65) ;
%Shape: Spring [id:dp7298602397998054] 
\draw   (476.58,132.75) .. controls (479.36,133.19) and (482.16,134.99) .. (482.22,138.63) .. controls (482.33,145.92) and (471.23,146.01) .. (471.17,142.36) .. controls (471.12,138.72) and (482.22,138.63) .. (482.33,145.92) .. controls (482.44,153.21) and (471.34,153.3) .. (471.28,149.65) .. controls (471.23,146.01) and (482.33,145.92) .. (482.44,153.21) .. controls (482.56,160.5) and (471.45,160.59) .. (471.4,156.94) .. controls (471.34,153.3) and (482.44,153.21) .. (482.56,160.5) .. controls (482.67,167.79) and (471.56,167.88) .. (471.51,164.23) .. controls (471.45,160.59) and (482.56,160.5) .. (482.67,167.79) .. controls (482.78,175.08) and (471.68,175.17) .. (471.62,171.52) .. controls (471.56,167.88) and (482.67,167.79) .. (482.78,175.08) .. controls (482.89,182.37) and (471.79,182.46) .. (471.73,178.81) .. controls (471.68,175.17) and (482.78,175.08) .. (482.89,182.37) .. controls (483,189.66) and (471.9,189.75) .. (471.84,186.1) .. controls (471.79,182.46) and (482.89,182.37) .. (483,189.66) .. controls (483.03,191.72) and (482.18,193.2) .. (480.93,194.19) ;
%Flowchart: Connector [id:dp8645880825795657] 
\draw  [fill={rgb, 255:red, 155; green, 155; blue, 155 }  ,fill opacity=1 ] (480.54,192.97) .. controls (483.06,192.95) and (485.11,194.16) .. (485.14,195.67) .. controls (485.16,197.18) and (483.14,198.42) .. (480.63,198.44) .. controls (478.12,198.46) and (476.06,197.25) .. (476.04,195.74) .. controls (476.01,194.23) and (478.03,192.99) .. (480.54,192.97) -- cycle ;
%Straight Lines [id:da700468653729901] 
\draw  [dash pattern={on 0.84pt off 2.51pt}]  (480.59,195.71) -- (439.62,245.93) ;
%Straight Lines [id:da3370541179968922] 
\draw  [dash pattern={on 0.84pt off 2.51pt}]  (480.59,195.71) -- (531.63,244.83) ;
%Straight Lines [id:da5820603327204441] 
\draw    (544.53,94.82) -- (519.94,117.11) ;
\draw [shift={(518.46,118.46)}, rotate = 317.81] [color={rgb, 255:red, 0; green, 0; blue, 0 }  ][line width=0.75]    (10.93,-3.29) .. controls (6.95,-1.4) and (3.31,-0.3) .. (0,0) .. controls (3.31,0.3) and (6.95,1.4) .. (10.93,3.29)   ;
%Straight Lines [id:da32415279685144804] 
\draw    (445.94,253.17) -- (465.54,230.04) ;
\draw [shift={(466.84,228.52)}, rotate = 130.28] [color={rgb, 255:red, 0; green, 0; blue, 0 }  ][line width=0.75]    (10.93,-3.29) .. controls (6.95,-1.4) and (3.31,-0.3) .. (0,0) .. controls (3.31,0.3) and (6.95,1.4) .. (10.93,3.29)   ;
%Straight Lines [id:da49097902837762875] 
\draw    (532.66,237.21) -- (512.32,215.37) ;
\draw [shift={(510.95,213.9)}, rotate = 47.04] [color={rgb, 255:red, 0; green, 0; blue, 0 }  ][line width=0.75]    (10.93,-3.29) .. controls (6.95,-1.4) and (3.31,-0.3) .. (0,0) .. controls (3.31,0.3) and (6.95,1.4) .. (10.93,3.29)   ;
\draw   (603.43,161.35) -- (614.09,161.35)(608.76,154.51) -- (608.76,168.2) ;
%Flowchart: Connector [id:dp4661807153730414] 
\draw  [fill={rgb, 255:red, 155; green, 155; blue, 155 }  ,fill opacity=1 ] (700.28,127.61) .. controls (700.28,125.39) and (701.72,123.58) .. (703.5,123.58) .. controls (705.28,123.58) and (706.73,125.39) .. (706.73,127.61) .. controls (706.73,129.84) and (705.28,131.64) .. (703.5,131.64) .. controls (701.72,131.64) and (700.28,129.84) .. (700.28,127.61) -- cycle ;
%Flowchart: Connector [id:dp6136921976175252] 
\draw  [fill={rgb, 255:red, 155; green, 155; blue, 155 }  ,fill opacity=1 ] (709.14,219.43) .. controls (709.14,217.2) and (710.77,215.4) .. (712.77,215.4) .. controls (714.77,215.4) and (716.39,217.2) .. (716.39,219.43) .. controls (716.39,221.65) and (714.77,223.46) .. (712.77,223.46) .. controls (710.77,223.46) and (709.14,221.65) .. (709.14,219.43) -- cycle ;
%Shape: Spring [id:dp5501332051428571] 
\draw   (643.87,72.38) .. controls (645.77,70.32) and (649.11,69.77) .. (652.95,73.8) .. controls (660.64,81.86) and (654.94,92.11) .. (651.1,88.08) .. controls (647.26,84.05) and (652.95,73.8) .. (660.64,81.86) .. controls (668.32,89.91) and (662.63,100.16) .. (658.79,96.13) .. controls (654.94,92.11) and (660.64,81.86) .. (668.32,89.91) .. controls (676,97.97) and (670.31,108.22) .. (666.47,104.19) .. controls (662.63,100.16) and (668.32,89.91) .. (676,97.97) .. controls (683.68,106.03) and (677.99,116.27) .. (674.15,112.25) .. controls (670.31,108.22) and (676,97.97) .. (683.68,106.03) .. controls (691.37,114.08) and (685.67,124.33) .. (681.83,120.3) .. controls (677.99,116.27) and (683.68,106.03) .. (691.37,114.08) .. controls (699.05,122.14) and (693.36,132.39) .. (689.52,128.36) .. controls (685.67,124.33) and (691.37,114.08) .. (699.05,122.14) .. controls (701.33,124.53) and (702.43,127.11) .. (702.75,129.45) ;
%Shape: Spring [id:dp47023663025830953] 
\draw   (714.14,218.84) .. controls (711.42,218.28) and (708.56,215.24) .. (708.21,208.62) .. controls (707.51,195.38) and (718.23,194.31) .. (718.58,200.93) .. controls (718.93,207.55) and (708.21,208.62) .. (707.51,195.38) .. controls (706.81,182.14) and (717.53,181.07) .. (717.88,187.69) .. controls (718.23,194.31) and (707.51,195.38) .. (706.81,182.14) .. controls (706.12,168.9) and (716.83,167.83) .. (717.18,174.45) .. controls (717.53,181.07) and (706.81,182.14) .. (706.12,168.9) .. controls (705.42,155.66) and (716.14,154.59) .. (716.48,161.21) .. controls (716.83,167.83) and (706.12,168.9) .. (705.42,155.66) .. controls (704.72,142.42) and (715.44,141.35) .. (715.79,147.97) .. controls (716.14,154.59) and (705.42,155.66) .. (704.72,142.42) .. controls (704.02,129.18) and (714.74,128.11) .. (715.09,134.73) .. controls (715.44,141.35) and (704.72,142.42) .. (704.02,129.18) .. controls (704,128.8) and (703.99,128.43) .. (703.99,128.07) ;
%Straight Lines [id:da12686156825494965] 
\draw  [dash pattern={on 0.84pt off 2.51pt}]  (712.77,219.43) -- (785.67,95.14) ;
%Shape: Arc [id:dp40782283463524827] 
\draw  [draw opacity=0][dash pattern={on 0.84pt off 2.51pt}] (736.01,159.94) .. controls (736.25,158.58) and (736.68,157.27) .. (737.3,156.08) .. controls (740.21,150.48) and (746.29,149.59) .. (750.88,154.09) .. controls (755.46,158.59) and (756.81,166.78) .. (753.9,172.38) .. controls (752.8,174.48) and (751.27,175.92) .. (749.54,176.66) -- (745.6,164.23) -- cycle ; \draw  [dash pattern={on 0.84pt off 2.51pt}] (736.01,159.94) .. controls (736.25,158.58) and (736.68,157.27) .. (737.3,156.08) .. controls (740.21,150.48) and (746.29,149.59) .. (750.88,154.09) .. controls (755.46,158.59) and (756.81,166.78) .. (753.9,172.38) .. controls (752.8,174.48) and (751.27,175.92) .. (749.54,176.66) ;  
%Straight Lines [id:da2566080247246729] 
\draw  [dash pattern={on 0.84pt off 2.51pt}]  (703.5,127.61) -- (735.91,160.53) ;
%Straight Lines [id:da9436484642446896] 
\draw  [dash pattern={on 0.84pt off 2.51pt}]  (748.91,176.89) -- (759.09,248.9) ;
%Straight Lines [id:da7957638625667275] 
\draw  [dash pattern={on 0.84pt off 2.51pt}]  (712.77,219.43) -- (657.59,257.75) ;
%Straight Lines [id:da9173368423686427] 
\draw    (663.23,70.49) -- (681.33,93.56) ;
\draw [shift={(682.56,95.14)}, rotate = 231.89] [color={rgb, 255:red, 0; green, 0; blue, 0 }  ][line width=0.75]    (10.93,-3.29) .. controls (6.95,-1.4) and (3.31,-0.3) .. (0,0) .. controls (3.31,0.3) and (6.95,1.4) .. (10.93,3.29)   ;
%Straight Lines [id:da23448948127856628] 
\draw    (774.92,127.28) -- (764.96,145.51) ;
\draw [shift={(764,147.27)}, rotate = 298.65] [color={rgb, 255:red, 0; green, 0; blue, 0 }  ][line width=0.75]    (10.93,-3.29) .. controls (6.95,-1.4) and (3.31,-0.3) .. (0,0) .. controls (3.31,0.3) and (6.95,1.4) .. (10.93,3.29)   ;
%Straight Lines [id:da6170454250864975] 
\draw    (764,231.87) -- (760.93,205.23) ;
\draw [shift={(760.7,203.24)}, rotate = 83.42] [color={rgb, 255:red, 0; green, 0; blue, 0 }  ][line width=0.75]    (10.93,-3.29) .. controls (6.95,-1.4) and (3.31,-0.3) .. (0,0) .. controls (3.31,0.3) and (6.95,1.4) .. (10.93,3.29)   ;
%Straight Lines [id:da7700064536283738] 
\draw    (666.45,261.86) -- (687.38,246.75) ;
\draw [shift={(689.01,245.58)}, rotate = 144.18] [color={rgb, 255:red, 0; green, 0; blue, 0 }  ][line width=0.75]    (10.93,-3.29) .. controls (6.95,-1.4) and (3.31,-0.3) .. (0,0) .. controls (3.31,0.3) and (6.95,1.4) .. (10.93,3.29)   ;
%Straight Lines [id:da4933404738143562] 
\draw  [dash pattern={on 0.84pt off 2.51pt}]  (475.75,133.42) -- (553.64,72.86) ;
%Straight Lines [id:da40866671869820426] 
\draw    (410.15,89.2) -- (434.21,118.17) ;
\draw [shift={(435.49,119.71)}, rotate = 230.29] [color={rgb, 255:red, 0; green, 0; blue, 0 }  ][line width=0.75]    (10.93,-3.29) .. controls (6.95,-1.4) and (3.31,-0.3) .. (0,0) .. controls (3.31,0.3) and (6.95,1.4) .. (10.93,3.29)   ;

% Text Node
\draw (135.48,67.54) node [anchor=north west][inner sep=0.75pt]    {$p_{1} ,\ a_{1}$};
% Text Node
\draw (145.39,220.48) node [anchor=north west][inner sep=0.75pt]    {$p_{2} ,\alpha _{2}$};
% Text Node
\draw (235.25,220.34) node [anchor=north west][inner sep=0.75pt]    {$p_{4} ,\alpha _{4}$};
% Text Node
\draw (240.79,84.31) node [anchor=north west][inner sep=0.75pt]    {$p_{3} ,\alpha _{3}$};
% Text Node
\draw (370.36,104.35) node [anchor=north west][inner sep=0.75pt]    {$p_{1} ,a_{1}$};
% Text Node
\draw (531.88,209.91) node [anchor=north west][inner sep=0.75pt]    {$p_{4} ,\alpha _{4}$};
% Text Node
\draw (403.24,208.87) node [anchor=north west][inner sep=0.75pt]    {$p_{2} ,\alpha _{2}$};
% Text Node
\draw (541.18,99.29) node [anchor=north west][inner sep=0.75pt]    {$p_{3} ,\alpha _{3}$};
% Text Node
\draw (672.64,59.67) node [anchor=north west][inner sep=0.75pt]    {$p_{1} ,\ a_{1}$};
% Text Node
\draw (781.22,120.66) node [anchor=north west][inner sep=0.75pt]    {$p_{3} ,\alpha _{3}$};
% Text Node
\draw (774.33,201.53) node [anchor=north west][inner sep=0.75pt]    {$p_{4} ,\alpha _{4}$};
% Text Node
\draw (635.58,216.76) node [anchor=north west][inner sep=0.75pt]    {$p_{2} ,\alpha _{2}$};

\end{tikzpicture}
\caption{ 4-point amplitudes with gluon as an internal propagator}
  \label{fig: 4-point amplitudes with gluon as an internal propagator}
\end{figure}
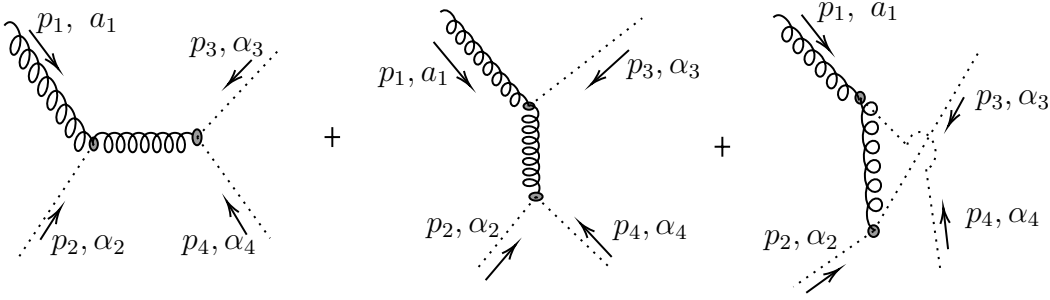

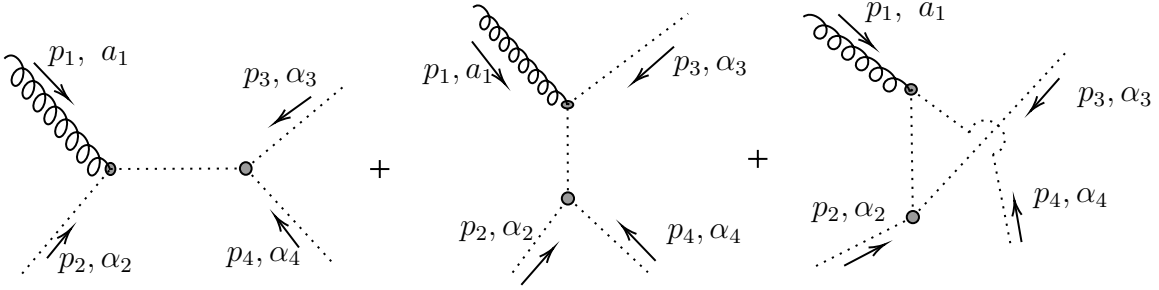
\begin{figure}

\tikzset{every picture/.style={line width=0.75pt}} %set default line width to 0.75pt        

\begin{tikzpicture}[x=0.55pt,y=0.55pt,yscale=-1,xscale=1]

%Flowchart: Connector [id:dp12247538475868958] 
\draw  [fill={rgb, 255:red, 155; green, 155; blue, 155 }  ,fill opacity=1 ] (169.06,173.1) .. controls (169.06,170.8) and (170.56,168.93) .. (172.42,168.93) .. controls (174.27,168.93) and (175.77,170.8) .. (175.77,173.1) .. controls (175.77,175.39) and (174.27,177.26) .. (172.42,177.26) .. controls (170.56,177.26) and (169.06,175.39) .. (169.06,173.1) -- cycle ;
%Shape: Spring [id:dp20066958342649854] 
\draw   (99.43,97.15) .. controls (101.85,94.55) and (105.69,93.6) .. (109.44,98.01) .. controls (116.94,106.82) and (109.11,119.42) .. (105.36,115.02) .. controls (101.61,110.61) and (109.44,98.01) .. (116.94,106.82) .. controls (124.44,115.63) and (116.61,128.23) .. (112.86,123.83) .. controls (109.11,119.42) and (116.94,106.82) .. (124.44,115.63) .. controls (131.94,124.44) and (124.11,137.04) .. (120.36,132.64) .. controls (116.61,128.23) and (124.44,115.63) .. (131.94,124.44) .. controls (139.44,133.25) and (131.61,145.85) .. (127.86,141.45) .. controls (124.11,137.04) and (131.94,124.44) .. (139.44,133.25) .. controls (146.94,142.06) and (139.11,154.66) .. (135.36,150.26) .. controls (131.61,145.85) and (139.44,133.25) .. (146.94,142.06) .. controls (154.44,150.87) and (146.61,163.47) .. (142.86,159.07) .. controls (139.11,154.66) and (146.94,142.06) .. (154.44,150.87) .. controls (161.94,159.68) and (154.11,172.28) .. (150.36,167.88) .. controls (146.61,163.47) and (154.44,150.87) .. (161.94,159.68) .. controls (169.44,168.49) and (161.61,181.09) .. (157.86,176.69) .. controls (154.11,172.28) and (161.94,159.68) .. (169.44,168.49) .. controls (170.86,170.15) and (171.73,171.95) .. (172.18,173.75) ;
%Straight Lines [id:da08167655370686389] 
\draw  [dash pattern={on 0.84pt off 2.51pt}]  (111.67,245.09) -- (153.48,195.54) -- (172.42,173.1) ;
%Straight Lines [id:da8133190598692511] 
\draw  [dash pattern={on 0.84pt off 2.51pt}]  (265.47,172.75) -- (286.41,195.52) -- (325.05,237.51) ;
%Straight Lines [id:da43307714979114054] 
\draw  [dash pattern={on 0.84pt off 2.51pt}]  (265.47,172.75) -- (332.45,116.54) ;
%Straight Lines [id:da8531460018688479] 
\draw    (119.75,99.97) -- (142.2,123.52) ;
\draw [shift={(143.58,124.97)}, rotate = 226.37] [color={rgb, 255:red, 0; green, 0; blue, 0 }  ][line width=0.75]    (10.93,-3.29) .. controls (6.95,-1.4) and (3.31,-0.3) .. (0,0) .. controls (3.31,0.3) and (6.95,1.4) .. (10.93,3.29)   ;
%Straight Lines [id:da24030518907483578] 
\draw    (128.79,234.5) -- (143.96,215.84) ;
\draw [shift={(145.23,214.29)}, rotate = 129.11] [color={rgb, 255:red, 0; green, 0; blue, 0 }  ][line width=0.75]    (10.93,-3.29) .. controls (6.95,-1.4) and (3.31,-0.3) .. (0,0) .. controls (3.31,0.3) and (6.95,1.4) .. (10.93,3.29)   ;
%Straight Lines [id:da5129840254149928] 
\draw    (302.06,227.12) -- (286.83,207.1) ;
\draw [shift={(285.62,205.51)}, rotate = 52.74] [color={rgb, 255:red, 0; green, 0; blue, 0 }  ][line width=0.75]    (10.93,-3.29) .. controls (6.95,-1.4) and (3.31,-0.3) .. (0,0) .. controls (3.31,0.3) and (6.95,1.4) .. (10.93,3.29)   ;
%Straight Lines [id:da24444216793378015] 
\draw    (310.41,123.37) -- (287.11,140.16) ;
\draw [shift={(285.49,141.33)}, rotate = 324.23] [color={rgb, 255:red, 0; green, 0; blue, 0 }  ][line width=0.75]    (10.93,-3.29) .. controls (6.95,-1.4) and (3.31,-0.3) .. (0,0) .. controls (3.31,0.3) and (6.95,1.4) .. (10.93,3.29)   ;
\draw   (351,173.14) -- (364.24,173.14)(357.62,166.95) -- (357.62,179.32) ;
\draw   (611,166.14) -- (624.24,166.14)(617.62,159.95) -- (617.62,172.32) ;
%Flowchart: Connector [id:dp8421212218608654] 
\draw  [fill={rgb, 255:red, 155; green, 155; blue, 155 }  ,fill opacity=1 ] (719,118.31) .. controls (719,116.3) and (720.79,114.67) .. (723,114.67) .. controls (725.21,114.67) and (727,116.3) .. (727,118.31) .. controls (727,120.32) and (725.21,121.95) .. (723,121.95) .. controls (720.79,121.95) and (719,120.32) .. (719,118.31) -- cycle ;
%Shape: Spring [id:dp4771394125569858] 
\draw   (648.96,68.39) .. controls (651.32,66.52) and (655.48,66.02) .. (660.25,69.66) .. controls (669.79,76.94) and (662.7,86.22) .. (657.93,82.58) .. controls (653.16,78.94) and (660.25,69.66) .. (669.79,76.94) .. controls (679.33,84.22) and (672.24,93.5) .. (667.47,89.86) .. controls (662.7,86.22) and (669.79,76.94) .. (679.33,84.22) .. controls (688.87,91.5) and (681.78,100.78) .. (677.01,97.14) .. controls (672.24,93.5) and (679.33,84.22) .. (688.87,91.5) .. controls (698.4,98.78) and (691.32,108.06) .. (686.55,104.42) .. controls (681.78,100.78) and (688.87,91.5) .. (698.4,98.78) .. controls (707.94,106.06) and (700.86,115.34) .. (696.09,111.7) .. controls (691.32,108.06) and (698.4,98.78) .. (707.94,106.06) .. controls (717.48,113.34) and (710.4,122.62) .. (705.63,118.98) .. controls (700.86,115.34) and (707.94,106.06) .. (717.48,113.34) .. controls (720.31,115.5) and (721.68,117.84) .. (722.08,119.96) ;
%Straight Lines [id:da61836109350637] 
\draw  [dash pattern={on 0.84pt off 2.51pt}]  (724.25,206.2) -- (829.25,93.1) ;
%Shape: Arc [id:dp6456078236872075] 
\draw  [draw opacity=0][dash pattern={on 0.84pt off 2.51pt}] (763.23,148.06) .. controls (763.51,146.64) and (764.07,145.27) .. (764.95,144.05) .. controls (768.57,138.98) and (776.12,138.18) .. (781.81,142.25) .. controls (787.5,146.31) and (789.18,153.72) .. (785.56,158.78) .. controls (784.04,160.9) and (781.84,162.27) .. (779.37,162.86) -- (775.25,151.41) -- cycle ; \draw  [dash pattern={on 0.84pt off 2.51pt}] (763.23,148.06) .. controls (763.51,146.64) and (764.07,145.27) .. (764.95,144.05) .. controls (768.57,138.98) and (776.12,138.18) .. (781.81,142.25) .. controls (787.5,146.31) and (789.18,153.72) .. (785.56,158.78) .. controls (784.04,160.9) and (781.84,162.27) .. (779.37,162.86) ;  
%Straight Lines [id:da35534389765233176] 
\draw  [dash pattern={on 0.84pt off 2.51pt}]  (723,118.31) -- (763.23,148.06) ;
%Straight Lines [id:da7295535085454792] 
\draw  [dash pattern={on 0.84pt off 2.51pt}]  (779.37,162.86) -- (792,227.95) ;
%Straight Lines [id:da21308133299730447] 
\draw  [dash pattern={on 0.84pt off 2.51pt}]  (724.25,206.2) -- (655.75,240.84) ;
%Straight Lines [id:da09652685471176092] 
\draw    (673,66.67) -- (695.53,87.59) ;
\draw [shift={(697,88.95)}, rotate = 222.88] [color={rgb, 255:red, 0; green, 0; blue, 0 }  ][line width=0.75]    (10.93,-3.29) .. controls (6.95,-1.4) and (3.31,-0.3) .. (0,0) .. controls (3.31,0.3) and (6.95,1.4) .. (10.93,3.29)   ;
%Straight Lines [id:da10780674032942761] 
\draw    (825,110.67) -- (806.32,131.9) ;
\draw [shift={(805,133.4)}, rotate = 311.34] [color={rgb, 255:red, 0; green, 0; blue, 0 }  ][line width=0.75]    (10.93,-3.29) .. controls (6.95,-1.4) and (3.31,-0.3) .. (0,0) .. controls (3.31,0.3) and (6.95,1.4) .. (10.93,3.29)   ;
%Straight Lines [id:da32391064261720803] 
\draw    (799,223.4) -- (794.34,196.37) ;
\draw [shift={(794,194.4)}, rotate = 80.22] [color={rgb, 255:red, 0; green, 0; blue, 0 }  ][line width=0.75]    (10.93,-3.29) .. controls (6.95,-1.4) and (3.31,-0.3) .. (0,0) .. controls (3.31,0.3) and (6.95,1.4) .. (10.93,3.29)   ;
%Straight Lines [id:da7037132177465658] 
\draw    (677,239.67) -- (703.23,225.88) ;
\draw [shift={(705,224.95)}, rotate = 152.27] [color={rgb, 255:red, 0; green, 0; blue, 0 }  ][line width=0.75]    (10.93,-3.29) .. controls (6.95,-1.4) and (3.31,-0.3) .. (0,0) .. controls (3.31,0.3) and (6.95,1.4) .. (10.93,3.29)   ;
%Flowchart: Connector [id:dp5969556897546876] 
\draw  [fill={rgb, 255:red, 155; green, 155; blue, 155 }  ,fill opacity=1 ] (486.68,126.29) .. controls (488.87,126.27) and (490.66,127.41) .. (490.68,128.83) .. controls (490.71,130.25) and (488.95,131.41) .. (486.76,131.43) .. controls (484.57,131.44) and (482.78,130.3) .. (482.75,128.89) .. controls (482.73,127.47) and (484.49,126.3) .. (486.68,126.29) -- cycle ;
%Shape: Spring [id:dp5463428106211838] 
\draw   (421.84,61.91) .. controls (423.76,60.07) and (426.92,59.62) .. (430.24,63.31) .. controls (436.9,70.71) and (430.9,79.93) .. (427.58,76.23) .. controls (424.25,72.54) and (430.24,63.31) .. (436.9,70.71) .. controls (443.55,78.11) and (437.55,87.33) .. (434.23,83.63) .. controls (430.9,79.93) and (436.9,70.71) .. (443.55,78.11) .. controls (450.2,85.51) and (444.21,94.73) .. (440.88,91.03) .. controls (437.55,87.33) and (443.55,78.11) .. (450.2,85.51) .. controls (456.85,92.91) and (450.86,102.13) .. (447.53,98.43) .. controls (444.21,94.73) and (450.2,85.51) .. (456.85,92.91) .. controls (463.5,100.31) and (457.51,109.53) .. (454.18,105.83) .. controls (450.86,102.13) and (456.85,92.91) .. (463.5,100.31) .. controls (470.15,107.7) and (464.16,116.92) .. (460.84,113.22) .. controls (457.51,109.53) and (463.5,100.31) .. (470.15,107.7) .. controls (476.81,115.1) and (470.81,124.32) .. (467.49,120.62) .. controls (464.16,116.92) and (470.15,107.7) .. (476.81,115.1) .. controls (483.46,122.5) and (477.47,131.72) .. (474.14,128.02) .. controls (470.81,124.32) and (476.81,115.1) .. (483.46,122.5) .. controls (485.4,124.66) and (486.27,126.98) .. (486.43,129.09) ;
%Straight Lines [id:da3978147039463672] 
\draw  [dash pattern={on 0.84pt off 2.51pt}]  (491.87,193.35) -- (487.02,193.38) -- (477.95,205.53) -- (448.24,245.34) ;
%Straight Lines [id:da23485496062518496] 
\draw  [dash pattern={on 0.84pt off 2.51pt}]  (487.67,193.35) -- (542.03,244.2) ;
%Straight Lines [id:da19555518291727525] 
\draw    (559.98,88.9) -- (533.7,112.05) ;
\draw [shift={(532.2,113.37)}, rotate = 318.62] [color={rgb, 255:red, 0; green, 0; blue, 0 }  ][line width=0.75]    (10.93,-3.29) .. controls (6.95,-1.4) and (3.31,-0.3) .. (0,0) .. controls (3.31,0.3) and (6.95,1.4) .. (10.93,3.29)   ;
%Straight Lines [id:da29871425060300183] 
\draw    (454.97,252.83) -- (475.91,228.82) ;
\draw [shift={(477.22,227.32)}, rotate = 131.09] [color={rgb, 255:red, 0; green, 0; blue, 0 }  ][line width=0.75]    (10.93,-3.29) .. controls (6.95,-1.4) and (3.31,-0.3) .. (0,0) .. controls (3.31,0.3) and (6.95,1.4) .. (10.93,3.29)   ;
%Straight Lines [id:da8221557196653658] 
\draw    (547.33,236.31) -- (525.6,213.63) ;
\draw [shift={(524.21,212.18)}, rotate = 46.23] [color={rgb, 255:red, 0; green, 0; blue, 0 }  ][line width=0.75]    (10.93,-3.29) .. controls (6.95,-1.4) and (3.31,-0.3) .. (0,0) .. controls (3.31,0.3) and (6.95,1.4) .. (10.93,3.29)   ;
%Straight Lines [id:da25328898878025485] 
\draw  [dash pattern={on 0.84pt off 2.51pt}]  (486.72,128.86) -- (569.68,66.16) ;
%Straight Lines [id:da8837949847116935] 
\draw    (421,85.3) -- (442.61,113.09) ;
\draw [shift={(443.84,114.66)}, rotate = 232.13] [color={rgb, 255:red, 0; green, 0; blue, 0 }  ][line width=0.75]    (10.93,-3.29) .. controls (6.95,-1.4) and (3.31,-0.3) .. (0,0) .. controls (3.31,0.3) and (6.95,1.4) .. (10.93,3.29)   ;
%Straight Lines [id:da9899208471932353] 
\draw  [dash pattern={on 0.84pt off 2.51pt}]  (486.76,131.42) -- (487.02,193.38) ;
%Shape: Circle [id:dp9135718381221207] 
\draw  [fill={rgb, 255:red, 155; green, 155; blue, 155 }  ,fill opacity=1 ] (482.82,193.38) .. controls (482.82,191.06) and (484.7,189.18) .. (487.02,189.18) .. controls (489.34,189.18) and (491.22,191.06) .. (491.22,193.38) .. controls (491.22,195.7) and (489.34,197.58) .. (487.02,197.58) .. controls (484.7,197.58) and (482.82,195.7) .. (482.82,193.38) -- cycle ;
%Straight Lines [id:da6945045715596146] 
\draw  [dash pattern={on 0.84pt off 2.51pt}]  (723,118.31) -- (724.25,206.2) ;
%Shape: Circle [id:dp18661315214642815] 
\draw  [fill={rgb, 255:red, 155; green, 155; blue, 155 }  ,fill opacity=1 ] (720.05,206.2) .. controls (720.05,203.88) and (721.93,202) .. (724.25,202) .. controls (726.57,202) and (728.45,203.88) .. (728.45,206.2) .. controls (728.45,208.52) and (726.57,210.4) .. (724.25,210.4) .. controls (721.93,210.4) and (720.05,208.52) .. (720.05,206.2) -- cycle ;
%Straight Lines [id:da30167208878009444] 
\draw  [dash pattern={on 0.84pt off 2.51pt}]  (172.42,173.1) -- (265.47,172.75) ;
%Shape: Ellipse [id:dp5882005101333959] 
\draw  [fill={rgb, 255:red, 155; green, 155; blue, 155 }  ,fill opacity=1 ] (261.45,172.75) .. controls (261.45,170.36) and (263.25,168.42) .. (265.47,168.42) .. controls (267.68,168.42) and (269.48,170.36) .. (269.48,172.75) .. controls (269.48,175.14) and (267.68,177.08) .. (265.47,177.08) .. controls (263.25,177.08) and (261.45,175.14) .. (261.45,172.75) -- cycle ;

% Text Node
\draw (127.87,86.84) node [anchor=north west][inner sep=0.75pt]    {$p_{1} ,\ a_{1}$};
% Text Node
\draw (134.82,229.36) node [anchor=north west][inner sep=0.75pt]    {$p_{2} ,\alpha _{2}$};
% Text Node
\draw (250.46,224.67) node [anchor=north west][inner sep=0.75pt]    {$p_{4} ,\alpha _{4}$};
% Text Node
\draw (261.98,101.54) node [anchor=north west][inner sep=0.75pt]    {$p_{3} ,\alpha _{3}$};
% Text Node
\draw (690,56.07) node [anchor=north west][inner sep=0.75pt]    {$p_{1} ,\ a_{1}$};
% Text Node
\draw (836,115.07) node [anchor=north west][inner sep=0.75pt]    {$p_{3} ,\alpha _{3}$};
% Text Node
\draw (805.5,184.21) node [anchor=north west][inner sep=0.75pt]    {$p_{4} ,\alpha _{4}$};
% Text Node
\draw (653,194.07) node [anchor=north west][inner sep=0.75pt]    {$p_{2} ,\alpha _{2}$};
% Text Node
\draw (385.67,99.07) node [anchor=north west][inner sep=0.75pt]    {$p_{1} ,a_{1}$};
% Text Node
\draw (553.49,208.98) node [anchor=north west][inner sep=0.75pt]    {$p_{4} ,\alpha _{4}$};
% Text Node
\draw (410.88,207.28) node [anchor=north west][inner sep=0.75pt]    {$p_{2} ,\alpha _{2}$};
% Text Node
\draw (557.81,93.83) node [anchor=north west][inner sep=0.75pt]    {$p_{3} ,\alpha _{3}$};

\end{tikzpicture}

\caption{ 4-point amplitudes with scalar as an internal propagator}
  \label{fig: 4-point amplitudes with scalar as an internal propagator}
\end{figure}

For simplicity, we restrict our attention to the soft factorisation of the four-point amplitudes of three scalars and one positive helicity gluon. This will be sufficient to show the corrections in the leading soft gluon theorem. At the tree level, there are two classes of Feynman diagrams.
\begin{enumerate}
    \item Diagrams with an internal gluon propagator (Fig \ref{fig: 4-point amplitudes with gluon as an internal propagator}). We denote the 4-point amplitude of this class as $M_4^g\left( 1^{+,a_1},2^{\alpha_2}_\phi,3^{\alpha_3}_\phi,4^{\alpha_4}_\phi \right)$.
    \item Diagrams with an internal scalar propagator (Fig \ref{fig: 4-point amplitudes with scalar as an internal propagator}). We denote the 4-point amplitude of this class as $M_4^\phi\left( 1^{+,a_1},2^{\alpha_2}_\phi,3^{\alpha_3}_\phi,4^{\alpha_4}_\phi \right)$.
\end{enumerate}
We begin with the diagrams in class~1, where the internal propagator is a gluon. This class includes the \(s\)-, \(t\)-, and \(u\)-channel contributions. The 4-point tree level amplitude is given by,
\begin{eqnarray}
    \begin{gathered}
      M_4^g\left( 1^{+,a_1},2^{\alpha_2}_\phi,3^{\alpha_3}_\phi,4^{\alpha_4}_\phi \right) = \varepsilon^+_{\mu_1}(p_1)M_4^{g,\mu_1}\left( 1^{+,a_1},2^{\alpha_2}_\phi,3^{\alpha_3}_\phi,4^{\alpha_4}_\phi \right)\\[4pt]
      = \varepsilon^+_{\mu_1}(p_1) \left( M^{g,\mu_1}_{4,s} + M^{g,\mu_1}_{4,t} + M^{g,\mu_1}_{4,u} \right)\,, 
    \end{gathered}
\end{eqnarray}
where
\begin{equation}
\begin{gathered}
    M^{g,\mu_1}_{4,s}=-2ig C^{\alpha_2a_1 a'}\Big(-p_1.(p_1+p_2)\eta^{\mu_1\mu'}+(p_1+p_2)^{\mu_1}p_1^{\mu'}\Big)\frac{i\delta^{a' b'}\eta_{\mu'\nu'}}{(p_1+p_2)^4}i g(T^{b'}_{\mathfrak{R}})^{\alpha_3\alpha_4}(p_3-p_4)^{\nu'}\\[4pt]
   M^{g,\mu_1}_{4,t}=-2ig C^{\alpha_3a_1 a'}\Big(-p_1.(p_1+p_3)\eta^{\mu_1\mu'}+(p_1+p_3)^{\mu_1}p_1^{\mu'}\Big)\frac{i\delta^{a' b'}\eta_{\mu'\nu'}}{(p_1+p_3)^4}i g(T^{b'}_{\mathfrak{R}})^{\alpha_2\alpha_4}(p_2-p_4)^{\nu'}\\[4pt]
    M^{g,\mu_1}_{4,u}=-2ig C^{\alpha_4a_1 a'}\Big(-p_1.(p_1+p_4)\eta^{\mu_1\mu'}+(p_1+p_4)^{\mu_1}p_1^{\mu'}\Big)\frac{i\delta^{a' b'}\eta_{\mu'\nu'}}{(p_1+p_4)^4}i g(T^{b'}_{\mathfrak{R}})^{\alpha_2\alpha_3}(p_2-p_3)^{\nu'}
\end{gathered}
\end{equation}
and $\varepsilon^{+}_{\mu}(p)$ is the polarisation vector for the positive helicity gluon.

\vspace{5pt}
After taking the soft limit, $p_1 \to 0$, we will find
\begin{equation}
\begin{gathered}
\lim_{p_1 \to 0} M_4^g\left( 1^{+,a_1},2^{\alpha_2}_\phi,3^{\alpha_3}_\phi,4^{\alpha_4}_\phi \right)\Big{|}_{\text{leading}}
    = -\frac{g}{2} \sum_{i=2}^4 \frac{\varepsilon^{+}(p_1) \cdot p_i}{p_1 \cdot p_i} \mathcal{F}_i^{a_1}M_3(2^{\alpha_2}_\phi,3^{\alpha_3}_{\phi},4^{\alpha_4}_{\phi})\,,
\end{gathered}
\label{4-point-amp-g-s-s-s}
\end{equation}
where $\mathcal{F}_2^{a_1}M_3(2^{\alpha_2}_\phi,3^{\alpha_3}_{\phi},4^{\alpha_4}_{\phi}) = C^{\alpha_2 a_1 a'}M_3(2^{+,a'},3^{\alpha_3}_{\phi},4^{\alpha_4}_{\phi})$ etc. and
 $$M_3(2^{+,a'},3^{\alpha_3}_\phi,4^{\alpha_4}_{\phi})=ig (T^{a'}_{\mathfrak{R}})^{\alpha_3\alpha_4}\Big(\tilde{\varepsilon}(p_2)^+ \cdot (p_3-p_4)\Big)$$ is a 3-point amplitude. The operator $\mathcal{F}_i^{a}$ acting on the amplitude transforms a scalar into a positive helicity gluon with polarisation
\begin{equation}
    \tilde{\varepsilon}^{+,\mu}(p_i)=\frac{1}{\varepsilon^+(p_1) \cdot p_i} \left[\varepsilon^{+,\mu}(p_1) - \frac{\varepsilon^+(p_1) \cdot p_i}{p_1\cdot p_i} p_1^\mu \right]  
\end{equation}
for $i\ne 1$. Note that this choice is a reasonable one for polarisation vector of a gluon with null momentum $p_i$ as it satisfies $p_i \cdot \tilde{\varepsilon}^{+}(p_i) =0$ and $\tilde{\varepsilon}^{+}(p_i) \cdot \tilde{\varepsilon}^{+}(p_i) =0$ provided $p_1 \cdot \varepsilon^{+} (p_1) = \varepsilon^{+} (p_1)\cdot \varepsilon^{+}(p_1) =0$ which we have assumed.

Let us now consider the diagrams (Fig \ref{fig: 4-point amplitudes with scalar as an internal propagator}) in class 2, where the internal propagator is a scalar.  The full amplitude is obtained by adding the $s$, $t$ and $u$-channel diagrams. After taking the soft limit $p_1 \to 0 $ in the full 4-point amplitude of this class, we get the following result:
\begin{equation}\label{sc_s}
\begin{aligned}
\lim_{p_1 \to 0} M^\phi_4(1^{+,a_1},2^{\alpha_2}_\phi,3^{\alpha_3}_\phi,4^{\alpha_4}_\phi)\Big{|}_{\text{leading}}
= -g \sum_{i=2}^4 \frac{\varepsilon^+(p_1) \cdot p_i}{p_1 \cdot p_i} \left(T^{a_1}_{\mathfrak{R},i}\right) M_3(2^{\alpha_2}_\phi,3^{\alpha_3}_\phi,4^{\alpha_4}_\phi)\,,
 \end{aligned}
\end{equation}
where $\left(T^{a_1}_{\mathfrak{R},2}\right) M_3(2^{\alpha_2}_\phi,3^{\alpha_3}_\phi,4^{\alpha_4}_\phi)=\left(T^{a_1}_{\mathfrak{R},2}\right)^{\alpha_i \alpha'} M_3(2^{\alpha'}_\phi,3^{\alpha_3}_\phi,4^{\alpha_4}_\phi)$ etc. Combining equations \eqref{4-point-amp-g-s-s-s} and \eqref{sc_s}, we get the leading soft gluon theorem for the full tree-level 4-point amplitude (including both the propagators). This is given by,
\begin{equation} \label{lead_soft_gluon}
\begin{aligned}
\lim_{p_1 \to 0}M_4(1^{+,a_1},2^{\alpha_2}_\phi,3^{\alpha_3}_\phi,4^{\alpha_4}_\phi)\Big{|}_{\text{leading}}
   = -g \sum_{i=2}^4 \frac{\varepsilon^+(p_1) \cdot p_i}{p_1 \cdot p_i} \left[\left(T^{a_1}_{\mathfrak{R},i}\right) + \frac{1}{2}\mathcal{F}_i^{a_1} \right] M_3(2^{\alpha_2}_\phi,3^{\alpha_3}_{\phi},4^{\alpha_4}_{\phi})\,.
\end{aligned}
\end{equation}
Examining \eqref{lead_soft_gluon}, we observe that, if we make a positive-helicity gluon soft in a 4-point amplitude in $(DF)^2$ theory, then at the leading order we get the standard soft factorisation and a correction term. The correction term, though it factorises into a 3-point amplitude, one of the external scalars in that amplitude effectively transforms into a positive-helicity gluon. By analysing the propagators and the three-point vertices, one can check that there is no particle change from a positive helicity gluon to a scalar at the leading order in the soft expansion of the gluon momentum. This indicates that, in the \(4\)-derivative gauge theory (such as the \(DF^2\) theory), the standard leading-order soft theorem for gluons is modified by additional contributions that involve particle transitions from a scalar to a gluon within the amplitude. 
Now, after Mellin transformation, one can write the leading soft gluon theorem \eqref{lead_soft_gluon} as the Ward identity of the leading soft gluon current for a positive helicity gluon on the celestial sphere, given by
\begin{equation}\label{ward_gl}
    \left< \mathcal{R}^{1,a_1}_0(z_1)  \prod_{i=2}^4 \phi^{\alpha_i}_{\Delta_i}(z_i,\bar z_i)\right> = - g \sum_{i=2}^4 \frac{\left(T^{a_1}_{\mathfrak{R},i}\right) + \frac{1}{2}\mathcal{F}_i^{a_1} }{z_1 - z_i}  \left< \prod_{i=2}^4 \phi^{\alpha_i}_{\Delta_i}(z_i,\bar z_i)\right>\,,
\end{equation}
where the leading soft gluon current for positive helicity gluon is defined by, $  \mathcal{R}^{1, a}_0(z) = \lim_{\Delta\to 1}(\Delta-1)\mathcal{O}^{+,a}_{\Delta}(z,\bar z)$. $\mathcal{O}^{+,a}_{\Delta}(z,\bar z), \phi^{\alpha_i}_{\Delta_i}(z_i,\bar z_i)$ are the celestial primary operators corresponding to the positive helicity gluon and $i$-th scalar in the bulk, respectively. The actions of the operators, $\left(T^{a_1}_{\mathfrak{R},i}\right), \mathcal{F}_i^{a_1} $ on the scalar primary operator are defined as $\left(T^{a_1}_{\mathfrak{R},i}\right) \phi^{\alpha_i}_{\Delta_i}(z_i,\bar z_i) = \left(T^{a_1}_{\mathfrak{R},i}\right)^{\alpha_i \alpha'} \phi^{\alpha'}_{\Delta_i}(z_i,\bar z_i)  , \mathcal{F}_i^{a_1} \phi^{\alpha_i}_{\Delta_i}(z_i,\bar z_i) = C^{\alpha_i a_1 a'} \mathcal{O}^{+,a'}_{\Delta_i}(z_i,\bar z_i)$, respectively. From \eqref{ward_gl}, one can read out the OPE between the leading soft gluon current for the positive helicity gluon and a scalar primary operator
\begin{equation}\label{ope_phi}
 \mathcal{R}^{1,a}_0(z)  \phi^{\alpha}_{\Delta}(w,\bar w) = g \left[\frac{ \left(T^{a}_{\mathfrak{R}}\right)^{\alpha \alpha'} \phi^{\alpha'}_{\Delta}(w,\bar w)  }{z-w} + \frac{1}{2}\frac{C^{\alpha a a'} \mathcal{O}^{+,a'}_{\Delta}(w,\bar w)}{z-w}\right]  + \cdots\,.
\end{equation}
Working with a 4-point amplitude with two positive helicity gluons in the external state, one can obtain the following OPE between $R^{1,a}_0(z)$ and a gluon primary operator
\begin{equation}\label{ope_gl}
    \mathcal{R}^{1,a}_0(z)  \mathcal{O}^{+,b}_{\Delta}(w,\bar w) = -i g \frac{f^{ a b c} \mathcal{O}^{+,c}_{\Delta}(w,\bar w)}{z-w}  + \cdots\,.
\end{equation}
This is the same OPE that we obtain in two-derivative gauge theory. Now, using the Jacobi identity \eqref{jacobi_id}, OPEs \eqref{ope_phi}, \eqref{ope_gl} and the following identities \cite{Johansson:2017srf},
\begin{equation}
    \begin{split}
    \left(T^a_{\mathfrak{R}}\right)^{\alpha\gamma} \left(T^b_{\mathfrak{R}}\right)^{\gamma\beta} -  \left(T^b_{\mathfrak{R}}\right)^{\alpha\gamma} \left(T^a_{\mathfrak{R}}\right)^{\gamma\beta} &= if^{abc}\left( T^c_{\mathfrak{R}}\right)^{\alpha\beta}\,,\\[4pt]
    f^{bae} C^{\alpha e c} + f^{cae}C^{\alpha b e} &= i \left( T^a_{\mathfrak{R}}\right)^{\alpha\beta}C^{\beta b c}\,.
    \end{split}
\end{equation}
One can derive the following mode algebra (again, up to a central term) for the leading soft gluon current associated with a positive helicity gluon:
\begin{eqnarray}
    \left[ \mathcal{R}^{1,a}_{m,0}, \mathcal{R}^{1,b}_{n,0}\right] = - i f^{abc} \mathcal{R}^{1,c}_{m+n,0}.
\end{eqnarray}
Thus, just as in the case of subleading soft theorems in the BW-theory case, although the representation of the leading soft gluon operator is modified, the commutation relations between two leading soft currents remain identical to those in the MHV gluon scattering in Yang-Mills theories \cite{Banerjee:2020vnt}.

\bibliographystyle{}
%\bibliography{aps01}
\providecommand{\href}[2]{#2}\begingroup\raggedright

\end{document}